\documentclass[onecolumn,prd,floatfix,showpacs,preprintnumbers,nofootinbib,%
superscriptaddress]{revtex4-2}
\usepackage{mathrsfs}
\usepackage{amssymb}	
\usepackage{amsmath}
\usepackage{graphicx}
\usepackage{subdepth}
\usepackage{xcolor}
\usepackage{slashed}
\definecolor{lcolor}{rgb}{0.5,0,0}
\definecolor{citcolor}{rgb}{0,0.3,0.0}

\usepackage[breaklinks,colorlinks,urlcolor=blue,citecolor=citcolor,linkcolor=lcolor]{hyperref}

\usepackage{tikz}

\newcommand{\Pt}{{\mathbf{P}}}

\newcommand{\rt}{{\mathbf{r}}}
\newcommand{\xt}{{\mathbf{x}}}
\newcommand{\bt}{{\mathbf{b}}}
\newcommand{\yt}{{\mathbf{y}}}

\newcommand{\pt}{{\mathbf{p}}}
\newcommand{\qt}{{\mathbf{q}}}
\newcommand{\kt}{{\mathbf{k}}}
\newcommand{\ktpzero}{{\mathbf{k}_{0'}}}

\newcommand{\Rt}{{\mathbf{R}}}

\newcommand{\Kt}{\mathbf{K}}

\newcommand{\Lt}{\mathbf{L}}

\newcommand{\ed}{\mathrm{ED}}

\newcommand{\epst}{\boldsymbol{\varepsilon}}

\newcommand{\kvec}{{\hat{k}}}

\newcommand{\qvec}{{\hat{q}}}

\newcommand{\xvec}{{\hat{x}}}

\newcommand{\lo}{{\textnormal{LO}}}
\newcommand{\nlo}{{\textnormal{NLO}}}

\newcommand{\epsl}{{\varepsilon\!\!\!/}}

\newcommand{\ksl}{{k\!\!\!/}}

\newcommand{\dk}{{\widetilde{\mathrm{d} k}}}

\newcommand{\dkpzero}{{\widetilde{\mathrm{d} k_{0'}}}}
\newcommand{\dkppzero}{{\widetilde{\mathrm{d} k_{0''}}}}
\newcommand{\dkpone}{{\widetilde{\mathrm{d} k_{1'}}}}

\newcommand{\ud}{\, \mathrm{d}}

\newcommand{\nc}{{N_\mathrm{c}}}

\newcommand{\cf}{C_\mathrm{F}}

\newcommand{\nr}[1]{(\ref{#1})}

\newcommand{\qs}{Q_\mathrm{s}}

\newcommand{\as}{\alpha_{\mathrm{s}}}

\newcommand{\fig}{Fig.~}
\newcommand{\figs}{Figs.~}
\newcommand{\eq}{Eq.~}

\newcommand{\eqs}{Eqs.~}

\newcommand{\xbj}{{x}}

\newcounter{diag}
\newcommand{\namediag}[1]{\refstepcounter{diag} \thediag \label{#1}}
\renewcommand{\thediag}{(\alph{diag})}

\begin{document}

\author{G. Beuf}
\affiliation{
Theoretical Physics Division, National Centre for Nuclear Research,
Pasteura 7, Warsaw 02-093, Poland
}

 \author{T. Lappi}
\affiliation{
Department of Physics, %
 P.O. Box 35, 40014 University of Jyv\"askyl\"a, Finland}
\affiliation{
Helsinki Institute of Physics, P.O. Box 64, 00014 University of Helsinki,
Finland}

\author{R. Paatelainen}
\affiliation{
Department of Physics, P.O. Box 64, 00014 University of Helsinki,
Finland}
\affiliation{
Helsinki Institute of Physics, P.O. Box 64, 00014 University of Helsinki,
Finland}

\title{Massive quarks in NLO dipole factorization for DIS: Longitudinal photon}
\preprint{HIP-2021-14/TH}


\begin{abstract}
In this work, we will present the first complete calculation of the one-loop longitudinal photon-to-quark-antiquark light cone wave function, with massive quarks. The quark masses are renormalized in the pole mass scheme. The result is used to calculate the next-to-leading order  correction to the high energy Deep Inelastic Scattering  longitudinal structure function on a dense target in the dipole factorization framework. For massless quarks the next-to-leading order correction was already known to be sizeable, and our result makes it possible to evaluate it also for massive quarks.
\end{abstract}

\maketitle

\tableofcontents 

\section{Introduction}

There are strong indications that high energy hadronic and nuclear collisions at present and future collider experiments can reach the regime of gluon saturation. This means that nonlinear interactions and unitarity play an important role even at short  distance scales where the QCD coupling constant is small. Such effects are expected to become increasingly important at higher energies, when the additional phase space available for radiation  leads to a growth of the gluon density. This is typically parametrized in terms of the \emph{saturation scale} $\qs(\xbj)$, where for resolution scales $Q^ 2 \sim \qs^ 2$, gluon saturation is important. The growth of the gluon density as the energy increases (i.e. as $\xbj \to 0$) leads to a growth of $\qs$. In the high energy limit, a convenient way to quantitatively analyze scattering in the saturation regime is provided by the Color Glass Condensate (CGC) effective theory formulation of high energy QCD~\cite{Gelis:2010nm}. In the CGC framework, one can understand the scattering of a dilute probe with the target hadron or nucleus in a picture of eikonal scattering~\cite{Bjorken:1970ah}. Here the gluonic structure of the target is parametrized in terms of Wilson lines, which are eikonal amplitudes for the scattering of the bare partonic constituents of the probe off the color field of the target.

Deep inelastic scattering (DIS) provides a clean and precise way to measure the partonic structure  of a hadron or a nucleus. Here the eikonal limit corresponds to the \emph{dipole picture} of DIS~\cite{Nikolaev:1990ja,Nikolaev:1991et,Mueller:1993rr,Mueller:1994jq,Mueller:1994gb}, where the virtual photon first splits to partonic constituents, which then eikonally interact with the target.   
Several fits combining the leading order dipole picture with the BK~\cite{Balitsky:1995ub,Kovchegov:1999yj,Kovchegov:1999ua} evolution equation  for the target, have achieved a good description of total small-$x$ inclusive cross sections at measured at HERA~\cite{Albacete:2009fh,Lappi:2013zma}. Noting that also calculations assuming collinear factorization start showing some tension with the data at the smallest values of $\xbj$ probed at HERA~\cite{Caola:2009iy}, it seems that total cross section measurements at HERA have provided some hints for gluon saturation, but the picture based on these results remains inconclusive. 

Fortunately, there are many avenues to improve our understanding of the QCD description of DIS in the high energy limit. One of these is to go to higher orders in perturbation theory. For the DIS process with light quarks, the dipole picture for inclusive scattering has been extended to next-to-leading order (NLO) accuracy in a series of works over the recent years~\cite{Beuf:2011xd,Balitsky:2012bs,Beuf:2016wdz,Beuf:2017bpd,Ducloue:2017ftk,Hanninen:2017ddy}. Combined with the NLO BK evolution equation~\cite{Balitsky:2008zza,Balitsky:2013fea,Kovner:2013ona,Balitsky:2014mca,Beuf:2014uia,Lappi:2015fma,Iancu:2015vea,Iancu:2015joa,Albacete:2015xza,Lappi:2016fmu,Lublinsky:2016meo,Ducloue:2019ezk,Ducloue:2019jmy} these enable a fully NLO calculation of the total light quark DIS cross section. The NLO impact factor for DIS with massless quarks, combined with approximate NLO evolution, has recently been shown to give a good description of inclusive HERA data~\cite{Beuf:2020dxl}. Another avenue is to simultaneously study more exclusive processes. Indeed, a good description of HERA data in the NLO dipole picture has recently been demonstrated in Ref.~\cite{Beuf:2020dxl}. Diffractive or exclusive DIS cross sections can provide a valuable separate experimental constraint for LO calculations~ \cite{Kowalski:2006hc,Watt:2007nr,Rezaeian:2012ji,Mantysaari:2018zdd}, and calculations for diffractive DIS processes are now also advancing to NLO accuracy~\cite{Boussarie:2014lxa,Boussarie:2016bkq,Boussarie:2016ogo,Escobedo:2019bxn}. Another important process, and the one that we will focus on here, is inclusive scattering with heavy quarks. The cross section for light quarks gets a significant contribution from ``aligned jet'' configurations of large dipoles~\cite{Mantysaari:2018nng}. In spite of gluon saturation, this brings in a systematical uncertainty to light quark inclusive cross sections in the dipole picture. The aligned jet contributions are, however, cut off by a finite quark mass. Thus inclusive heavy quark DIS cross  sections can be  more reliably perturbative probes of gluon saturation. In LO fits with BK (or JIMWLK) evolution it has not been obvious~ \cite{Albacete:2010sy,Mantysaari:2018zdd} how to achieve a good simultaneous description of the final combined heavy and light quark data from HERA~\cite{Aaron:2009aa,Abramowicz:1900rp,Abramowicz:2015mha,H1:2018flt}. It is therefore important to extend also the calculations of heavy quark inclusive DIS cross sections in the dipole picture to next to leading order accuracy in  the QCD coupling. Doing so is our purpose here. 

The calculational tool of choice for this situation is light cone perturbation theory (LCPT)~\cite{Kogut:1969xa,Bjorken:1970ah,Lepage:1980fj,Brodsky:1997de}.
While LCPT can be used to understand the partonic structure of the proton (see
\cite{Dumitru:2020gla,Dumitru:2021tvw,Dumitru:2018vpr} for recent advances in this direction), in our case we use it to quantize the virtual photon of DIS, a perturbatively controllable object. In an LCPT calculation one follows a set of diagrammatic rules to calculate the coefficients of the  expansion of an interacting theory Fock state in terms of the bare Fock states. These coefficients are known as light cone wave functions (LCWF's). Recent work~\cite{Beuf:2016wdz,Lappi:2016oup,Lublinsky:2016meo,Beuf:2017bpd,Hanninen:2017ddy,Iancu:2018hwa} has led to important technical advances in performing LCPT calculations at loop level. However, the introduction of quark masses introduces some additional issues that must be dealt with. The typical way of regularizing these recent LCPT loop calculations has been to use a cutoff for longitudinal and dimensional regularization for transverse momentum integrals. It has long been known~\cite{Mustaki:1990im,Zhang:1993is,Zhang:1993dd,Harindranath:1993de} that using such a cutoff procedure causes a problem for the fermion\footnote{Gauge boson mass renormalization and thus gauge invariance is also affected, but this is not an issue for the process we are calculating here.} mass renormalization. At a fundamental level the  problem is associated with the well-known fact that the regularization procedure should respect the symmetries of the underlying theory. In the case of LCPT, the fermion mass appears in two different places in the Hamiltonian that one quantizes. Firstly there is the free fermion term, where the ``kinetic'' mass determines the relation between the energy and momentum of a free fermion. There is also a quark mass in the quark-gluon interaction term, where the quark-gauge boson vertex consists of two parts. Out of these two the light-cone helicity conserving part is independent of the quark mass. The light-cone helicity flip term, on the other hand, is proportional to the ``vertex mass'' of the fermion. The equality of the kinetic and vertex masses is due to the rotational invariance of the underlying theory at the Lagrangian level. In LCPT, however, one first derives from the Lagrangian the Hamiltonian formulation of the theory and only then starts to quantize it. If this quantization uses a regularization procedure that breaks (3-dimensional) rotational invariance, it can happen that one needs to separately renormalize the two masses at each order in perturbation theory with an additional renormalization condition restoring rotational invariance~\cite{Burkardt:1991tj}. Both this conceptual issue, and the more complicated structure of the basic quark-gluon vertex due to the light-cone helicity~\cite{Soper:1972xc} flip-term, make the calculation of the DIS cross section for massive quarks more involved than the corresponding one for massless quarks~\cite{Beuf:2016wdz,Beuf:2017bpd,Hanninen:2017ddy}. The factorization of high energy divergences into BK evolution of the target, on the other hand, is rather orthogonal to these additional complications from the quark masses. Thus, the high energy factorization aspect of the calculation is quite similar to the massless case, and will only be discussed briefly in this work, although it of course needs to be treated carefully in order to eventually compare the calculations to experimental data.

This is the first in a set of papers, where we will fully analyze the calculation of the DIS cross section in the dipole picture to NLO accuracy with massive quarks. In this first paper we will concentrate on the case of a longitudinal virtual photon. For the longitudinal polarization the numerator algebra is simpler, making the calculation slightly less lengthy. More importantly, only the relatively simple ``propagator correction'' diagrams lead to a renormalization of the quark (kinetic) mass. Therefore, the mass renormalization in a pole scheme can be performed in a relatively straightforward way, without encountering the intricacies discussed above. Transverse photons will be addressed in a separate paper. There the helicity and polarization algebra is somewhat more complicated, and also the renormalization of both the kinetic and vertex mass needs to be addressed. Separately from these calculations, we plan to address more formal aspects of mass renormalization in LCPT, its relation to the regularization  procedure and the treatment of so called ``self-induced inertia'' or ``seagull'' diagrams ~\cite{Pauli:1985pv,Tang:1991rc,Brodsky:1997de} in more detail in  separate future work.

The rest of the paper is structured as follows: first, in Section~\ref{sec:notations}, in order to keep the paper self-contained, we give some basic background notions of light cone perturbation theory and explain the regularization approach used in this paper, although relatively briefly since this is very similar to the previous calculations in Refs.~\cite{Beuf:2016wdz,Beuf:2017bpd,Hanninen:2017ddy}. In Section~\ref{sec:dipfact} we recall how the DIS cross section is, in the dipole picture, factorized into light cone wave functions encoding the Fock state of the virtual photon, and Wilson line operators describing the interactions of these states with the target. We then derive in Sec.~\ref{sec:lo} the leading order virtual photon-to-quark-antiquark light cone wavefunction in $D$ spacetime dimensions as a warmup. In Sec.~\ref{sec:loop} we calculate the  (mass renormalized) one loop corrections to this wavefunction in momentum space, which are then transformed into mixed transverse coordinate-longitudinal momentum space in Sec.~\ref{sec:fourier}. The tree level gluon emission diagrams needed for the real corrections to the cross section are calculated in Sec.~\ref{sec:qqbg}. We then derive in Sec.~\ref{sec:xs}  the cross section from these light cone wave functions, including the subtractions needed to cancel divergences between the real and virtual contributions. Finally we summarize our result for the cross section in Sec.~\ref{sec:full} and briefly discuss future steps in Sec.~\ref{sec:conc}. Many technical details on the calculations are explained in the Appendices.

\section{Preliminaries: notation and regularization}
\label{sec:notations}


\subsection{Light cone coordinates and conventions}

This section will be rather brief, as our notations are a combination of the conventions used in  Refs.~\cite{Beuf:2016wdz,Beuf:2017bpd,Hanninen:2017ddy}. We refer the reader there for a more thorough explanation. For an arbitrary Minkowskian four-vector $x^{\mu}  = (x^0, \vec x)$ with $\vec x = (x^1,x^2,x^3)$ we define the light cone coordinates as 
\begin{equation}
x^{\mu} = (x^+,x^-,\xt),
\end{equation}
where $x^+$ is the light cone time along which the states are evolved, $x^-$ is the longitudinal coordinate, and $\xt =(x^1,x^2)$ is the transverse position with $\xt^2 = \vert \xt\vert^2 = \xt \cdot \xt$. In this paper, the spatial (in the light cone sense) three-vectors are denoted by 
$\xvec  \equiv (x^+,\xt)$. The components $x^+$ and $x^-$ are related to the Minkowski coordinates by 
\begin{equation}
x^{\pm} = \frac{1}{\sqrt{2}}(x^0 \pm x^3).
\end{equation}
The inner product of two four-vectors is 
\begin{equation}
\label{eq:metric}
x\cdot y = x^+y^- + x^-y^+ - \xt\cdot \yt, 
\end{equation}
from which one sees that $x^{\pm} = x_{\mp}$. The canonical conjugate of the longitudinal coordinate $x^-$ is the longitudinal momentum $p^+$, and the evolution in light cone time  $x^+$ is generated by the light cone energy $p^-$. With the form of the metric in \eq\nr{eq:metric}, the on-shell light cone energy becomes
\begin{equation}
p^- = \frac{\pt^2 + m^2}{2p^+}.
\end{equation}

\subsection{Regularization}

In loop calculations one needs to integrate over internal (on-shell) momenta. Here (contrary to e.g. the review~\cite{Brodsky:1997de}) we use conventional relativistic field theory conventions for the normalization. Thus, the momentum space integral measure  is given by 
\begin{equation}
\label{eq:momspacedef}
\int \dk = \int \frac{\ud^4k	}{(2\pi)^4}(2\pi)\Theta(k^0)\delta(k^2-m^2) = \int \frac{\ud k^+ \Theta(k^+)}{2k^+(2\pi)}\int\frac{\ud^2\kt}{(2\pi)^2}.
\end{equation}
In the evaluation of loop or final state phase space integrals over longitudinal and transverse momenta, we encounter divergences which have to be properly regularized. In this work, following the same procedure as the one described in Refs.~\cite{Beuf:2016wdz,Beuf:2017bpd,Hanninen:2017ddy},  we regularize the ultraviolet (UV) divergent ($\kt \rightarrow \infty$) transverse momentum integrals via dimensional regularization by evaluating them in $(D-2)$ transverse dimensions  and regularize (if needed) the soft divergences ($k^+ \rightarrow 0$) with a cutoff. We consider two variants of dimensional regularization and present our results in both schemes, checking explicitly that the final result for the cross section is scheme-independent.  These variants are the conventional dimensional regularization (CDR) scheme~\cite{Collins:1984xc} that was used in~\cite{Beuf:2016wdz,Beuf:2017bpd}, and the four dimensional helicity (FDH) scheme \cite{Bern:1991aq,Bern:2002zk} that was used in Ref.~\cite{Hanninen:2017ddy}. The precise implementation of these schemes is carefully explained in \cite{Hanninen:2017ddy}. Thus, in here, we only give an small overview of these two approaches. 

Both regularization schemes involve continuing space-time from four to $D$ dimensions, but differ in the way how they treat the momenta and the polarization vectors of unobserved and observed particles. Here, the unobserved particles are either virtual ones which circulate in internal loops or particles which are external but soft or collinear with other external particles. All the rest are observed particles. The following rules listed below will be used to compute the LCWF's in $D$ dimensions. 
\begin{itemize}
\item In the CDR scheme, the momenta and polarization vectors of the observed  and the unobserved particles are continued to $D$ spacetime dimensions. 
\item In the FDH scheme, the momenta and polarization vectors of observed particles are kept in four dimensions (i.e. observed gluons have 2 helicity states) and the momenta of unobserved particles are continued to $D > 4$. The helicities (spinor structures) of unobserved internal states are treated as $D_s$-dimensional, where $D_s > D$ at all intermediate steps in the calculation. Once all the helicity (Dirac) and Lorentz algebra is done, one sets $D_s=4$ and analytically continues to $D<4$ dimensions. 
\end{itemize}
In order to perform intermediate computations for both schemes at the same time,  we will use the following rules:
 \begin{itemize}
\item Any factor of spacetime dimension arising from the Dirac and Lorentz algebra for spin or polarization vectors should be labelled as $D_s$ and should be distinct from the dimension of the momentum vectors $D$.
\item The vertices are proportional to $(D_s-2)$-dimensional gluon polarization vectors $\epst^{\ast i}_{\lambda}$, and summing over the helicity states of gluons yields 
\begin{equation}
\sum_{\lambda}\epst^{\ast i}_{\lambda}\epst^{j}_{\lambda} = \delta^{ij}_{(D_s)},
\end{equation}
where by $\delta^{ij}_{(D_s)}$ we denote a Kronecker delta in  $(D_s-2)$-dimensional transverse space.
\item The tensoral structures resulting from  transverse momentum integrals are kept in $(D-2)$ dimensions. For example, if the integrand in the transverse momentum integral is proportional to $\kt^i\kt^j$, then the value of the dimensionally regulated integral is proportional to a $(D-2)$-dimensional Kronecker delta $\delta^{ij}_{(D)}$.
\item Since $D_s > D$, we have 
\begin{equation}
\delta^{ij}_{(D_s)}\delta^{ij}_{(D_s)} = D_s-2, \quad \delta^{ij}_{(D)}\delta^{ij}_{(D)} = D-2, \quad \delta^{ij}_{(D_s)}\delta^{ij}_{(D)}=D-2, \quad 
\delta^{ij}_{(D_s)}\delta^{jk}_{(D)}=\delta^{ik}_{(D)}.
\end{equation}
\item Since both $D>4$ and $D_s>4$ when the algebra is done, contractions of Kronecker deltas with fixed momentum $\pt$ or polarization vectors $\epst^{i}_{\lambda}$ of \emph{external} particles preserve these vectors
\begin{equation}
\delta^{ij}_{(D_s)} p^j =\delta^{ij}_{(D)} p^j = p^i, \quad 
\delta^{ij}_{(D_s)} \epst^{j}_{\lambda} =\delta^{ij}_{(D)} \epst^{j}_{\lambda} =  \epst^{i}_{\lambda}.
\end{equation}
\item Only after the spin and tensor algebra is done, one can analytically continue the obtained result to $D<4$ and take the limit $D_s \to 4$ in the FDH scheme or  $D_s \to D$ in the CDR scheme. If the calculation was done only in one of the two schemes, we would not need the notation $D_s$ at any intermediate step; this could be replaced by $4$ or $D$. Here we will, however, present the results in both schemes simultaneously, and for this it is necessary to keep $D_s$ general. 
\end{itemize}

\section{Dipole factorization for DIS: Cross section at NLO}
\label{sec:dipfact}

We use here the standard procedure where the DIS cross section is expressed in terms of the cross section of a virtual photon scattering on the hadronic target.
The virtual photon cross section  can be obtained by the optical theorem as twice the real part of the forward elastic scattering amplitude 
\begin{equation}
\sigma^{\gamma^{\ast}}_{\lambda} = 2\mathrm{Re} \left[(-i) \mathcal{M}^{\text{fwd}}_{\gamma^{\ast}_{\lambda} \rightarrow \gamma^{\ast}_{\lambda}}\right] .
\end{equation} 
where the forward elastic amplitude $\mathcal{M}^{\text{fwd}}_{\gamma^{\ast}_{\lambda} \rightarrow \gamma^{\ast}_{\lambda}} $ is defined in light cone quantization from the $S$-matrix element as~\cite{Bjorken:1970ah}
\begin{equation}
\label{eq:fwdamplitude}
{}_{i}\langle \gamma^{\ast}_{\lambda}(\qvec',Q^2)\vert (\hat{\mathcal{S}}_{E} - \mathbf{1})\vert \gamma^{\ast}_{\lambda}(\qvec,Q^2)\rangle_i = 2q^+(2\pi)\delta(q'^+ - q^+)i\mathcal{M}^{\text{fwd}}_{\gamma^{\ast}_{\lambda} \rightarrow \gamma^{\ast}_{\lambda}}.
\end{equation} 
At high energy (or, equivalently, at small Bjorken $x$) the interactions with the target described by the operator $\hat{\mathcal{S}}_{E}$ are eikonal interactions with a classical color field.  These interactions are represented by Wilson lines, which are the scattering matrix elements of bare partonic states in mixed space: transverse position -- longitudinal momentum. Thus in order to calculate the amplitude we need to develop the incoming virtual photon state in terms of these bare states. 

The state $\vert \gamma^{\ast}_{\lambda}(\qvec,Q^2)\rangle_i$ is the physical one-photon-state in the interaction picture. We start by its perturbative Fock state decomposition in momentum space, which is given by
\begin{equation}
\label{eq:FSexpansionmspace}
\begin{split}
\vert \gamma^{\ast}_{\lambda}(\qvec,Q^2)\rangle_i = \sqrt{Z_{\gamma^{\ast}}}\Biggl [\text{Non-QCD Fock states} & + \sum_{q\bar{q}~ \text{F. states}}\Psi^{\gamma^{\ast}_{\lambda}\rightarrow q\bar{q}}\vert q(\kvec_0,h_0,\alpha_0)\bar{q}(\kvec_1,h_1,\alpha_1)\rangle\\
& +  \sum_{q\bar{q}g~\text{F. states}}\Psi^{\gamma^{\ast}_{\lambda}\rightarrow q\bar{q}g}\vert q(\kvec_0,h_0,\alpha_0)\bar{q}(\kvec_1,h_1,\alpha_1)g(\kvec_2,\sigma,a)\rangle + \cdots\Biggr ],
\end{split}
\end{equation}
where $\qvec = (q^+,\qt)$, $Q$ and $\lambda$   are the spatial three momentum,  virtuality and polarization of the virtual photon, respectively. The explicit phase space sum over the quark-antiquark $(q\bar{q})$ and the quark-antiquark-gluon $(q\bar{q}g)$ Fock states are defined in $D$ dimensions as:
\begin{equation}
\begin{split}
\sum_{q\bar{q}~ \text{F. states}} & \! \! = \!
\sum_{\substack{h_0,h_1\\\lambda\\\alpha_0,\alpha_1}}
\left (\prod_{i=0}^{1}\int_{0}^{\infty} \frac{\ud k^+_i \Theta(k^+_i)}{2k^+_i(2\pi)}\right )
2\pi\delta\left(q^+ - \sum_{j=0}^{1}k^+_j\right)
\left (\prod_{i=0}^{1}\int\!\!\frac{\ud^{D-2}\kt_i}{(2\pi)^{D-2}}\right )(2\pi)^{D-2}\delta^{(D-2)}\!\left(\qt-\sum_{j=0}^{1}\kt_j\right),
\\
\sum_{q\bar{q}g~ \text{F. states}} & \! \!\!\! = \!
\sum_{\substack{h_0,h_1\\\lambda,\sigma\\\alpha_0,\alpha_1,a}}
\left (\prod_{i=0}^{2}\int_{0}^{\infty} \frac{\ud k^+_i \Theta(k^+_i)}{2k^+_i(2\pi)}\right )
2\pi\delta\left(q^+ - \sum_{j=0}^{2}k^+_j\right)
\left (\prod_{i=0}^{2}\int\!\!\frac{\ud^{D-2}\kt_i}{(2\pi)^{D-2}}\right )(2\pi)^{D-2}\delta^{(D-2)}\!\left(\qt - \sum_{j=0}^{2} \kt_j\right).
\end{split}
\end{equation}
From now on we will leave the  helicity $h_0,h_1$, polarization $\sigma,\lambda$  and color $\alpha_1,\alpha_1,a$ indices as well as the quark flavor implicit, summed over when appropriate.  While the leading order cross section is of order $\alpha_{em}$, for NLO accuracy in this context we want to calculate the  cross section  to order\footnote{Here $\alpha_{em} = e^2/(4\pi)$ and $\alpha_s=g^2/(4\pi)$ are the QED and QCD coupling constants, respectively.}  $\alpha_{em}\alpha_s$. This requires the knowledge of the quark-antiquark wave function  $\Psi^{\gamma^{\ast}_{\lambda}\rightarrow q\bar{q}}$ to one loop order, and that of the quark-antiquark-gluon one $\Psi^{\gamma^{\ast}_{\lambda}\rightarrow q\bar{q}g}$ at tree level. The non-QCD Fock states containing electromagnetic interactions via photons or leptons, higher order Fock states (represented by the dots) and the photon wave function renormalization constant $Z_{\gamma^{\ast}} = 1 + \mathcal{O}(\alpha_{em})$ can be ignored since they do not contribute at the order considered in the present calculation.

In the dipole factorization picture with eikonal scattering, we need to switch from the full momentum space representation to mixed space, in which the kinematics of a particle is described by its light cone longitudinal momentum and transverse position. In this case, the Fock state expansion in \eq\nr{eq:FSexpansionmspace} reduces to the following form
\begin{equation}
\label{eq:FSexpansionmixspace}
\begin{split}
\vert \gamma^{\ast}_{\lambda}(q^+,Q^2)\rangle_i = \sqrt{Z_{\gamma^{\ast}}}\Biggl [&\text{Non-QCD Fock states}  + \widetilde{\sum_{q\bar{q}~ \text{F. states}}}\widetilde{\Psi}^{\gamma^{\ast}_{\lambda}\rightarrow q\bar{q}}
\vert q(k^+_0,\xt_0,h_0,\alpha_0)\bar{q}(k^+_1,\xt_1,h_1,\alpha_1)\rangle\\
& +  \widetilde{\sum_{q\bar{q}g~\text{F. states}}}\widetilde{\Psi}^{\gamma^{\ast}_{\lambda}\rightarrow q\bar{q}g}\vert q(k^+_0,\xt_0,h_0,\alpha_0)\bar{q}(k^+_1,\xt_1,h_1,\alpha_1)g(k^+_2,\xt_2,\sigma,a)\rangle + \cdots\Biggr ],
\end{split}
\end{equation}
in terms of the mixed space states defined as transverse Fourier transforms
\begin{equation}
\vert q(k^+,\xt,h,\alpha)\rangle
= \int\frac{\ud^{D-2}\kt}{(2\pi)^{D-2}} e^{-i\kt \cdot \xt}
\vert q(k^+,\kt,h,\alpha)\rangle
\end{equation}
and so forth.
The phase space sums over the mixed space $q\bar q$ and $q\bar q g$ Fock states are:
\begin{equation}
\begin{split}
\widetilde{\sum_{q\bar{q}~ \text{F. states}}} & = 
\left (\prod_{i=0}^{1}  \int_{0}^{\infty} \frac{\ud k^+_i \Theta(k^+_i)}{2k^+_i(2\pi)}\right )
2\pi\delta\left(q^+ - \sum_{j=0}^{1}k^+_j\right)\int \ud^{D-2}\xt_0\int \ud^{D-2}\xt_1,
\\
\widetilde{\sum_{q\bar{q}g~ \text{F. states}}} & = 
\left (\prod_{i=0}^{2}\int_{0}^{\infty} \frac{\ud k^+_i \Theta(k^+_i)}{2k^+_i(2\pi)}\right )
2\pi\delta\left(q^+ - \sum_{j=0}^{2}k^+_j\right)\int \ud^{D-2}\xt_0\int \ud^{D-2}\xt_1\int \ud^{D-2}\xt_2  .
\end{split}
\end{equation}
The  Fourier transforms to the mixed space wavefunctions  $\widetilde{\Psi}^{\gamma^{\ast}_{\lambda}\rightarrow q\bar{q}}$ and $\widetilde{\Psi}^{\gamma^{\ast}_{\lambda}\rightarrow q\bar{q}g}$ are defined as:
\begin{eqnarray}
\label{eq:FTqqbar}
\widetilde{\Psi}^{\gamma^{\ast}_L\rightarrow q\bar{q}} &=& \int \frac{\ud^{D-2}\kt_0}{(2\pi)^{D-2}}\int \frac{\ud^{D-2}\kt_1}{(2\pi)^{D-2}} e^{i\kt_0\cdot \xt_0 + i\kt_1\cdot \xt_1 }
(2\pi)^{D-2}\delta^{(D-2)}(\qt- \kt_0 - \kt_1) \Psi^{\gamma^{\ast}_L\rightarrow q\bar{q}}, 
\\
\label{eq:FTqqbarg}
\widetilde{\Psi}^{\gamma^{\ast}_L\rightarrow q\bar{q}g} &=& \int \frac{\ud^{D-2}\kt_0}{(2\pi)^{D-2}}\int \frac{\ud^{D-2}\kt_1}{(2\pi)^{D-2}} \int \frac{\ud^{D-2}\kt_2}{(2\pi)^{D-2}}e^{i\kt_0\cdot \xt_0 + i\kt_1\cdot \xt_1 + i\kt_2\cdot \xt_2}
\\ \nonumber & & \quad \quad \quad\quad
\times (2\pi)^{D-2}\delta^{(D-2)}(\qt- \kt_0 - \kt_1-\kt_2) \Psi^{\gamma^{\ast}_L\rightarrow q\bar{q}g}. 
\end{eqnarray}
It is convenient to factorize out the dependence on the  ``center of mass''\footnote{In LCPT the ``mass'' that serves as a weight in this linear combination is the longitudinal momentum.} coordinate  and the color structure of the partonic system out from the LCWFs as: 
\begin{equation}
\label{eq:LCWFgeneral}
\begin{split}
\widetilde{\Psi}^{\gamma^{\ast}_{\lambda}\rightarrow q\bar{q}} & = \delta_{\alpha_0\alpha_1}
e^{i(\qt/q^+)\cdot \left (k^+_0\xt_0 + k^+_1\xt_1  \right )}\widetilde{\psi}^{\gamma^{\ast}_{\lambda}\rightarrow q\bar{q}},\\
\widetilde{\Psi}^{\gamma^{\ast}_{\lambda}\rightarrow q\bar{q}g} & = t^{a}_{\alpha_0\alpha_1}
e^{i(\qt/q^+)\cdot \left (k^+_0\xt_0 + k^+_1\xt_1 + k^+_2\xt_2 \right )}\widetilde{\psi}^{\gamma^{\ast}_{\lambda}\rightarrow q\bar{q}g}.
\end{split}
\end{equation}
The reduced LCWFs $\widetilde{\psi}^{\gamma^{\ast}_{\lambda}\rightarrow q\bar{q}}$ and $\widetilde{\psi}^{\gamma^{\ast}_{\lambda}\rightarrow q\bar{q}g}$ are independent of the photon transverse momentum $\qt$ and cannot depend on the absolute transverse positions of the Fock state partons, just on their differences. The color structures $\delta_{\alpha\beta}$ and $t^{a}_{\alpha\beta}$ are the only invariant SU$(\nc)$ color tensors available for the $q\bar{q}$ and the $q\bar{q}g$ Fock states, respectively. 

Using these notations, the total NLO cross section for a virtual photon scattering from a classical gluon field takes the following general form
\begin{equation}
\label{eq:crosssectionformula}
\sigma^{\gamma^{\ast}}_{\lambda} = 2\nc\widetilde{\sum_{q\bar{q}~ \text{F. states}}} \frac{1}{2q^+}
\left| \widetilde{\psi}^{\gamma^{\ast}_{\lambda}\rightarrow q\bar{q}} \right|^2 \mathrm{Re}[1-\mathcal{S}_{01}] +  2\nc\cf\widetilde{\sum_{q\bar{q}g~ \text{F. states}}} \frac{1}{2q^+}
\left| \widetilde{\psi}^{\gamma^{\ast}_{\lambda}\rightarrow q\bar{q}g} \right|^2 \mathrm{Re}[1-\mathcal{S}_{012}] + \mathcal{O}(\alpha_{em}\alpha_s^2),
\end{equation}
where we have introduced the notation\footnote{Here, $\nc$ is the number of colors, and the color factor $\cf = (\nc^2-1)/(2\nc)$.}
\begin{equation}
\begin{split}
\mathcal{S}_{01} &= 
\frac{1}{\nc} \mathrm{Tr} \left(U_F(\xt_0)U^{\dagger}_{F}(\xt_1) \right),
\\
\mathcal{S}_{012} &= 
\frac{1}{\nc\cf}\mathrm{Tr} \left(t^{b}U_F(\xt_0)t^{a}U^{\dagger}_{F}(\xt_1) \right) U_{A}(\xt_2)_{ba}, \\
\end{split}
\end{equation}
for the quark-antiquark $(\mathcal{S}_{01})$ and quark-antiquark-gluon $(\mathcal{S}_{012})$ amplitudes.  Here, the fundamental (F) and the adjoint (A) Wilson lines are defined as light-like path ordered exponentials for a classical gluon target
\begin{equation}
\begin{split}
U_{F}(\xt) & = \mathcal{P}\exp\biggl [-ig\int \ud x^{+} t^{a} A^{-}_{a}(x^{+},0,\xt) \biggr ],\\
U_{A}(\xt) & = \mathcal{P}\exp\biggl [-ig\int \ud x^{+} T^{a} A^{-}_{a}(x^{+},0,\xt) \biggr ],\\
\end{split}
\end{equation}
where $t^a$ and $T^a$ are the generators of the fundamental and adjoint representations, respectively.

\section{Leading order longitudinal photon wave function}
\label{sec:lo}

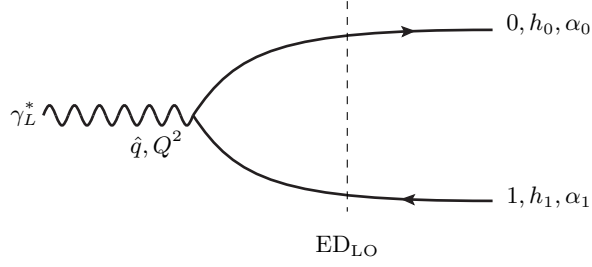
\begin{figure}[tb!]
\centerline{
\includegraphics[width=6.4cm]{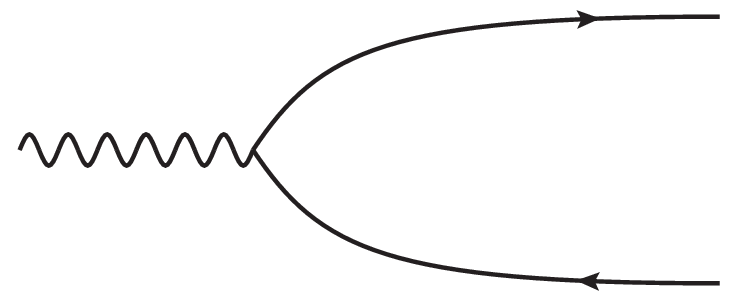}
\begin{tikzpicture}[overlay]
\draw [dashed] (-2.3,2.8) -- (-2.3,0);
\node[anchor=north] at (-2.3cm,-0.2cm) {$\ed_{\rm LO}$};
\node[anchor=west] at (-0.3,2.5) {$0,h_0,\alpha_0$};
\node[anchor=west] at (-0.3,0.2) {$1,h_1,\alpha_1$};
\node[anchor=west] at (-5.3,0.9) {$\qvec, Q^2 $};
\node[anchor=east] at (-6.3,1.3) {$\gamma^{\ast}_{L}$};
 \end{tikzpicture}
}
\rule{0pt}{1ex}
\caption{Time ordered light cone diagram (momenta flows from left to right) contributing to the longitudinal virtual photon wave function at leading order. Here, the quark(antiquark) helicity and color index are denoted as $h_0(h_1)$ and $\alpha_0(\alpha_1)$, respectively. In the vertex, the momentum is conserved $\qvec = \kvec_0 + \kvec_1$, where the spatial three momentum vectors are denoted as $\qvec= (q^+,\qt)$, $\kvec_0 = (k^{+}_0,\kt_0)$ and $\kvec_1 = (k^{+}_1,\kt_1)$, 
}
\label{fig:lovertex}
 \end{figure}

We now recall the well known leading order light cone wave function for the longitudinal virtual photon splitting into a quark antiquark dipole. The labeling of the kinematics for this process is shown in \fig\ref{fig:lovertex}.  Using the light cone perturbation theory (LCPT) rules as presented in \cite{Beuf:2016wdz, Hanninen:2017ddy}, the leading order $\gamma^{\ast}_L\rightarrow q\bar{q}$ light cone wave function can be written as 
\begin{equation}
\label{eq:LOlcwf}
\Psi^{\gamma^{\ast}_L\rightarrow q\bar{q}}_{\lo}  = \frac{\delta_{\alpha_0\alpha_1}}{\ed_{\lo}}V^{\gamma^{\ast}_L\rightarrow q\bar{q}}_{h_0;h_1}.
\end{equation}
Here, the function
\begin{equation}
\label{eq:LOvertex}
V^{\gamma^{\ast}_L\rightarrow q\bar{q}}_{h_0;h_1}= +ee_f\frac{Q}{q^+}\bar{u}(0)\gamma^+v(1)
\end{equation}
is an effective QED photon splitting vertex into a $q\bar{q}$ pair. In the physical DIS process with a longitudinal photon this is strictly speaking a part of an instantaneous interaction vertex with the lepton. However, as discussed in Refs.~\cite{Beuf:2011xd,Beuf:2016wdz}, it can in practice be treated as a separate vertex. As a remnant of this nature as a part of an instantaneous interaction, the longitudinal photon does not couple to instantanous quarks.
Following Ref.~\cite{Beuf:2016wdz}, we use a condensed notation $\bar{u}(k_0,h_0) = \bar{u}(0)$, $v(k_1,h_1) = v(1)$  etc. to shorten the expressions.   In \eq\nr{eq:LOvertex} the parameter $e_f$ is the fractional charge  of quark flavor $f$ and $e$ is the QED coupling constant. In the LO diagram, there is only one intermediate state, which is the  $q\bar{q}$ state\footnote{Recall that in the case of a LCWF computation the state at the final end of the diagram is an intermediate state with its energy denominator. It is the state that interacts with the shockwave of the target and thus not an actual external final state.}. The corresponding light cone energy denominator $\ed_{\rm LO}$ appearing in \eq\nr{eq:LOlcwf} is given by the differences of the light cone energies of the states
\begin{equation}
\label{eq:LOEDv1}
\ed_{\lo} = q^{-} - (k^{-}_0 + k^{-}_1) + i\delta. 
\end{equation}

Following the discussion of \cite{Beuf:2016wdz}, it is possible to generalize the notation of LCWF to the case of an off-shell (virtual) photon by assigning the off-shell value 
\begin{equation}
\label{eq:qminusdef}
q^{-} = \frac{\qt^2 + q^2}{2q^+},\quad \text{with}\quad q^2 = -Q^2 < 0
\end{equation}
to the light cone energy of the photon.  The virtuality of the photon in LCPT actually corresponds to the light cone energy difference between the incoming electron and the outgoing electron+photon state. The off-shell value of $q^{-}$ is then used in each light cone energy denominator appearing in the perturbative expansion of the LCWF. The quarks energies are given by the mass shell relation
\begin{equation}
\label{eq:k0minusk1minusdefs}
\begin{split}
 k^{-}_0 = \frac{\kt^2_0 + m^2}{2k^{+}_0}, \quad k^{-}_1 = \frac{\kt^2_1 + m^2}{2k^{+}_1},
\end{split}
\end{equation}
where $m$ is the quark mass. Thus, using \eqs\nr{eq:qminusdef} and \nr{eq:k0minusk1minusdefs}, the energy denominator given in \eq\nr{eq:LOEDv1} can be written as 
\begin{equation}
\label{eq:LOED}
\begin{split}
\ed_{\lo} & = \frac{\qt^2 - Q^2}{2q^+} - \biggl [\frac{\kt^2_0 + m^2}{2k^{+}_0} + \frac{\kt^2_1 + m^2}{2k^{+}_1}\biggr ] + i\delta\\
& = -\frac{q^+}{2k^{+}_0k^{+}_1}\biggl [\left (\kt_0 - \frac{k^{+}_0}{q^+}\qt\right )^2 + m^2 + \frac{k^{+}_0k^{+}_1}{(q^+)^2}Q^2\biggr ] + i\delta. 
\end{split}
\end{equation}

At this stage, it is convenient to introduce the relative transverse momentum $\Pt$ and the normalized photon virtuality squared $\overline{Q}^2$ as 
\begin{equation}
\begin{split}
\label{eq:PtdefandQ}
\Pt & = \kt_0 - \frac{k^{+}_0}{q^+}\qt = -\kt_1 +  \frac{k^{+}_1}{q^+}\qt , \quad\quad \overline{Q}^2  = \frac{k^{+}_0k^{+}_1}{(q^+)^2}Q^2,\\
\Pt &= \kt_0 - z\qt = -\kt_1 +  (1-z)\qt , \quad\quad \overline{Q}^2  = z(1-z)Q^2,
\end{split}
\end{equation}
where, in the second line, the variables $\Pt$ and $\overline{Q}^2$ are expressed in terms of the longitudinal momentum fraction $z = k^{+}_0/q^+$ with $z\in [0,1]$. Using these notations, the LO energy denominator becomes
\begin{equation}
\label{eq:LOEDv2}
\begin{split}
\ed_{\lo} = \frac{1}{(-2q^+)z(1-z)}\biggl [\Pt^2 +  \overline{Q}^2  + m^2 \biggr ],
\end{split}
\end{equation}
where we have dropped the factor $i\delta$ since we only consider the case $Q^2 > 0$ in which the energy denominator is strictly negative.

For the DIS cross section (given in \eq\nr{eq:crosssectionformula}), we need to Fourier transform momentum space expression of LCWF into mixed space. Using the leading order expression in \eq\nr{eq:LOlcwf}, we find for the reduced LCWF (see \eq\nr{eq:LCWFgeneral}) in mixed space the following expression
\begin{equation}
\widetilde{\psi}^{\gamma^{\ast}_L\rightarrow q\bar{q}}_{\lo} = (-2q^+)z(1-z)V^{\gamma^{\ast}_L\rightarrow q\bar{q}}_{h_0;h_1} \int \frac{\ud^{D-2}\Pt}{(2\pi)^{D-2}} \frac{e^{i\Pt \cdot \xt_{01}}}{[\Pt^2 + \overline{Q}^2 + m^2]}.
\end{equation}
Here we have introduced the notation $\xt_{01} = \xt_0 - \xt_1$. Performing the remaining $(D-2)$-dimensional transverse integral by using the result in \eq\nr{eq:F1} yields 
\begin{equation}
\label{eq:LOlcwfmixfact}
\widetilde{\psi}^{\gamma^{\ast}_L\rightarrow q\bar{q}}_{\lo} = \frac{-2ee_fQ}{2\pi}z(1-z)\bar{u}(0)\gamma^+v(1) \left ( \frac{\sqrt{\overline{Q}^2 + m^2}}{2\pi\vert \xt_{01}\vert }\right )^{\frac{D}{2}-2}K_{\frac{D}{2} -2}\left (\vert \xt_{01}\vert\sqrt{\overline{Q}^2 + m^2}\right ),
\end{equation}
where the function $K_{\nu}(z)$ is the modified Bessel function of the second kind. Setting $D=4$, and calculating explicitly (see e.g. \cite{Pauli:2000gw}) the matrix element $\bar{u}(0)\gamma^+v(1)$ one recovers the conventional result for the wavefunction~\cite{Nikolaev:1990ja}.

\section{NLO corrections to the $\gamma^{\ast}_L\rightarrow q\bar{q}$ wave function}
\label{sec:loop}

\subsection{Spinor structures and energy denominators}
In the longitudinal photon case, at NLO accuracy in QCD, one finds that the initial-state LCWF for $\gamma^{\ast}_L\rightarrow q\bar{q}$ can be written as a linear combination of two spinor structures. Using a convenient choice of basis for that space of spinor structures, one can write
\begin{equation}
\label{NLOformfactorsL}
\begin{split}
\Psi^{\gamma^{\ast}_L\rightarrow q\bar{q}}_{\nlo} & = \frac{\delta_{\alpha_0\alpha_1}}{\ed_{\lo}}ee_f \frac{Q}{q^+}\biggl \{\bar u(0) \gamma^+ v(1)\left (1 + \frac{\alpha_s\cf}{2\pi}\mathcal{V}^L\right ) +  \frac{(q^+)^2}{2k^+_0k^+_1} \Pt^j m \bar u(0)\gamma^+\gamma^j v(1) \left (\frac{\alpha_s\cf}{2\pi}\right )\mathcal{S}^L \biggr \}, 
\end{split}
\end{equation}
where the NLO form factors $\mathcal{V}^L$ and $\mathcal{S}^L$ are the light cone helicity non-flip (h.nf.) and helicity flip (h.f.) contributions, respectively. Note that while the transverse photon wave function has both light cone helicity flip and nonflip terms already at LO, for the longitudinal photon the flip term only appears at NLO. This term is related to the quark Pauli form factor, or the quark anomalous magnetic moment, and is discussed in more detail in Appendix~\ref{app:formfactorSL}.

\begin{figure}[tb!]
\centerline{
\includegraphics[width=6.4cm]{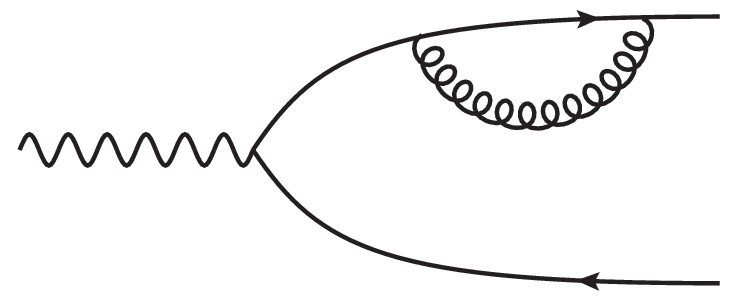}
\hspace*{1.7cm}
\includegraphics[width=6.4cm]{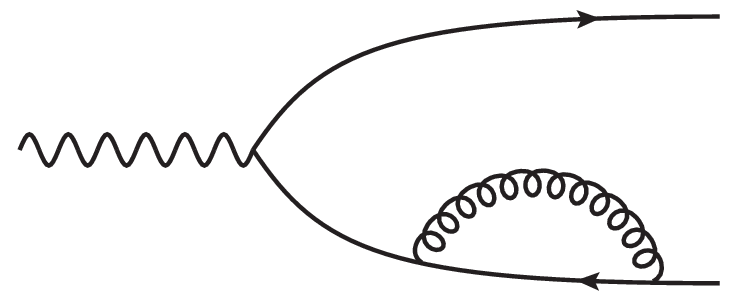}
\begin{tikzpicture}[overlay]
\draw [dashed] (-11.5,2.8) -- (-11.5,-0.3);
\node[anchor=north] at (-11.5cm,-0.3cm) {$\ed_{\lo}$};
\draw [dashed] (-10.0,2.8) -- (-10.0,-0.3);
\node[anchor=north] at (-10.0cm,-0.3cm) {$\ed_{\rm a}$};
\draw [dashed] (-8.9,2.8) -- (-8.9,-0.3);
\node[anchor=north] at (-8.9cm,-0.3cm) {$\ed_{\lo}$};
\node[anchor=east] at (-14.7,1.3) {$\gamma^{\ast}_{L}$};
\node[anchor=west] at (-14.0,0.9) {$\qvec, Q^2 $};
\node[anchor=west] at (-8.7,2.5) {$0,h_0,\alpha_0$};
\node[anchor=west] at (-8.7,0.2) {$1,h_1,\alpha_1$};
\node[anchor=west] at (-11.0,1.2) {$\kvec,\sigma,a$};
\node[anchor=west] at (-10.8,2.7) {$0'$};
\node[anchor=west] at (-12.4,2.3) {$0''$};
\draw [dashed] (-3.3,2.8) -- (-3.3,-0.3);
\node[anchor=north] at (-3.3cm,-0.3cm) {$\ed_{\lo}$};
\draw [dashed] (-2.0,2.8) -- (-2.0,-0.3);
\node[anchor=north] at (-2.0cm,-0.3cm) {$\ed_{\rm b}$};
\draw [dashed] (-0.6,2.8) -- (-0.6,-0.3);
\node[anchor=north] at (-0.6cm,-0.3cm) {$\ed_{\lo}$};
\node[anchor=west] at (-5.3,0.9) {$\qvec, Q^2 $};
\node[anchor=east] at (-6.3,1.3) {$\gamma^{\ast}_{L}$};
\node[anchor=south west] at (-15.0cm,0cm) {\namediag{diag:oneloopSEUPL}};
\node[anchor=south west] at (-6.5cm,0cm) {\namediag{diag:oneloopSEDOWNL}};
 \end{tikzpicture}
}
\rule{0pt}{1ex}
\caption{Time ordered (momenta flows from left to right) one-gluon-loop quark self-energy diagrams \ref{diag:oneloopSEUPL} and \ref{diag:oneloopSEDOWNL} contributing to the longitudinal virtual photon wave function at NLO. In diagram \ref{diag:oneloopSEUPL} imposing the spatial three momentum conservation at each vertex gives: $\qvec = \kvec_{0''} + \kvec_1$, $\kvec_{0''} = \kvec_{0'} + \kvec$ and $\kvec_{0'} + \kvec = \kvec_0$.} 
\label{fig:selfenergyL}
 \end{figure}

\begin{figure}[tb!]
\centerline{
\includegraphics[width=6.4cm]{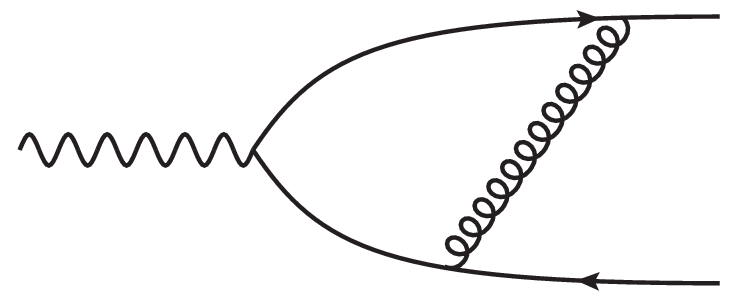}
\hspace*{1.7cm}
\includegraphics[width=6.4cm]{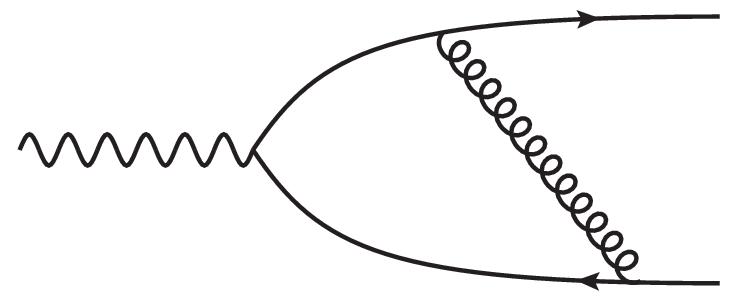}
\begin{tikzpicture}[overlay]
\draw [dashed] (-11.5,2.8) -- (-11.5,-0.3);
\node[anchor=north] at (-11.5cm,-0.3cm) {$\ed_{\rm v}$};
\draw [dashed] (-10.0,2.8) -- (-10.0,-0.3);
\node[anchor=north] at (-10.0cm,-0.3cm) {$\ed_{\rm a}$};
\draw [dashed] (-8.9,2.8) -- (-8.9,-0.3);
\node[anchor=north] at (-8.9cm,-0.3cm) {$\ed_{\lo}$};
\node[anchor=east] at (-14.7,1.3) {$\gamma^{\ast}_{L}$};
\node[anchor=west] at (-14.0,0.9) {$\qvec, Q^2 $};
\node[anchor=west] at (-8.7,2.5) {$0,h_0,\alpha_0$};
\node[anchor=west] at (-8.7,0.2) {$1,h_1,\alpha_1$};
\node[anchor=west] at (-10.0,1.2) {$\kvec,\sigma,a$};
\node[anchor=west] at (-12.0,2.6) {$0'$};
\node[anchor=west] at (-12.3,0.0) {$1'$};
\draw [dashed] (-3.3,2.8) -- (-3.3,-0.3);
\node[anchor=north] at (-3.3cm,-0.3cm) {$\ed_{\rm v}$};
\draw [dashed] (-2.0,2.8) -- (-2.0,-0.3);
\node[anchor=north] at (-2.0cm,-0.3cm) {$\ed_{\rm b}$};
\draw [dashed] (-0.6,2.8) -- (-0.6,-0.3);
\node[anchor=north] at (-0.6cm,-0.3cm) {$\ed_{\lo}$};
\node[anchor=west] at (-5.3,0.9) {$\qvec, Q^2$};
\node[anchor=east] at (-6.3,1.3) {$\gamma^{\ast}_{L}$};
\node[anchor=south west] at (-15.0cm,0cm) {\namediag{diag:vertexqbaremL}};
\node[anchor=south west] at (-6.5cm,0cm) {\namediag{diag:vertexqemL}};
 \end{tikzpicture}
}
\rule{0pt}{1ex}
\caption{Time ordered (momenta flows from left to right) one-gluon-loop vertex diagrams \ref{diag:vertexqbaremL} and \ref{diag:vertexqemL} contributing to the longitudinal virtual photon wave function at NLO. In diagram \ref{diag:vertexqbaremL} imposing the spatial three momentum conservation at each vertex gives: $\qvec = \kvec_{0'} + \kvec_{1'} $, $\kvec_{0'}  + \bar{k} = \kvec_{0} $ and $\kvec_{1'}  = \kvec +  \kvec_{1} $.}
\label{fig:vertexL}
\end{figure}
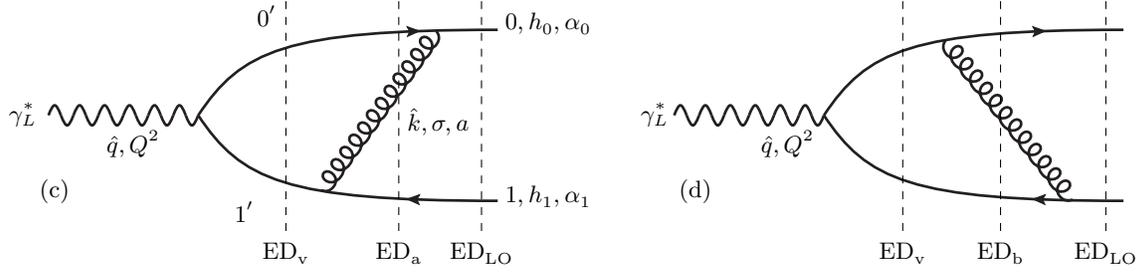

\begin{figure}[tb!]
\centerline{
\includegraphics[width=6.4cm]{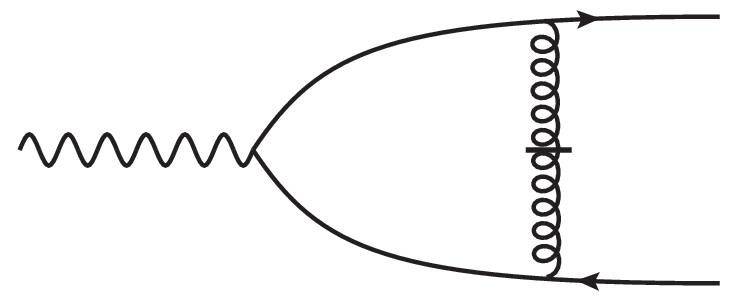}
\begin{tikzpicture}[overlay]
\draw [dashed] (-3.3,2.8) -- (-3.3,-0.3);
\node[anchor=north] at (-3.3cm,-0.3cm) {$\ed_{\rm v}$};
\draw [dashed] (-0.6,2.8) -- (-0.6,-0.3);
\node[anchor=north] at (-0.6cm,-0.3cm) {$\ed_{\lo}$};
\node[anchor=west] at (-0.3,2.5) {$0,h_0,\alpha_0$};
\node[anchor=west] at (-4.0,2.4) {$0'$};
\node[anchor=west] at (-4.0,0.1) {$1'$};
\node[anchor=west] at (-0.3,0.2) {$1,h_1,\alpha_1$};
\node[anchor=west] at (-5.3,0.9) {$\qvec, Q^2 $};
\node[anchor=east] at (-6.3,1.3) {$\gamma^{\ast}_{L}$};
\node[anchor=south west] at (-6.5cm,0cm) {\namediag{diag:vertexqemLinst}};
 \end{tikzpicture}
}
\rule{0pt}{1ex}
\caption{Time ordered (momenta flows from left to right) one-gluon-loop instantaneous diagram \ref{diag:vertexqemLinst} contributing to the the longitudinal virtual photon wave function at NLO. Diagram \ref{diag:vertexqemLinst}: Imposing the spatial three momentum conservation gives: $\qvec = \kvec_{0'} + \kvec_{1'}$. For the case, in which the gluon momentum flows to the quark, the spatial three momentum conservation also gives: $\kvec_{0'} = \kvec + \kvec_{0}$ and  $\kvec_{1'} = \kvec_{1} - \kvec_{0}$}
\label{fig:vertexqemLinst}
\end{figure}
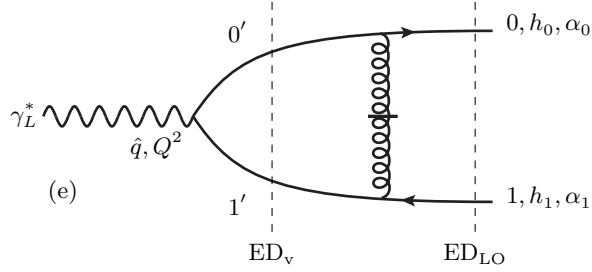

The LCPT diagrams relevant for the  calculation of the $\gamma^{\ast}_L\rightarrow q\bar{q}$ LCWF at NLO with massive quarks are shown in \figs\ref{fig:selfenergyL}, \ref{fig:vertexL} and \ref{fig:vertexqemLinst}. There are two ``propagator correction'' diagrams, \ref{diag:oneloopSEUPL} and \ref{diag:oneloopSEDOWNL} in \fig\ref{fig:selfenergyL}, with a gluon loop attached to the quark or the antiquark. Then, in \fig\ref{fig:vertexL}, there are two ``vertex correction'' diagrams  \ref{diag:vertexqbaremL} and \ref{diag:vertexqemL}, corresponding to two different kinematical possibilities, with longitudinal momentum (which is always positive) flowing either up from the antiquark to the quark or vice versa. Finally, in \fig\ref{fig:vertexqemLinst}, there is an instantaneous gluon exchange \ref{diag:vertexqemLinst} between the quark and the antiquark, in which the gluon momentum can either flow upwards into the quark or downwards into the antiquark. It is convenient to split  up the contribution from this diagram into terms contributing to one or the other of the diagrams in \fig\ref{fig:vertexL} according to the direction of the momentum flow.  Due to the symmetry of the kinematics by exchange of the quark and the antiquark between the two classes of graphs, only the calculation of half of the diagrams is necessary; in this case we will calculate the ones labeled \ref{diag:oneloopSEUPL}, \ref{diag:vertexqbaremL} and the part of  \ref{diag:vertexqemLinst} where the momentum flows to the quark as in \ref{diag:vertexqbaremL}. Note that since the longitudinal photon is really fundamentally a part of an instantaneous interaction, there is no diagram with an instantaneous quark line.

In order to set the stage for our NLO (one-loop) computation, we start by writing down all the energy denominators appearing in the first class diagrams. In  diagram \ref{diag:oneloopSEUPL}, there are two energy denominators $\ed_{\lo}$ and $\ed_{\rm a}$. Following the notation in \fig\ref{fig:selfenergyL}\ref{diag:oneloopSEUPL} and imposing the plus and transverse momentum conservation, it is straightforward to show that 
\begin{equation}
\label{eq:EDa}
\begin{split}
\ed_{\rm a} &= q^{-} - \biggl [k^{-}_{0'} + k^ {-} + k^{-}_1\biggr ] = \frac{\qt^2 -Q^2}{2q^+} - \Biggl [\frac{\kt^2_{0'} + m^2}{2k^{+}_{0'}} + \frac{\kt^2}{2k^+} + \frac{\kt^2_1+m^2}{2k^{+}_1}\biggr ]\\
& = \frac{-k^+_0}{2k^+(k^+_0-k^+)}\biggl [\left (\kt - \frac{k^+}{k^+_0}\kt_0\right )^2 + \frac{k^+(k^+_0-k^+)q^+}{(k^+_0)^2k^+_1}\Biggl \{\Pt^2 + \overline{Q}^2 + \frac{k^+_0(q^+-k^+)}{q^+(k^+_0-k^+)}m^2\Biggr \}\Biggr ]
\end{split}
\end{equation}
and the LO energy denominator $\ed_{\lo}$ is given by \eq\nr{eq:LOED}. 

In diagram \ref{diag:vertexqbaremL}, there are three distinct energy denominators $\ed_{\rm v}, \ed_{\rm a}$ and $\ed_{\lo}$. Again, following the notation in \fig\ref{fig:vertexL}\ref{diag:vertexqbaremL} and imposing the momentum conservation, we obtain  
\begin{equation}
\label{eq:EDv}
\begin{split}
\ed_{\rm v}  &= q^{-} - \biggl [k^{-}_{0'} + k^{-}_{1'}\biggr ] = \frac{\qt^2 -Q^2}{2q^+} - \biggl [\frac{\kt^2_{0'} + m^2}{2k^{+}_{0'}} + \frac{\kt^2_{1'} + m^2}{2k^{+}_{1'}} \biggr ]\\ 
& = \frac{-q^+}{2(k^+_0-k^+)(k^++k^+_1)}\Biggl [\left (\left (\kt - \frac{k^+}{k^+_0}\kt_0\right ) + \Lt\right )^2 + \frac{(k^+_0-k^+)(k^++k^+_1)}{k^+_0k^+_1}\overline{Q}^2 + m^2 \Biggr ],
\end{split}
\end{equation}
where the notation $\Lt = -(k^+_0-k^+)\Pt/k^+_0$ has been introduced.

Finally, for the instantaneous diagram \ref{diag:vertexqemLinst}, in which the gluon momentum flows to the quark, there are two distinct energy denominators $\ed_{\rm v}$ and $\ed_{\lo}$.

For all three diagrams, it is convenient to change variables from the momentum of the gluon $\kt$ and $k^+$ to the relative transverse momentum $\Kt$ and the longitudinal momentum fraction $\xi$ of the gluon with respect to the final state quark. These are the natural variables of the gluon emission and absorption vertices in diagram \ref{diag:oneloopSEUPL} and the gluon absorption vertex in diagram \ref{diag:vertexqbaremL}. They are defined as
\begin{equation}
\label{eq:KTdef}
\Kt = \kt - \frac{k^+}{k^{+}_0}\kt_0 = \kt - \xi\kt_0 \quad \text{with} \quad k^+ = \xi zq^+.
\end{equation}
In these notations, the energy denominators in \eqs\nr{eq:EDa} and \nr{eq:EDv} can be cast into the following form
\begin{equation}
\label{eq:EDdef 
}
\begin{split}
\ed_{\rm a} & = \frac{1}{(-2q^+)z\xi(1-\xi)}\biggl [\Kt^2 + \Delta_1 \biggr ],\\
\ed_{\rm v} & = \frac{1}{(-2q^+)z(1-\xi)(1-z(1-\xi))}\biggl [\left (\Kt + \Lt\right )^2 + \Delta_2 \biggr ],\\
\Lt & = -(1-\xi)\Pt,
\end{split}
\end{equation}
where the coefficients $\Delta_1$ and $\Delta_2$ are given by
\begin{equation}
\label{eq:delta1and2}
\begin{split}
\Delta_1 & = \frac{\xi(1-\xi)}{(1-z)}\biggl [\Pt^2 + \overline{Q}^2 + m^2\biggr ] + \xi^2m^2,\\
\Delta_2 & = (1-\xi)\left (1 + \frac{\xi z}{(1-z)}\right )\overline{Q}^2  + m^2.
\end{split}
\end{equation}

In the following subsections, we present the detailed computation of the NLO form factors $\mathcal{V}^L$ and $\mathcal{S}^L$ coming from the one-loop self-energy, vertex and instantaneous diagrams. 

\subsection{One-loop quark self-energy }

We start by computing the  contribution to the $\gamma^{\ast}_L\rightarrow q\bar{q}$ LCWF in \eq\nr{NLOformfactorsL} from the one-loop massive one-loop quark self-energy diagrams shown in  \fig\ref{fig:selfenergyL}\ref{diag:oneloopSEUPL} and \ref{diag:oneloopSEDOWNL}. Applying the diagrammatic LCPT rules formulated in  momentum space yields the following expression for the diagram \fig\ref{fig:selfenergyL}\ref{diag:oneloopSEUPL}
\begin{equation}
\label{eq:diagaLCWF}
\begin{split}
\Psi^{\gamma^{\ast}_L\rightarrow q\bar{q}}_{\ref{diag:oneloopSEUPL}} & =  \int \dk \int \dkpzero \int \dkppzero (2\pi)^{D-1}\delta^{(D-1)}(\kvec_{0'} + \kvec - \kvec_0)(2\pi)^{D-1}\delta^{(D-1)}(\kvec_{0''} - \kvec_{0'} -  \kvec)\frac{N^{L}_{\ref{diag:oneloopSEUPL}}}{\ed_{\rm a}(\ed_{\lo})^2}\\
& = \frac{1}{16\pi}\int_{0}^{k^+_0}\frac{\ud k^+}{k^+k^+_0(k^+_0-k^+)}\int\frac{\ud^{D-2}\kt}{(2\pi)^{D-2}}\frac{N^{L}_{\ref{diag:oneloopSEUPL}}}{\ed_{\rm a}(\ed_{\lo})^2},
\end{split}
\end{equation}
where the energy denominators are written down in \eqs\nr{eq:LOED} and \nr{eq:EDa}. The product of light cone vertices (the summation is implicit over the internal helicities, gluon polarization and color) is given by the numerator
\begin{equation}
\label{eq:numa}
N^{L}_{\ref{diag:oneloopSEUPL}} = \frac{+ee_fg^2Qt^{a}_{\alpha_0\bar{\alpha}_0}t^{a}_{\bar{\alpha}_0\alpha_1}}{q^+}\biggl [\bar{u}(0)\epsl_{\sigma}(k)u(0')\biggr ]\biggl [\bar{u}(0')\epsl^{\ast}_{\sigma}(k)u(0'')\biggr ]\biggl [\bar{u}(0'')\gamma^{+}v(1)\biggr ].
\end{equation}
We make the change of variables $(\kt,k^+)\mapsto (\Kt,\xi)$ as defined in \eq\nr{eq:KTdef} and regulate the small $k^+\rightarrow 0$ (or $\xi \rightarrow 0)$ divergences by an explicit cutoff $k^+>\alpha q^+$ (or $\xi > \alpha/z$) with the dimensionless parameter $\alpha>0$. Using \eq\nr{eq:EDdef 
} together with \eq\nr{eq:delta1and2}, we can simplify the expression in \eq\nr{eq:diagaLCWF} to    
\begin{equation}
\label{eq:lcwfaint}
\begin{split}
\Psi^{\gamma^{\ast}_L\rightarrow q\bar{q}}_{\ref{diag:oneloopSEUPL}} & = \frac{-1}{8\pi q^+(\ed_{\lo})^2}\int_{\alpha/z}^{1}\frac{\ud \xi}{z}\int\frac{\ud^{D-2}\Kt}{(2\pi)^{D-2}}\frac{N^{L}_{\ref{diag:oneloopSEUPL}} }{[\Kt^2 + \Delta_1]}.
\end{split}
\end{equation}
The detailed calculation of the numerator in \eq\nr{eq:numa} is performed in Appendix~\ref{app:numa} and gives 
\begin{equation}
\label{eq:numav2}
N^{L}_{\ref{diag:oneloopSEUPL}} = \frac{2g^2\cf}{\xi^2(1-\xi)}\delta_{\alpha_0\alpha_1}V^{\gamma^{\ast}_L\rightarrow q\bar{q}}_{h_0;h_1}\Biggl \{\biggl [1 + (1-\xi)^2\biggr ]\Kt^2  + m^2\xi^4 + \frac{(D_s-4)}{2}\xi^2\biggl [\Kt^2 + m^2\xi^2\biggr ] \Biggr \}.
\end{equation}
We remind the reader that from this expression one obtains both the FDH scheme result by taking the limit $D_s \rightarrow 4$, and the CDR one by setting $D_s = D$.

At this point, it is convenient to define the UV divergent one-loop scalar integral as 
\begin{equation}
\mathcal{A}_0(\Delta_1) = 4\pi(\mu^2)^{2-D/2}\int\frac{\ud^{D-2}\Kt}{(2\pi)^{D-2}}\frac{1}{[\Kt^2 + \Delta_1]},
\end{equation}
where $\mu^2$ is the scale introduced by transverse dimensional regularization. Note that in the framework of transverse dimensional regularization   
\begin{equation}
4\pi(\mu^2)^{2-D/2}\int\frac{\ud^{D-2}\Kt}{(2\pi)^{D-2}}\frac{\Kt^2}{[\Kt^2 + \Delta_1]} =-\Delta_1 \mathcal{A}_0(\Delta_1) .
\end{equation}
Under these simplifications \eq\nr{eq:lcwfaint} reduces to 
\begin{equation}
\label{eq:finalURSEa}
\begin{split}
\Psi^{\gamma^{\ast}_L\rightarrow q\bar{q}}_{\ref{diag:oneloopSEUPL}}  = \Psi^{\gamma^{\ast}_L\rightarrow q\bar{q}}_{\lo}\left (\frac{\alpha_s\cf}{2\pi} \right ) \Biggl \{-\int_{\alpha/z}^{1}\frac{\ud \xi}{\xi}\biggl [1 + (1-\xi)^2\biggr ]\mathcal{A}_0(\Delta_1) &- \frac{2(1-z)m^2}{[\Pt^2 + \overline{Q}^2 + m^2]}\int_{0}^{1}\ud \xi \mathcal{A}_0(\Delta_1)\\
& - \frac{(D_s-4)}{2}\int_{0}^{1}\ud \xi  \xi \mathcal{A}_0(\Delta_1)\Biggr \},
\end{split}
\end{equation}
where the result for the integral $\mathcal{A}_0(\Delta_1)$ can be written as~\cite{Beuf:2016wdz} 
\begin{equation}
\label{eq:A0exp}
\mathcal{A}_0(\Delta_1) = \Gamma\left (2 - \frac{D}{2}\right ) \biggl [\frac{\Delta_1}{4\pi\mu^2}\biggr ]^{\frac{D}{2}-2} =  \frac{(4\pi)^{2-\frac{D}{2}}}{(2-\frac{D}{2})}\Gamma\left (3-\frac{D}{2}\right ) - \log\left (\frac{\Delta_1}{\mu^2}\right ) + \mathcal{O}(D-4).
\end{equation}
We note that this result is the same for any variant of dimensional regularization.

\subsection{Quark mass renormalization}

From the expression \eqref{eq:A0exp} for the integral $\mathcal{A}_0(\Delta_1)$, it is clear that the one-loop quark self energy diagram \ref{diag:oneloopSEUPL} is UV divergent. UV divergences have been already found for that diagram and for the other ones in the massless quark case \cite{Beuf:2016wdz,Beuf:2017bpd,Hanninen:2017ddy}. In that case, the UV divergences cancel each other at the cross section level. UV divergences with such a behavior are expected to occur as well in the massive quark case studied in the present paper. However, since the expression for the longitudinal photon wave function (and thus the one for the longitudinal photon cross section) involves the quark mass already at LO, NLO corrections are expected to also include UV divergences associated with quark mass renormalization.
Hence, one expect two types of UV divergences, which need to be disentangled. 

In the LO expression \eqref{eq:LOlcwf} for the longitudinal photon wave function, the quark mass appears in the energy denominator $\ed_{\lo}$, \eq\eqref{eq:LOEDv2}, but not in the numerator \eqref{eq:LOvertex}. In the bare perturbation theory approach, one uses the bare mass $m_0$ when writing the expression of the diagrams from the Feynman rules in light-front perturbation theory, in particular in the energy denominators. Then, the bare mass is rewritten as $m_0^2 = m^2-\delta m^2$ in the result, and a Taylor expansion at small mass shift $\delta m^2$  is performed, assuming $\delta m^2\sim \as$. Mass renormalization then amounts to imposing an extra condition in order to determine $\delta m^2$.
When following such a bare perturbation theory approach for the longitudinal photon wave function, the energy denominator $\ed_{\lo}$ becomes
\begin{eqnarray}
 \frac{1}{\ed_{\lo}(m_0^2)}  &=& \frac{1}{\ed_{\lo}(m^2)}  +  
 \frac{\frac{\partial}{\partial m^2}\ed_{\lo}(m^2)}{\ed_{\lo}(m^2)^2}\delta m^2 
 + \mathcal{O}\left((\delta m^2)^2\right)
 \nonumber
\\ &=&
\frac{1}{\ed_{\lo}(m^2)}  \left\{ 1 
-\left(\frac{\delta m^2}{2k_0^+}+\frac{\delta m^2}{2k_1^+}\right)
 \frac{1}{\ed_{\lo}(m^2)}
\right\}
+ \mathcal{O}\left((\delta m^2)^2\right).
\label{eq:ctexpansion}
\end{eqnarray}
By contrast, in the renormalized perturbation approach, one  uses the renormalized mass $m$ when writing down the diagrams, and instead includes mass counter-terms in the light-front Hamiltonian, corresponding to additional two-point vertices. In that context, the contributions $\delta m^2/2k_0^+$ and $\delta m^2/2k_1^+$ in Eq.~\eqref{eq:ctexpansion} are reinterpreted as coming from the NLO diagrams obtained by inserting such a mass counterterm on the quark or antiquark line respectively in the LO diagram in \fig\ref{fig:lovertex}.   
The important point to note is the doubling of the energy denominator for these terms. Hence, these terms are enhanced in the limit $\ed_{\lo}\rightarrow 0$, corresponding to the on-shell limit for the quark-antiquark Fock state. In the bare perturbation theory approach, the doubling of the energy denominator comes naturally from the Taylor expansion. In the renormalized perturbation theory approach, on the other hand, it comes from the fact that one should include an energy denominator both before and after the insertion of the mass counterterm. Only NLO corrections coming with an extra copy of the energy denominator $\ed_{\lo}$ can thus be absorbed into the quark mass in the energy denominator, via mass renormalization.

In the massless quark case, the extra copy of the energy denominator $\ed_{\lo}$ disappears in the course of the evaluation of diagram \ref{diag:oneloopSEUPL} (see for example section IV. in Ref.~\cite{Beuf:2016wdz}), showing that no quark mass can be radiatively generated by such a self-energy diagram, and that the UV divergences encountered in the massless case have nothing to do with the renormalization of the quark mass in the energy denominator.

Let us now come back to the expression \eqref{eq:finalURSEa} for the quark self-energy diagram \ref{diag:oneloopSEUPL}. Remember that one copy of the energy denominator $\ed_{\lo}$ is already included in the LO wavefunction, and that $\ed_{\lo}$ is proportional to $(\Pt^2 + \overline{Q}^2 + m^2)$ (see Eq.~\eqref{eq:LOEDv2}). One finds that only the second term of \eqref{eq:finalURSEa} exhibits a doubling of the energy denominator and can thus be associated with mass renormalization. Note also from Eq.~\eqref{eq:delta1and2} that $\Delta_1$ depends on $\ed_{\lo}$, and that $\Delta_1\rightarrow \xi^2m^2$ in the on-shell limit $(\Pt^2 + \overline{Q}^2 + m^2) \rightarrow 0$. We now add and subtract a term with $\Delta_1\rightarrow \xi^2m^2$ to rewrite  \eq\eqref{eq:finalURSEa}  as 
\begin{equation}
\label{eq:finalRESEav1}
\begin{split}
\Psi^{\gamma^{\ast}_L\rightarrow q\bar{q}}_{\ref{diag:oneloopSEUPL}}  = \Psi^{\gamma^{\ast}_L\rightarrow q\bar{q}}_{\lo}&\left (\frac{\alpha_s\cf}{2\pi} \right ) \Biggl \{
- \frac{2(1-z)m^2}{[\Pt^2 + \overline{Q}^2 + m^2]}  \int_{0}^{1}\ud \xi \mathcal{A}_0(\xi^2m^2)
-\int_{\alpha/z}^{1}\frac{\ud \xi}{\xi}\biggl [1 + (1-\xi)^2\biggr ]\mathcal{A}_0(\Delta_1)\\
& - 2(1-z)m^2\int_{0}^{1}\ud \xi 
\frac{[\mathcal{A}_0(\Delta_1)-\mathcal{A}_0(\xi^2m^2)  
]}{[\Pt^2 + \overline{Q}^2 + m^2]}  - \frac{(D_s-4)}{2}\int_{0}^{1}\ud \xi  \xi \mathcal{A}_0(\Delta_1)\Biggr \},
\end{split}
\end{equation}
where now only the first term can contribute to mass renormalization. By contrast, the UV divergence in the second term is one that should cancel at the cross section level, like in the massless case. The terms on the second line in \eqref{eq:finalRESEav1} are UV finite. 

Other contributions to mass renormalization in the energy denominators come from the ``self-induced inertia'' or ``seagull'' diagrams \cite{Pauli:1985pv,Tang:1991rc,Brodsky:1997de}. These diagrams, which correspond to instantaneous self-energy loops, are highly sensitive to the details of the UV regularization procedure. They bring a pure extra power of the energy denominator $\ed_{\lo}$, and can thus be entirely absorbed into the quark mass in the energy denominator, via mass renormalization.
We plan to present a complete analysis of mass renormalization in light-front perturbation theory at one loop in a separate future work, with a special emphasis on UV regularization issues and on ``self-induced inertia'' contributions. 

Here, in order to avoid entering further into these issues, we choose the on-shell scheme for mass renormalization. In the context of the present calculation, this scheme is defined by imposing a strict cancellation between all the enhanced contributions in the on-shell limit $\ed_{\lo} \rightarrow0$, or equivalently  $(\Pt^2 + \overline{Q}^2 + m^2) \rightarrow0$.\footnote{Note that the total energy of the quark-antiquark state is positive. This means that the renormalization condition is determined at a time-like virtuality for the  photon $Q^2<0$, away from the physical (spacelike) region for the DIS process.} Hence, the quark mass counterterm is chosen in order to cancel exactly the ``self-induced inertia'' insertions on the quark line and the first term in Eq.~\eqref{eq:finalRESEav1} from the diagram \ref{diag:oneloopSEUPL}. 
Adding the ``self-induced inertia'' contributions and the quark counterterm to the diagram \ref{diag:oneloopSEUPL} and choosing 
the on-shell scheme for mass renormalization thus simply leads to a result where the first term of \eq\eqref{eq:finalRESEav1} is absent
\begin{equation}
\label{eq:finalRESEav2}
\begin{split}
\Psi^{\gamma^{\ast}_L\rightarrow q\bar{q}}_{\ref{diag:oneloopSEUPL}}  = \Psi^{\gamma^{\ast}_L\rightarrow q\bar{q}}_{\lo}&\left (\frac{\alpha_s\cf}{2\pi} \right ) \Biggl \{
-\int_{\alpha/z}^{1}\frac{\ud \xi}{\xi}\biggl [1 + (1-\xi)^2\biggr ]\mathcal{A}_0(\Delta_1)\\
& - 2(1-z)m^2\int_{0}^{1}\ud \xi 
\frac{[\mathcal{A}_0(\Delta_1)-\mathcal{A}_0(\xi^2m^2)  
]}{[\Pt^2 + \overline{Q}^2 + m^2]}  - \frac{(D_s-4)}{2}\int_{0}^{1}\ud \xi  \xi \mathcal{A}_0(\Delta_1)\Biggr \}.
\end{split}
\end{equation}
We keep calling this expression the (mass renormalized) contribution of diagram \ref{diag:oneloopSEUPL}, by a slight abuse of language.
The same treatment is done on the antiquark line, with diagram \ref{diag:oneloopSEDOWNL}.
 
In the case of the transverse photon wavefunction, which we will study in a future publication, the quark mass also appears in the numerator in the LO expression, in the transverse photon to quark-antiquark vertex. Hence, at NLO, one has to deal with mass renormalization in the numerator as well. This involves vertex correction diagrams analog to diagrams \ref{diag:vertexqbaremL} and \ref{diag:vertexqemL}, and another quark mass counterterm which is a three-point vertex. In light-front perturbation theory, mass renormalization corrections in either the denominator or the numerator thus occur in a very different pattern. In principle, the final result for the mass shift should be the same in both case, but this property is typically lost if the UV regularization procedure does not preserve the full Poincar\'e symmetry \cite{Mustaki:1990im,Burkardt:1991tj,Perry:1992sw,Zhang:1993dd,Harindranath:1993de}.

The expression~\eqref{eq:finalRESEav2} for the mass renormalized contribution from the self-energy diagram \ref{diag:oneloopSEUPL} can now be written as 
\begin{equation}
\label{eq:finalRESEafinal}
\begin{split}
\Psi^{\gamma^{\ast}_L\rightarrow q\bar{q}}_{\ref{diag:oneloopSEUPL}}  = \Psi^{\gamma^{\ast}_L\rightarrow q\bar{q}}_{\lo}\left (\frac{\alpha_s\cf}{2\pi} \right )\mathcal{V}^{L}_{\ref{diag:oneloopSEUPL}}.
\end{split}
\end{equation}
Using Eq.~\eqref{eq:A0exp}, the form factor $\mathcal{V}^{L}_{\ref{diag:oneloopSEUPL}}$ can be evaluated as
\begin{equation}
\label{eq:Va}
\begin{split}
\mathcal{V}^{L}_{\ref{diag:oneloopSEUPL}} = \biggl [\frac{3}{2} &+ 2\log\left (\frac{\alpha}{z}\right )\biggr ]\Biggl \{ \frac{(4\pi)^{2-\frac{D}{2}}}{(2-\frac{D}{2})}\Gamma\left (3-\frac{D}{2}\right )  + \log\left (\frac{(1-z)\mu^2}{\overline{Q}^2 + m^2}\right ) - \log\left (\frac{\Pt^2 + \overline{Q}^2 + m^2}{\overline{Q}^2 + m^2}\right )\Biggr \}\\
&- \log^2\left (\frac{\alpha}{z}\right ) -\frac{\pi^2}{3} + 3 + \frac{1}{2}\frac{(D_s-4)}{(D-4)}  + \mathcal{I}_{\mathcal{V}_{\ref{diag:oneloopSEUPL}}}(z,\Pt) + \mathcal{O}(D-4).
\end{split}
\end{equation}
In the above expression, the factor $(D_s-4)/2(D-4)$ is the regularization scheme dependent coefficient coming from the following integral
\begin{equation}
\label{eq:Kscheme}
- \frac{(D_s-4)}{2}\int_{0}^{1}\ud \xi  \xi A_0(\Delta_1) = \frac{1}{2}\frac{(D_s-4)}{(D-4)} + \mathcal{O}(D_s-4),
\end{equation}
and the function $\mathcal{I}_{\mathcal{V}_{\ref{diag:oneloopSEUPL}}}$ is defined as 
\begin{equation}
\label{eq:IVaint}
\mathcal{I}_{\mathcal{V}_{\ref{diag:oneloopSEUPL}}}(z,\Pt) = (\Pt^2 + \overline{Q}^2 + m^2)\int_{0}^{1}\frac{\ud \xi}{\xi} \left [-\frac{2\log(\xi)}{(1-\xi)} + \frac{(1+\xi)}{2}\right ]\biggl \{\frac{1}{\Pt^2 + \overline{Q}^2 + m^2} -  \frac{1}{\Pt^2 + \overline{Q}^2 + m^2 + \frac{\xi(1-z)}{(1-\xi)}m^2}\biggr \}.
\end{equation}
This integral is $\xi\rightarrow 0$ finite and we could in principle perform the integration over $\xi$ analytically. However, it turns out that it is more convenient to Fourier transform first and then perform the remaining integral numerically.

The mass renormalized LCWF for the quark self-energy diagram in diagram \ref{diag:oneloopSEDOWNL} shown in  \fig\ref{fig:selfenergyL} can be now easily obtained by using the symmetry between the diagrams \ref{diag:oneloopSEUPL} and \ref{diag:oneloopSEDOWNL}, i.e. by  making the substitution $z \mapsto 1-z$ and $\Pt \mapsto -\Pt$ in \eq\nr{eq:finalRESEafinal} simultaneously. This yields
\begin{equation}
\label{eq:finalRESEbfinal}
\begin{split}
\Psi^{\gamma^{\ast}_L\rightarrow q\bar{q}}_{\ref{diag:oneloopSEDOWNL}}  = \Psi^{\gamma^{\ast}_L\rightarrow q\bar{q}}_{\lo}\left (\frac{\alpha_s\cf}{2\pi} \right )\mathcal{V}^{L}_{\ref{diag:oneloopSEDOWNL}},
\end{split}
\end{equation}
where 
\begin{equation}
\label{eq:Vb}
\begin{split}
\mathcal{V}^{L}_{\ref{diag:oneloopSEDOWNL}} = \biggl [\frac{3}{2} & + 2\log\left (\frac{\alpha}{1-z}\right )\biggr ]\Biggl \{ \frac{(4\pi)^{2-\frac{D}{2}}}{(2-\frac{D}{2})}\Gamma\left (3-\frac{D}{2}\right )   + \log\left (\frac{z \mu^2}{\overline{Q}^2 + m^2}\right )  - \log\left (\frac{\Pt^2 + \overline{Q}^2 + m^2}{\overline{Q}^2 + m^2}\right )\Biggr \}\\
& - \log^2\left (\frac{\alpha}{1-z}\right ) -\frac{\pi^2}{3} + 3 + \frac{1}{2}\frac{(D_s-4)}{(D-4)}  + \mathcal{I}_{\mathcal{V}_{\ref{diag:oneloopSEDOWNL}}}(z,\Pt) + \mathcal{O}(D-4)
\end{split}
\end{equation}
and 
\begin{equation}
\label{eq:IVbint}
\mathcal{I}_{\mathcal{V}_{\ref{diag:oneloopSEDOWNL}}}(z,\Pt) = (\Pt^2 + \overline{Q}^2 + m^2)\int_{0}^{1}\frac{\ud \xi}{\xi} \left [-\frac{2\log(\xi)}{(1-\xi)} + \frac{(1+\xi)}{2}\right ]\biggl \{\frac{1}{\Pt^2 + \overline{Q}^2 + m^2} - \frac{1}{\Pt^2 + \overline{Q}^2 + m^2 + \frac{z \xi }{(1-\xi)}m^2}\biggr \}.
\end{equation}

Summing the expressions in \eqs\nr{eq:finalRESEafinal} and \nr{eq:finalRESEbfinal}, we obtain for the full contribution of the one-loop quark self-energy to the $\gamma^{\ast}_{L}\rightarrow q\qvec$ LCWF the result 
\begin{equation}
\label{eq:finalSE}
\begin{split}
\Psi^{\gamma^{\ast}_L\rightarrow q\bar{q}}_{\ref{diag:oneloopSEUPL} + \ref{diag:oneloopSEDOWNL}}  = \Psi^{\gamma^{\ast}_L\rightarrow q\bar{q}}_{\lo}\left (\frac{\alpha_s\cf}{2\pi} \right )\mathcal{V}^{L}_{\ref{diag:oneloopSEUPL} + \ref{diag:oneloopSEDOWNL}},
\end{split}
\end{equation}
where the NLO form factor $\mathcal{V}^{L}_{\ref{diag:oneloopSEUPL} + \ref{diag:oneloopSEDOWNL}}$ can be written as 
\begin{equation}
\label{eq:Vselfenergy}
\begin{split}
\mathcal{V}^{L}_{\ref{diag:oneloopSEUPL} + \ref{diag:oneloopSEDOWNL}}  =   2\biggl [\frac{3}{2}  &+ \log\left (\frac{\alpha}{z}\right ) + \log\left (\frac{\alpha}{1-z}\right )\biggr ]\Biggl \{  \frac{(4\pi)^{2-\frac{D}{2}}}{(2-\frac{D}{2})}\Gamma\left (3-\frac{D}{2}\right )   + \log\left (\frac{\mu^2}{\overline{Q}^2 + m^2}\right )  - \log\left (\frac{\Pt^2 + \overline{Q}^2 + m^2}{\overline{Q}^2 + m^2}\right )\Biggr \} \\
& + \biggl [\log(z) + \log(1-z)  \biggr ]\left (\frac{3}{2} + 2\log(\alpha) \right ) - 4\log(z)\log(1-z) - \log^2\left (\frac{\alpha}{z}\right ) - \log^2\left (\frac{\alpha}{1-z}\right )\\
& -\frac{2\pi^2}{3} + 6 + \frac{(D_s-4)}{(D-4)} + \mathcal{I}_{\mathcal{V}_{\ref{diag:oneloopSEUPL}}}(z,\Pt)  +  \mathcal{I}_{\mathcal{V}_{\ref{diag:oneloopSEDOWNL}}}(z,\Pt) + \mathcal{O}(D-4).
\end{split}
\end{equation}

\subsection{Vertex and instantaneous contributions}
\label{sec:vertexcor}

We now proceed to calculate the one-loop vertex correction diagram \ref{diag:vertexqbaremL} shown in \fig\ref{fig:vertexL} and the instantaneous diagram \ref{diag:vertexqemLinst} shown in \fig\ref{fig:vertexqemLinst}. For diagram \ref{diag:vertexqbaremL}, the momentum space expression of the LCWF can be written as  
\begin{equation}
\label{eq:vertexc}
\begin{split}
\Psi^{\gamma^{\ast}_L\rightarrow q\bar{q}}_{\ref{diag:vertexqbaremL}} & =  \int \dk \int \dkpzero \int \dkpone (2\pi)^{D-1}\delta^{(D-1)}(\kvec_{0'} + \kvec - \kvec_0)(2\pi)^{D-1}\delta^{(D-1)}(\kvec_{1'} - \kvec -  \kvec_1)\frac{N^L_{\ref{diag:vertexqbaremL}}}{\ed_{\rm v}\ed_{\rm a}\ed_{\lo}}\\
& = \frac{1}{16\pi}\int_{0}^{k^+_0}\frac{\ud k^+}{k^+(k^+_0 - k^+)(k^+ + k^+_1)}\int\frac{\ud^{D-2}\kt}{(2\pi)^{D-2}}\frac{N^L_{\ref{diag:vertexqbaremL}}}{\ed_{\rm v}\ed_{\rm a}\ed_{\lo}},
\end{split}
\end{equation}
where the light cone energy denominators $\ed_{\rm v}$,  $\ed_{\rm a}$ and $\ed_{\lo}$ are given in  \eqs\nr{eq:EDv}, \nr{eq:EDa} and \nr{eq:LOED}, respectively. The numerator (again the summation is implicit over the internal helicities, gluon polarization and color) is given by 
\begin{equation}
\label{eq:numc}
N^L_{\ref{diag:vertexqbaremL}} = \frac{-ee_fg^2Q\delta_{\alpha_0\alpha_1}\cf}{q^+}\biggl [\bar{u}(0)\epsl_{\sigma}(k)u(0')\biggr ]\biggl [\bar{u}(0')\gamma^{+}v(1')\biggr ]\biggl [\bar{v}(1')\epsl^{\ast}_{\sigma}(k)v(1)\biggr ].
\end{equation}

Applying again the change of variables $(\kt,k^+) \mapsto (\Kt,\xi)$, we obtain
\begin{equation}
\label{eq:vertexcv2}
\begin{split}
\Psi^{\gamma^{\ast}_L\rightarrow q\bar{q}}_{\ref{diag:vertexqbaremL}} & =  \frac{1}{4\pi}\frac{z}{\ed_{\lo}}\int_{\alpha/z}^{1}\ud \xi (1-\xi) 
\int\frac{\ud^{D-2}\Kt}{(2\pi)^{D-2}}\frac{N^L_{\ref{diag:vertexqbaremL}}}{[\Kt^2 + \Delta_1][\left (\Kt + \Lt\right )^2 + \Delta_2]},
\end{split}
\end{equation}
where the detailed calculation of the numerator in \eq\nr{eq:numc} is found in Appendix \ref{app:numc}, and gives 
\begin{equation}
\label{numcappfinaltext}
\begin{split}
N^L_{\ref{diag:vertexqbaremL}} = &-2\delta_{\alpha_0\alpha_1}V^{\gamma^{\ast}_{L}\rightarrow q\bar{q}}_{h_0;h_1}(g^2\cf)\Biggl \{\frac{1}{\xi^2}\Biggl [\frac{(1-z)}{z} + \frac{(1-\xi)(1-z(1-\xi))}{z} - \frac{(D_s-4)}{2}\xi^2\Biggr ]\\
& \times \delta^{ik}_{(D_s)}\Kt^i\left (\Kt^{k} + \frac{\xi}{1-z}\Pt^{k}\right ) + \frac{z\xi^2}{1-z}m^2 f_{(D_s)} \Biggr \}\\
& +2\delta_{\alpha_0\alpha_1}\frac{ee_fQ}{q^+}(g^2\cf) m\bar{u}(0)\gamma^+\gamma^iv(1)\frac{1}{z}\Biggl \{\Biggl [\frac{z}{1-z} - 1 - \frac{z\xi}{1-z}f_{(D_s)}\Biggr ]\Kt^i - \xi\biggl [1 + \frac{z\xi}{1-z}f_{(D_s)}\biggr ]\Pt^i\Biggr \}
\end{split}
\end{equation}
with $f_{(D_s)} = 1 + (D_s-4)/2$.

At tree level, the longitudinal photon to quark-antiquark splitting~\eq\nr{eq:LOlcwfmixfact} is proportional to the light cone helicity conserving Dirac structure $\bar{u}(0)\gamma^+v(1) \sim \delta_{h_0,-h_1}$. At one loop level, the helicity structure is more complicated, since in addition to a correction to the light cone helicity conserving structure, the result also contains a light cone helicity flip term $\sim  \delta_{h_0,h_1}$. However, to calculate the NLO cross section the one-loop wavefunction in the amplitude will be convoluted with the tree level one in the conjugate amplitude. Since also the eikonal interactions with the target conserve light cone helicity, the two different helicity structures do not interfere. Thus, in fact, the light cone helicity flip term does not contribute to the cross section at this order in perturbation theory. At NNLO the square of this term would contribute to the cross section. It will therefore be convenient to separate out the two helicity structures at this point in the calculation. To do this, we split the result~\eq\nr{eq:vertexcv2} into two parts
\begin{equation}
\label{eq:psidiagcfull}
\Psi^{\gamma^{\ast}_L\rightarrow q\bar{q}}_{\ref{diag:vertexqbaremL}} = \Psi^{\gamma^{\ast}_L\rightarrow q\bar{q}}_{\lo}\left (\frac{\alpha_s\cf}{2\pi}\right )\mathcal{V}^{L}_{\ref{diag:vertexqbaremL}} + \Psi^{\gamma^{\ast}_L\rightarrow q\bar{q}}_{\text{h.f.};\ref{diag:vertexqbaremL}},
\end{equation}
where the NLO form factor $\mathcal{V}^{L}_{\ref{diag:vertexqbaremL}}$, which factorizes from the LO contribution, is given by
\begin{equation}
\label{eq:factnloc}
\mathcal{V}^{L}_{\ref{diag:vertexqbaremL}} = -\int_{\alpha/z}^{1}\ud \xi (1-\xi)\mathcal{I}^{L}_{\ref{diag:vertexqbaremL}}
\end{equation}
with 
\begin{equation}
\begin{split}
\mathcal{I}^{L}_{\ref{diag:vertexqbaremL}} = 4\pi(\mu^2)^{2-d/2}\int\frac{\ud^{d-2}\Kt}{(2\pi)^{d-2}}\frac{1}{[\Kt^2 + \Delta_1][\left (\Kt + \Lt\right )^2 + \Delta_2]}\Biggl \{&\frac{(1-z)}{\xi^2}\Biggl [1 + (1-\xi)\left (1 + \frac{z\xi}{1-z}\right ) - \frac{(D_s-4)}{2}\frac{z\xi^2}{1-z}\Biggr ]\\
& \times \delta^{ik}_{(D_s)}\Kt^i\left (\Kt^k + \frac{\xi}{1-z}\Pt^k\right )
 + \frac{z^2\xi^2}{1-z}m^2f_{(D_s)}\Biggr \}.
\end{split}
\end{equation}
The second term in \eq\nr{eq:psidiagcfull} is the light cone helicity flip term that appears only in the massive quark case, and it contributes to the form factor $\mathcal{S}^L$ in \eq\nr{NLOformfactorsL}. This term can be simplified to 
\begin{equation}
\label{eq:hfc}
\Psi^{\gamma^{\ast}_L\rightarrow q\bar{q}}_{\text{h.f.};\ref{diag:vertexqbaremL}} = \delta_{\alpha_0\alpha_1}\frac{ee_f}{\ed_{\lo}}\frac{Q}{q^+}m \bar{u}(0)\gamma^+\gamma^iv(1)\left ( \frac{\alpha_s\cf}{2\pi}\right )\int_{0}^{1}\ud \xi (1-\xi) \mathcal{I}^{L;i}_{\text{h.f.};\ref{diag:vertexqbaremL}}
\end{equation}
with 
\begin{equation}
\begin{split}
\mathcal{I}^{L;i}_{\text{h.f.};\ref{diag:vertexqbaremL}} = 4\pi(\mu^2)^{2-d/2}\int\frac{\ud^{d-2}\Kt}{(2\pi)^{d-2}}\frac{1}{[\Kt^2 + \Delta_1][\left (\Kt + \Lt\right )^2 + \Delta_2]}\Biggl \{&\Biggl [\frac{z}{1-z} - 1 - \frac{z\xi}{1-z}f_{(D_s)}\Biggr ]\Kt^i\\
& - \xi\biggl [1 + \frac{z\xi}{1-z}f_{(D_s)}\biggr ]\Pt^i\Biggr \}.
\end{split}
\end{equation}
It turns out that the most efficient way to compute the integrals $\mathcal{I}^{L}_{\ref{diag:vertexqbaremL}}$ and $\mathcal{I}^{L;i}_{\text{h.f.};\ref{diag:vertexqbaremL}}$ is to rewrite them as a linear combination of tensor and scalar integrals\footnote{Here, we follow the notation and discussion presented in Ref.~\cite{Beuf:2016wdz}, see Appendix D.}. This procedure, in both regularization (FDH and CDR) schemes,  gives 
\begin{equation}
\label{eq:ILc}
\begin{split}
\mathcal{I}^{L}_{\ref{diag:vertexqbaremL}} = & \frac{1}{\xi}\Biggl [1 + (1-\xi)\left (1 + \frac{z\xi}{1-z}\right ) - \frac{(D_s-4)}{2}\frac{z\xi^2}{1-z}\Biggr ]\Biggl \{\frac{(1-z)}{\xi}\mathcal{A}_0(\Delta_2) + \Pt^i\mathcal{B}^i(\Delta_1,\Delta_2,\Lt)\\
& - \biggl [(1-\xi)[\Pt^2 + \overline{Q}^2 + m^2] + \xi(1-z)m^2\biggr ]\mathcal{B}_0(\Delta_1,\Delta_2,\Lt) \Biggr \} + \frac{z^2\xi^2}{1-z}m^2f_{(D_s)} \mathcal{B}_0(\Delta_1,\Delta_2,\Lt)
\end{split}
\end{equation}
and 
\begin{equation}
\label{eq:ILchf}
\begin{split}
\mathcal{I}^{L;i}_{\text{h.f.};\ref{diag:vertexqbaremL}} =\Biggl [\frac{z}{1-z} - 1 - \frac{z\xi}{1-z}f_{(D_s)}\Biggr ]\mathcal{B}^i(\Delta_1,\Delta_2,\Lt) - \xi \biggl [1 + \frac{z\xi}{1-z}f_{(D_s)}\biggr ] \Pt^i \mathcal{B}_0(\Delta_1,\Delta_2,\Lt),
\end{split}
\end{equation}
where the one-loop vector integral $\mathcal{B}^i(\Delta_1,\Delta_2,\Lt) = \Lt^i \mathcal{B}_1(\Delta_1,\Delta_2,\Lt)$ and the scalar integrals $\mathcal{B}_0$ and $\mathcal{B}_1$ are given in Ref.~\cite{Beuf:2016wdz}. The integrals $\mathcal{B}_0$ and $\mathcal{B}^i$  in \eqs\nr{eq:ILc} and \nr{eq:ILchf} are both UV finite, and therefore all the UV divergences are carried by the scalar integral $\mathcal{A}_0(\Delta_2)$, where the coefficient $\Delta_2$ (defined in \eq\nr{eq:delta1and2}) is independent of $\Pt$. This is particularly convenient, since we have to later Fourier transform the $\Pt$-dependence of the LCWF in \eq\nr{eq:psidiagcfull}. In addition, since only the integral $\mathcal{A}_0$ is UV divergent, the scheme dependent part comes from the term proportional to $\mathcal{I}^{L}_{\ref{diag:vertexqbaremL}}  \sim (D_s -4)\mathcal{A}_0(\Delta_2)$ in \eq\nr{eq:ILc}, and in the other terms we can set $D_s=4$ and $f_{(D_s)} = 1$. 

Since the vector integral $\mathcal{B}^i(\Delta_1,\Delta_2,\Lt)$ is proportional to $\Pt^i$, the helicity flip contribution can be written as
\begin{equation}
\label{eq:hfc_2}
\Psi^{\gamma^{\ast}_L\rightarrow q\bar{q}}_{\text{h.f.};\ref{diag:vertexqbaremL}} = \delta_{\alpha_0\alpha_1}\frac{ee_f}{\ed_{\lo}}\; \frac{Q}{q^+}\;
\left ( \frac{\alpha_s\cf}{2\pi}\right )\;
\frac{\Pt^i}{2z(1\!-\!z)}\;
m \bar{u}(0)\gamma^+\gamma^iv(1)\;
\mathcal{S}^L_{\ref{diag:vertexqbaremL}}
\, ,
\end{equation}
with
\begin{align}
{\cal S}^{L}_{\ref{diag:vertexqbaremL}} & = 2z \int_{0}^{1} d\xi\, (1\!-\!\xi) \left\{\big[2z\!-\!1\!-\!z\xi f_{(D_s)}\big]\; \frac{\Pt^j\, \mathcal{B}^j}{\Pt^2} 
+\big[z\!-\!1\!-\!z\xi f_{(D_s)}\big]\; \xi\, \mathcal{B}_0 \right\} 
\label{SL1_gen_Ds}.
\end{align}
The (UV and IR finite) light cone helicity flip contribution~\nr{eq:hfc_2} will not contribute to the cross section at NLO, since it cannot interfere with the LO LFWF which is 
helicity non-flip.
Hence, we will from here on concentrate only on the helicity conserving part that does contribute to the cross section at NLO. The helicity flip term can be used, as discussed in Appendix \ref{app:formfactorSL}, to rederive the known result for the one-loop Pauli form factor of a massive quark, serving as an additional cross check of our result.

To proceed with the helicity conserving part, we first write the expression in \eq\nr{eq:factnloc} as a sum of two terms
\begin{equation}
\label{eq:vcsplit}
\mathcal{V}^{L}_{\ref{diag:vertexqbaremL}} = \mathcal{V}^{L}_{\ref{diag:vertexqbaremL}}\bigg\vert_{\mathcal{A}} + \mathcal{V}^{L}_{\ref{diag:vertexqbaremL}}\bigg\vert_{\mathcal{B}},
\end{equation}
where the first and the second term in r.h.s. of \eq\nr{eq:vcsplit} are the UV divergent and the UV finite pieces of $\mathcal{V}_{\ref{diag:vertexqbaremL}}$, respectively. Following the notation of \eq\nr{eq:ILc}, these two terms can be written as 
\begin{equation}
\label{eq:vcA}
\mathcal{V}^{L}_{\ref{diag:vertexqbaremL}}\bigg\vert_{\mathcal{A}} = -(1-z)\int_{\alpha/z}^{1}\ud \xi \frac{(1-\xi)}{\xi^2}\Biggl [1 + (1-\xi)\left (1 + \frac{z\xi}{1-z}\right ) - \frac{(D_s-4)}{2}\frac{z\xi^2}{1-z}\Biggr ] \mathcal{A}_0(\Delta_2)
\end{equation}
and
\begin{equation}
\label{eq:vcB}
\begin{split}
\mathcal{V}^{L}_{\ref{diag:vertexqbaremL}}\bigg\vert_{\mathcal{B}} = \int_{\alpha/z}^{1}\ud \xi &\Biggl \{-\frac{z^2\xi^2(1-\xi)}{1-z}m^2 \mathcal{B}_0(\Delta_1,\Delta_2,\Lt)  +
\frac{(1-\xi)}{\xi}\Biggl [1 + (1-\xi)\left (1 + \frac{z\xi}{1-z}\right )\Biggr ]\biggl \{-\Pt^i \mathcal{B}^i(\Delta_1,\Delta_2,\Lt)\\
& + \biggl [(1-\xi)[\Pt^2 + \overline{Q}^2 + m^2] + \xi(1-z)m^2 \biggr ]\mathcal{B}_0(\Delta_1,\Delta_2,\Lt) \biggr \}\Biggr \}.
\end{split}
\end{equation}

At this stage, we could carry on and evaluate explicitly the UV divergent term above, but it turns out to be convenient to first add a contribution coming from the instantaneous diagram \ref{diag:vertexqemLinst}. 
Using the notation shown in \fig\ref{fig:vertexqemLinst}, the contribution of the  instantaneous diagram \ref{diag:vertexqemLinst} to the LCWF is given by 
\begin{equation}
\label{eq:vertexinst}
\begin{split}
\Psi^{\gamma^{\ast}_L\rightarrow q\bar{q}}_{\ref{diag:vertexqemLinst}}   = \frac{1}{8\pi}\int_{0}^{q^+}\frac{\ud k^+_{0'}}{k^+_{0'}(q^+-k^+_{0'})}\int\frac{\ud^{d-2}\ktpzero}{(2\pi)^{d-2}}\frac{N^ L_{\ref{diag:vertexqemLinst}}}{\ed_{\rm v}\ed_{\lo}},
\end{split}
\end{equation}
where the numerator can be simplified to
\begin{equation}
N^L_{\ref{diag:vertexqemLinst}} = -\delta_{\alpha_0\alpha_1}V^{\gamma^{\ast}_{L}\rightarrow q\qvec}_{h_0;h_1}(g^2\cf)\frac{4k^+_{0'}(q^+-k^+_{0'})}{(k^+_{0'}-k^+_0)^2}.
\end{equation}
As discussed in \cite{Beuf:2016wdz, Hanninen:2017ddy} (but now in the massive quark case), the LCWF can be split up into two UV divergent contributions according to the direction of the momentum flow along the instantaneous gluon line.  It is straightforward to show that these two contributions take the following form \cite{Beuf:2016wdz}:
\begin{equation}
\begin{split}
\Psi^{\gamma^{\ast}_L\rightarrow q\bar{q}}_{\ref{diag:vertexqemLinst}}  = \Psi^{\gamma^{\ast}_L\rightarrow q\bar{q}}_{\lo}\left (\frac{\alpha_s\cf}{2\pi}\right )\biggl [\mathcal{V}^{L}_{\ref{diag:vertexqemLinst}_1}\bigg\vert_{\mathcal{A}} + \mathcal{V}^{L}_{\ref{diag:vertexqemLinst}_2}\bigg\vert_{\mathcal{A}}\biggr ],
\end{split}
\end{equation}
where, in the first term $\ref{diag:vertexqemLinst}_1$, the light cone momentum of the gluon line is flowing upwards into the quark line and in the second term $\ref{diag:vertexqemLinst}_2$ in the opposite direction. The UV divergent coefficient corresponding to the first term simplifies to 
\begin{equation}
\label{eq:v1eA}
\mathcal{V}^{L}_{\ref{diag:vertexqemLinst}_1}\bigg\vert_{\mathcal{A}} = 2\int_{0}^{k^+_0} \frac{\ud k^+}{q^+}\biggl [\frac{k^+_0k^+_1}{(k^+)^2} + \frac{(k^+_0-k^+_1)}{k^+} -1\biggr ]\mathcal{A}_0(\Delta_2) = 2(1-z)\int^{1}_{\alpha/z}\ud \xi\frac{(1-\xi)}{\xi^2}\biggl [1 + \frac{z\xi}{(1-z)}\biggr ]\mathcal{A}_0(\Delta_2).
\end{equation}
The coefficient $\mathcal{V}^{L}_{\ref{diag:vertexqemLinst}_2}$, where the light cone momentum of the gluon line is flowing downwards into the antiquark line, is obtained by making the substitution $z \mapsto 1-z$ in \eq\nr{eq:v1eA}.

We can now calculate the sum of diagrams \ref{diag:vertexqbaremL} and \ref{diag:vertexqemLinst} in  momentum space. First, summing the UV divergent pieces of the vertex correction diagram in \eqs\nr{eq:vcA} and the $\ref{diag:vertexqemLinst}_1$ part of the instantaneous diagram in \nr{eq:v1eA} together we obtain
\begin{equation}
\label{eq:sumvceA}
\biggl [\mathcal{V}^{L}_{\ref{diag:vertexqbaremL}} + \mathcal{V}^{L}_{\ref{diag:vertexqemLinst}_1}\biggr ]\bigg\vert_{\mathcal{A}} = \int^{1}_{\alpha/z} \ud \xi \frac{(1-\xi)}{\xi}\left (1 + z\xi\right )\mathcal{A}_0(\Delta_2) + \frac{(D_s-4)}{2}z\int^{1}_{0} \ud \xi (1-\xi)\mathcal{A}_0(\Delta_2).
\end{equation}
Here, the unphysical term $\ud \xi/\xi^2$ cancels out between the $A_0(\Delta_2)$ terms in the diagrams \ref{diag:vertexqbaremL} and \ref{diag:vertexqemLinst}, and the remaining integrals in \eq\nr{eq:sumvceA} can be performed analytically. Using the expression in \eq\nr{eq:A0exp} for  $A_0(\Delta_2)$, we can rewrite the expression above as
\begin{equation}
\label{eq:sumvceAv2}
\begin{split}
\biggl [\mathcal{V}^{L}_{\ref{diag:vertexqbaremL}} + \mathcal{V}^{L}_{\ref{diag:vertexqemLinst}_1}\biggr ]\bigg\vert_{\mathcal{A}} = \biggl [-1 + \frac{z}{2} & - \log\left (\frac{\alpha}{z}\right ) \biggr ] \frac{(4\pi)^{2-\frac{D}{2}}}{(2-\frac{D}{2})}\Gamma\left (3-\frac{D}{2}\right )   - \frac{z}{2}\frac{(D_s-4)}{(D-4)}\\
 &+ \int^{1}_{\alpha/z} \ud \xi \frac{(1-\xi)}{\xi}\left (1 + z\xi\right )\biggl [-\log\left (\frac{\Delta_2}{\mu^2}\right )\biggr ] + \mathcal{O}(D-4),
\end{split}
\end{equation}
For the remaining $\xi$ integral it is convenient to first factorize the $\Pt$ independent coefficient $\Delta_2$ with respect to $\xi$ as 
\begin{equation}
\Delta_2 = \frac{z}{(1-z)}\overline{Q}^2\left (\xi_{(+)} - \xi\right )\left (\xi - \xi_{(-)}\right ),
\end{equation}
where the zeroes in $\xi$ are given by
\begin{equation}
\label{eq:xizeros}
\xi_{(\pm)} = 1 - \frac{1}{2z} \pm \frac{1}{2z}\sqrt{1 + \frac{4z(1-z)m^2}{\overline{Q}^2}} =  1 - \frac{1}{2z} \pm \frac{1}{2z}\gamma,\quad  \text{with}\quad \gamma = \sqrt{1 + \frac{4m^2}{Q^2}}.
\end{equation}
The square root in \eq\nr{eq:xizeros} is associated with the threshold for massive quark pair creation if we were interested
in the timelike photon case. In the spacelike case of interest note that $\xi_{(+)} > 1$ and $\xi_{(-)} < 0$ (which can become
equalities for massless quarks) so that both zeros of $\Delta_2$ sit outside of the integration range in $\xi$. Utilizing these observations, we find the result 
\begin{equation}
\label{eq:sumvceAv3}
\biggl [\mathcal{V}^{L}_{\ref{diag:vertexqbaremL}} + \mathcal{V}^{L}_{\ref{diag:vertexqemLinst}_1}\biggr ]\bigg\vert_{\mathcal{A}} = \biggl [-1 + \frac{z}{2} - \log\left (\frac{\alpha}{z}\right ) \biggr ] \frac{(4\pi)^{2-\frac{D}{2}}}{(2-\frac{D}{2})}\Gamma\left (3-\frac{D}{2}\right )   - \frac{z}{2}\frac{(D_s-4)}{(D-4)}  + \mathcal{I}_{\xi;1} - \mathcal{I}_{\xi;2}  -z\left (\mathcal{I}_{\xi;3} -\mathcal{I}_{\xi;2}\right )+ \mathcal{O}(D-4),	
\end{equation}
where closed analytical expressions for the integrals $\mathcal{I}_{\xi;1}$, $\mathcal{I}_{\xi;2}$ and $\mathcal{I}_{\xi;3}$ are given in Appendix \ref{app:logintegrals} (see \eqs\nr{eq:xiint1}, \nr{eq:xiint2} and \nr{eq:xiint3}).

The computation of the UV finite part of the coefficient in \eq\nr{eq:vcB} is a bit more tricky. It is possible to directly compute the $\mathcal{B}_0$ and $\mathcal{B}^i$ integrals, but the result would be too complicated for further analytical integration, in particular over the required Fourier transform. Instead, we Feynman parametrize the denominator appearing in the $\mathcal{B}_0$ and $\mathcal{B}^i$ as
\begin{equation}
\label{eq:feynmanparam}
\left[ \begin{array}{c}
\mathcal{B}_0  \\
\mathcal{B}^j
\end{array} \right] = \int_{0}^{1} \ud x \left[ \begin{array}{c}
1  \\
-x\Lt^i
\end{array} \right]\frac{1}{[x(1-x)\Lt^2 + (1-x)\Delta_1 + x\Delta_2]},
\end{equation}  
where the denominator can be rewritten as 
\begin{equation}
\label{eq:denomfeynman}
\begin{split}
x(1-x)\Lt^2 + (1-x)\Delta_1 + x\Delta_2 = &(1-\xi)\biggl [(1-x)\Pt^2 + \overline{Q}^2 + m^2\biggr ]\biggl [x(1-\xi) + \frac{\xi}{(1-z)}\biggr ]\\
& + \xi m^2\biggl [\xi(1-x) + x\left (1 - \frac{z(1-\xi)}{(1-z)}\right )\biggr ].
\end{split}
\end{equation}
Now, from \eqs\nr{eq:feynmanparam} and \nr{eq:denomfeynman}, we find\footnote{Here, we have condensed the notation of $\mathcal{B}_0(\Delta_1,\Delta_2,\Lt)$ and $\mathcal{B}^i(\Delta_1,\Delta_2,\Lt)$ to simply $\mathcal{B}_0$ and $\mathcal{B}^i$.}
\begin{equation}
\label{eq:vcBidentity}
\begin{split}
-\Pt^i \mathcal{B}^i + (1-\xi)[\Pt^2 + \overline{Q}^2 + m^2]\mathcal{B}_0 & = \int_{0}^{1} \!\! \ud x \frac{(1-\xi)[(1-x)\Pt^2 + \overline{Q}^2 + m^2]}{[x(1-x)\Lt^2 + (1-x)\Delta_1 + x\Delta_2]}\\
& = \frac{\mathcal{I}_{+} }{(1-\xi)}- \xi m^2\int_{0}^{1} \!\!\! \ud x  \frac{ \xi(1-x) + x\left (1 - \frac{z(1-\xi)}{(1-z)}\right ) }{\biggl [x(1-\xi) + \frac{\xi}{(1-z)}\biggr ][x(1-x)\Lt^2 + (1-x)\Delta_1 + x\Delta_2]},
\end{split}
\end{equation}
in which the integral $\mathcal{I}_+$ is defined as 
\begin{equation}
\mathcal{I}_+ = \int_{0}^{1}\ud x \frac{(1-\xi)}{\biggl [x(1-\xi) + \frac{\xi}{(1-z)}\biggr ]} = -\log(\xi) + \log\left (1 - z(1-\xi)\right ). 
\end{equation}
Substituting the expression in \eq\nr{eq:vcBidentity} into \eq\nr{eq:vcB} yields
\begin{equation}
\label{eq:vcBsplit}
\mathcal{V}^{L}_{\ref{diag:vertexqbaremL}}\bigg\vert_{\mathcal{B}} = \int^{1}_{\alpha/z} \frac{\ud \xi}{\xi}\biggl [1 + (1-\xi)\left (1 + \frac{z\xi}{1-z)}\right )\biggr ]\mathcal{I}_{+} + \mathcal{I}_{\mathcal{V}_{\ref{diag:vertexqbaremL}}}(z,\Pt),
\end{equation}
where we have defined the function $\mathcal{I}_{\mathcal{V}_{\ref{diag:vertexqbaremL}}}$ as
\begin{equation}
\label{eq:IVcint}
\mathcal{I}_{\mathcal{V}_{\ref{diag:vertexqbaremL}}}(z,\Pt) = m^2\int^{1}_{0}\ud \xi \int^{1}_{0}\ud x \frac{C^{L}_{m}}{[x(1-x)\Lt^2 + (1-x)\Delta_1 + x\Delta_2]}
\end{equation}
with the coefficient
\begin{equation}\label{eq:defClm}
C^{L}_{m} = \frac{z^2(1-\xi)}{(1-z)}\Biggl \{-\xi^2 + x(1-\xi)\frac{\biggl [1 + (1-\xi)\left (1 + \frac{z\xi}{(1-z)}\right )\biggr ]}{\biggl [x(1-\xi) + \frac{\xi}{(1-z)} \biggr ]} \Biggr \}.
\end{equation}
The first term in r.h.s. of \eq\nr{eq:vcBsplit} is the same integral appearing in the massless case \cite{Beuf:2016wdz}. This integral can be done analytically and it contains both single and double logs in $\alpha$, but no dependence on $\Pt$ (thus this contribution factors out of the Fourier transform). The second term in r.h.s. of \eq\nr{eq:vcBsplit} is an additional UV and $\xi\rightarrow 0$ finite contribution coming from the massive quarks.  In \eq\nr{eq:IVcint}  we could in principle perform the integration over the $\xi$ and the Feynman parameter $x$ analytically, but it is more convenient to Fourier transform first and then perform the remaining integrals numerically.

Collecting now the contributions from \eqs\nr{eq:sumvceAv3} and \nr{eq:vcBsplit} together, we obtain the result
\begin{equation}
\label{eq:fullc1e}
\begin{split}
\Psi^{\gamma^{\ast}_L\rightarrow q\bar{q}}_{\ref{diag:vertexqbaremL} + \ref{diag:vertexqemLinst}_1} = \Psi^{\gamma^{\ast}_L\rightarrow q\bar{q}}_{\lo}\left (\frac{\alpha_s\cf}{2\pi}\right )\mathcal{V}^{L}_{\ref{diag:vertexqbaremL} + \ref{diag:vertexqemLinst}_1}  + \Psi^{\gamma^{\ast}_L\rightarrow q\bar{q}}_{\text{h.f.};\ref{diag:vertexqbaremL}},
\end{split}
\end{equation}
where the NLO form factor $\mathcal{V}^{L}_{\ref{diag:vertexqbaremL} + \ref{diag:vertexqemLinst}_1}$ can be simplified to
\begin{equation}
\label{eq:Vc1emomspace}
\begin{split}
\mathcal{V}^{L}_{\ref{diag:vertexqbaremL} + \ref{diag:vertexqemLinst}_1} & =  \biggl [-1 + \frac{z}{2}  - \log\left (\frac{\alpha}{z}\right ) \biggr ]\frac{(4\pi)^{2-\frac{D}{2}}}{(2-\frac{D}{2})}\Gamma\left (3-\frac{D}{2}\right )  - \frac{z}{2}\frac{(D_s-4)}{(D-4)}  + \mathcal{I}_{\xi;1} - \mathcal{I}_{\xi;2}-z\left (\mathcal{I}_{\xi;3}-\mathcal{I}_{\xi;2}\right ) - \frac{1}{2} \\
&   + \log^2\left (\frac{\alpha}{z}\right ) + \frac{1-3z}{2z}\log(1-z) - 2\log(1-z)\log\left (\frac{\alpha}{z}\right )  - 2\mathrm{Li}_2\left (-\frac{z}{1-z}\right ) + \mathcal{I}_{\mathcal{V}_{\ref{diag:vertexqbaremL}}}(z,\Pt) + \mathcal{O}(D-4).
\end{split}
\end{equation} 
Here, the function $\mathrm{Li}_2(z)$ is the dilogarithm function, defined as 
\begin{equation}
\mathrm{Li}_2(z) =  -\int_{0}^{z} \frac{\ud \xi}{\xi} \log(1-\xi).
\end{equation}

The final result for the full NLO vertex and instantaneous contribution in the momentum space can be obtained by first computing the contribution from the diagrams \ref{diag:vertexqemL} + $\ref{diag:vertexqemLinst}_2$, and then adding the obtained result together with the contribution in \eq\nr{eq:fullc1e}. The first part of the given task is most easily obtained by using the symmetry between diagrams  \ref{diag:vertexqbaremL} + $\ref{diag:vertexqemLinst}_1$ and \ref{diag:vertexqemL} + $\ref{diag:vertexqemLinst}_2$, i.e. by  making the substitution  $z \mapsto 1-z$ and $\Pt \mapsto -\Pt$ simultaneously in \eq\nr{eq:fullc1e}. This yields
\begin{equation}
\label{eq:fulld2e}
\begin{split}
\Psi^{\gamma^{\ast}_L\rightarrow q\bar{q}}_{\ref{diag:vertexqemL} + \ref{diag:vertexqemLinst}_2} = \Psi^{\gamma^{\ast}_L\rightarrow q\bar{q}}_{\lo}\left (\frac{\alpha_s\cf}{2\pi}\right )\mathcal{V}^{L}_{\ref{diag:vertexqemL} + \ref{diag:vertexqemLinst}_2}  + \Psi^{\gamma^{\ast}_L\rightarrow q\bar{q}}_{\text{h.f.};\ref{diag:vertexqemL}},
\end{split}
\end{equation}
where the NLO form factor $\mathcal{V}^{L}_{\ref{diag:vertexqemL} + \ref{diag:vertexqemLinst}_2}$ can be written as 
\begin{equation}
\label{eq:Vd1emomspace}
\begin{split}
\mathcal{V}^{L}_{\ref{diag:vertexqemL} + \ref{diag:vertexqemLinst}_2} =  & \biggl [-1 + \frac{1-z}{2}  - \log\left (\frac{\alpha}{1-z}\right ) \biggr ]\frac{(4\pi)^{2-\frac{D}{2}}}{(2-\frac{D}{2})}\Gamma\left (3-\frac{D}{2}\right )  - \frac{(1-z)}{2}\frac{(D_s-4)}{(D-4)}\\
& + \biggl [\mathcal{I}_{\xi;1} - \mathcal{I}_{\xi;2} -z\left (\mathcal{I}_{\xi;3}-\mathcal{I}_{\xi;2}\right ) \biggr ]\bigg\vert_{z \mapsto 1-z} - \frac{1}{2} + \log^2\left (\frac{\alpha}{1-z}\right )\\
& + \frac{1-3(1-z)}{2(1-z)}\log(z) - 2\log(z)\log\left (\frac{\alpha}{1-z}\right )  - 2\mathrm{Li}_2\left (-\frac{1-z}{z}\right ) + \mathcal{I}_{\mathcal{V}_{\ref{diag:vertexqemL}}}(z,\Pt) + \mathcal{O}(D-4).
\end{split}
\end{equation} 
Here, the function $\mathcal{I}_{\mathcal{V}_{\ref{diag:vertexqemL}}}$ is given by
\begin{equation}
\label{eq:IVdint}
\mathcal{I}_{\mathcal{V}_{\ref{diag:vertexqemL}}}(z,\Pt) = 
m^2\int^{1}_{0}\ud \xi \int^{1}_{0}\ud x \frac{\overline C^{L}_{m}}{[x(1-x)\Lt^2 + (1-x) \overline \Delta_1 + x \overline \Delta_2]},
\end{equation}
where the coefficients $\overline \Delta_1, \overline \Delta_2$ and $\overline C^{L}_{m}$ reduce to 
\begin{equation}
\begin{split}
\overline \Delta_1 & = \frac{\xi(1-\xi)}{z}\biggl [\Pt^2 + \overline{Q}^2 + m^2\biggr ] + \xi^2m^2,\\
\overline \Delta_2 & = (1-\xi)\left (1 + \frac{\xi (1-z)}{z}\right )\overline{Q}^2  + m^2,
\end{split}
\end{equation}
and
\begin{equation}\label{eq:defClmbar}
\overline C^{L}_{m} = \frac{(1-z)^2(1-\xi)}{z}\Biggl \{-\xi^2 + x(1-\xi)\frac{\biggl [1 + (1-\xi)\left (1 + \frac{(1-z)\xi}{z}\right )\biggr ]}{\biggl [x(1-\xi) + \frac{\xi}{z} \biggr ]} \Biggr \}.
\end{equation}

All in all, the final result for the full NLO vertex and instantaneous contribution in the momentum space is obtained by summing \eqs\nr{eq:fullc1e} and \nr{eq:fulld2e} together. After some amount of algebra, we obtain
\begin{equation}
\label{eq:finalVERTEX}
\Psi^{\gamma^{\ast}_L\rightarrow q\bar{q}}_{\ref{diag:vertexqemL} + \ref{diag:vertexqemL} + \ref{diag:vertexqemLinst}} = \Psi^{\gamma^{\ast}_L\rightarrow q\bar{q}}_{\lo}\left (\frac{\alpha_s\cf}{2\pi}\right )\mathcal{V}^{L}_{\ref{diag:vertexqemL} + \ref{diag:vertexqemL} + \ref{diag:vertexqemLinst}}  + \Psi^{\gamma^{\ast}_L\rightarrow q\bar{q}}_{\text{h.f.};\ref{diag:vertexqbaremL}+\ref{diag:vertexqemL}},
\end{equation}
where the NLO form factor $\mathcal{V}^{L}_{\ref{diag:vertexqemL} + \ref{diag:vertexqemL} + \ref{diag:vertexqemLinst}}$ can be written as 
\begin{equation}
\begin{split}
\mathcal{V}^{L}_{\ref{diag:vertexqemL} + \ref{diag:vertexqemL} + \ref{diag:vertexqemLinst}}  & = -\biggl [\frac{3}{2} + \log\left (\frac{\alpha}{z}\right ) + \log\left (\frac{\alpha}{1-z}\right ) \biggr ]\Biggl \{\frac{(4\pi)^{2-\frac{D}{2}}}{(2-\frac{D}{2})}\Gamma\left (3-\frac{D}{2}\right )  + \log\left (\frac{\mu^2}{\overline{Q}^2 + m^2}\right )\Biggr \}\\
&  -\frac{7}{2}  - \frac{1}{2}\frac{(D_s-4)}{(D-4)} - \biggl [\log(z) + \log(1-z)
\biggr ]\left (\frac{3}{2} + 2\log(\alpha) \right ) + 4\log(1-z)\log(z) 
+ \log^2\left (\frac{\alpha}{z}\right )\\
& + \log^2\left (\frac{\alpha}{1-z}\right )  -2\mathrm{Li}_{2}\left (-\frac{z}{1-z} \right ) - 2\mathrm{Li}_{2}
\left (-\frac{1-z}{z} \right ) + L(\gamma;z)\\
& + \Omega_{\mathcal{V}}(\gamma;z) +  \mathcal{I}_{\mathcal{V}_{\ref{diag:vertexqbaremL}}}(z,\Pt) + \ \mathcal{I}_{\mathcal{V}_{\ref{diag:vertexqemL}}}(z,\Pt) + \mathcal{O}(D-4).
\end{split}
\end{equation}
Here, we have again used a compact notation by introducing the functions $\Omega_{\mathcal{V}}(\gamma;z)$ and $L(\gamma;z)$, which are given by
\begin{equation}
\label{eq:GammaLm}
\begin{split}
\Omega_{\mathcal{V}}(\gamma;z)  =   \frac{1}{2z}\biggl [\log(1-z) &+ \gamma \log\left (\frac{1+\gamma}{1+\gamma - 2z}\right ) \biggr ]  + \frac{1}{2(1-z)}\biggl [\log(z) + \gamma \log\left (\frac{1+\gamma}{1+\gamma - 2(1-z)}\right ) \biggr ] \\
& + \frac{1}{4z(1-z)}\left ( \gamma - 1\right )\log\left (\frac{\overline{Q}^2 + m^2}{m^2}\right ) + \frac{m^2}{2\overline{Q}^2}\log\left (\frac{\overline{Q}^2 + m^2}{m^2}\right ),
\end{split}
\end{equation}
and
\begin{equation}
\label{eq:Lfunction}
L(\gamma;z) =  \mathrm{Li}_{2}\left (\frac{1}{1-\frac{1}{2z}(1-\gamma)} \right )  + \mathrm{Li}_{2}\left (\frac{1}{1-\frac{1}{2(1-z)}(1-\gamma)} \right )  
+ \mathrm{Li}_{2}\left (\frac{1}{1-\frac{1}{2z}(1+\gamma)} \right )  + \mathrm{Li}_{2}\left (\frac{1}{1-\frac{1}{2(1-z)}(1+\gamma)}  \right ),
\end{equation}
with
\begin{equation}
\gamma = \sqrt{1 + \frac{4m^2}{Q^2}}.
\end{equation}

In the massless limit (i.e. $\gamma \rightarrow 1$), the coefficient $\Omega_{\mathcal{V}}$, integrals $\mathcal{I}_{\mathcal{V}_{\ref{diag:vertexqbaremL}}},  \mathcal{I}_{\mathcal{V}_{\ref{diag:vertexqemL}}}$ and the light cone helicity flip term vanish, and the function $L$ satisfies 
\begin{equation}\label{eq:Lmassless}
L(1;z) = \frac{\pi^2}{3} + \mathrm{Li}_{2}\left (-\frac{z}{1-z} \right ) + \mathrm{Li}_{2}\left (-\frac{1-z}{z} \right ),
\end{equation}
where the sum of two dilogarithm function can be simplified by using the following identity
\begin{equation}
\mathrm{Li}_2\left (-\frac{z}{1-z} \right ) + \mathrm{Li}_2\left (-\frac{1-z}{z} \right ) = -\frac{\pi^2}{6} - \frac{1}{2}\log^2\left (\frac{z}{1-z}\right ),  \quad\quad \frac{z}{1-z} > 0.
\end{equation}
Using these observations one sees that in the massless limit \eq\nr{eq:finalVERTEX} simplifies to the result obtained in \cite{Beuf:2016wdz} and \cite{Hanninen:2017ddy}.

\section{The $\gamma^{\ast}_{L} \rightarrow q\bar{q}$ wave function in  coordinate space} 
\label{sec:fourier}

\subsection{Result in momentum space}\label{sec:qqbargmomspace}

Having all these results at hand, we can now write down the final result for the mass renormalized one-loop corrected $\gamma^{\ast}_{L} \rightarrow q\bar q$ LCWF in momentum space. Adding all the contributions in \eq\nr{eq:finalSE} and \eq\nr{eq:finalVERTEX} together, we obtain the final result
\begin{equation}
\label{eq:finalNLOmomspace}
\Psi^{\gamma^{\ast}_L\rightarrow q\bar{q}}_{\nlo} = \Psi^{\gamma^{\ast}_L\rightarrow q\bar{q}}_{\lo}\left (\frac{\alpha_s\cf}{2\pi}\right )\mathcal{V}^{L}  + \Psi^{\gamma^{\ast}_L\rightarrow q\bar{q}}_{\text{h.f.}},
\end{equation}
where the full form factor $\mathcal{V}^{L}$ at NLO simplifies to 
\begin{equation}
\label{eq:Vnlo}
\begin{split}
\mathcal{V}^{L} =  & \biggl [\frac{3}{2}  + \log\left (\frac{\alpha}{z}\right ) + \log\left (\frac{\alpha}{1-z}\right ) \biggr ]\Biggl \{ \frac{(4\pi)^{2-\frac{D}{2}}}{(2-\frac{D}{2})}\Gamma\left (3-\frac{D}{2}\right ) + \log\left (\frac{\mu^2}{\overline{Q}^2 + m^2}\right ) - 2\log\left (\frac{\Pt^2 + \overline{Q}^2 + m^2}{\overline{Q}^2 + m^2}\right )\Biggr \}\\
&+ \frac{5}{2} + \frac{1}{2}\frac{(D_s-4)}{(D-4)} - \frac{\pi^2}{3} + \log^2\left (\frac{z}{1-z}\right ) + \Omega_{\mathcal{V}}(\gamma;z) + L(\gamma;z) + \mathcal{I}_{\mathcal{V}}(z,\Pt)  + \mathcal{O}(D-4)
\end{split}
\end{equation}
with
\begin{equation}
\label{eq:IVsum}
\mathcal{I}_{\mathcal{V}}(z,\Pt) =  \mathcal{I}_{\mathcal{V}_{\ref{diag:oneloopSEUPL} + \ref{diag:oneloopSEDOWNL}}}(z,\Pt)  +  \mathcal{I}_{\mathcal{V}_{\ref{diag:vertexqbaremL} + \ref{diag:vertexqemL}}}(z,\Pt).
\end{equation}
Here, the functions $\mathcal{I}_{\mathcal{V}_{\ref{diag:oneloopSEUPL} + \ref{diag:oneloopSEDOWNL}}}$ and $\mathcal{I}_{\mathcal{V}_{\ref{diag:vertexqbaremL} + \ref{diag:vertexqemL}}}$ are given by the sum of integral expressions in \eqs\nr{eq:IVaint}, \nr{eq:IVbint} and \eqs\nr{eq:IVcint}, \nr{eq:IVdint}, respectively.
In the massless quark limit $m=0$ the functions 
$\mathcal{I}_{\mathcal{V}_{\ref{diag:oneloopSEUPL} + \ref{diag:oneloopSEDOWNL}}}$, $\mathcal{I}_{\mathcal{V}_{\ref{diag:vertexqbaremL} + \ref{diag:vertexqemL}}}$
and $\Omega_{\mathcal{V}}(\gamma;z)$ vanish. Using the massless limit of $L(\gamma;z)$ in \eq\nr{eq:Lmassless} it is then easy to see that the LCWF reduces to the known result of Ref.~\cite{Beuf:2016wdz,Hanninen:2017ddy}.
For clarity, in here we do not show explicitly the light cone helicity flip term $\Psi^{\gamma^{\ast}_L\rightarrow q\bar{q}}_{\text{h.f.}}$, since this contribution vanishes at the cross section level. 

\subsection{Fourier transformation to mixed space}

We now Fourier transform the full NLO result of $\gamma^{\ast}_L\rightarrow q\bar{q}$ LCWF in \eq\nr{eq:finalNLOmomspace} into mixed space by first using the explicit expression \eq\nr{eq:LOlcwf} for the leading order LCWF in the factorized form of the NLO result~\nr{eq:fullc1e}. We then factor out the exponential dependence on the center-of-mass coodinate of the dipole and the momentum of the photon as in \eq\nr{eq:LCWFgeneral}. This yields
\begin{equation}
\label{eq:finalNLOmixspace}
\widetilde{\Psi}^{\gamma^{\ast}_L\rightarrow q\bar{q}}_{\nlo} = \delta_{\alpha_0\alpha_1}e^{\frac{i\qt}{q^{+}}\cdot \left (k^+_0\xt_0 + k^+_1\xt_1\right )}\widetilde{\psi}^{\gamma^{\ast}_L\rightarrow q\bar{q}}_{\nlo}  + \widetilde{\Psi}^{\gamma^{\ast}_L\rightarrow q\bar{q}}_{\text{h.f.}},
\end{equation}
where the reduced NLO LCWF in mixed space simplifies to 
\begin{equation}
\label{eq:finalNLOmixspacereduced}
\widetilde{\psi}^{\gamma^{\ast}_L\rightarrow q\bar{q}}_{\nlo} = \frac{-2ee_fQ}{2\pi}z(1-z)\bar{u}(0)\gamma^{+}v(1)\left (\frac{\alpha_s\cf}{2\pi} \right )\widetilde{\mathcal{V}}^{L}  
\end{equation}
and the Fourier transformed form factor $\widetilde{\mathcal{V}}^{L}$ is given by 
\begin{equation}
\label{eq:VnloFT}
\widetilde{\mathcal{V}}^{L} = 2\pi\int \frac{\ud^{D-2}\Pt}{(2\pi)^{D-2}}\frac{e^{i\Pt\cdot \xt_{01}}}{\Pt^2 + \overline{Q}^2 + m^2}\times \mathcal{V}^{L}.
\end{equation}
Before writing down the final result for \eq\nr{eq:VnloFT}, we need to clarify some important points. Firstly, the UV finite terms appearing in \eq\nr{eq:Vnlo} can be Fourier transformed in four dimensions and only the UV regularization dependent terms (including the $D\rightarrow 4$ pole term) need to be Fourier transformed in $D$ dimensions. Secondly, most of the NLO correction terms appearing in \eq\nr{eq:Vnlo} are independent of  the transverse momentum $\Pt$ and thus factor out from the Fourier transform. In these cases, \eq\nr{eq:VnloFT} reduces to the LO Fourier transform \eq\nr{eq:F1} given in Appendix~\ref{app:FTSqqbarcase}. Thirdly, in the NLO correction \eq\nr{eq:Vnlo}, there are only three different types of $\Pt$ dependent terms. All the corresponding transverse Fourier integrals needed for these terms are given in \eqs\nr{eq:F2}, \nr{eq:F3} and \nr{eq:F4}.

Putting these points together, we find the following result
\begin{equation}
\label{eq:VnloFTfinal}
\begin{split}
\widetilde{\mathcal{V}}^{L}  =  &\Biggl \{ \Biggl [\frac{3}{2} + \log\left (\frac{\alpha}{z}\right ) + \log\left (\frac{\alpha}{1-z}\right ) \Biggr ] \Biggl \{ \frac{(4\pi)^{2-\frac{D}{2}} }{ (2-\frac{D}{2})} \Gamma \left (3-\frac{D}{2} \right ) + \log\left (\frac{\xt^2_{01}\mu^2}{4}\right ) + 2\gamma_E\Biggr \} + \frac{1}{2}\frac{(D_s-4)}{(D-4)}  \Biggr \}\\
& \times  \left (\frac{\sqrt{\overline{Q}^2 + m^2}}{2\pi \vert \xt_{01}\vert} \right )^{\frac{D}{2}-2}K_{\frac{D}{2}-2}\left (\vert \xt_{01}\vert\sqrt{\overline{Q}^2 + m^2}\right ) \\
&+ \Biggl \{\log^2\left (\frac{z}{1-z}\right ) - \frac{\pi^2}{3} + \frac{5}{2}  + \Omega_{\mathcal{V}}(\gamma;z)  +  L(\gamma;z)\biggr \}K_0\left (\vert \xt_{01}\vert\sqrt{\overline{Q}^2 + m^2}\right ) + \widetilde{\mathcal{I}}_{\mathcal{V}}(z,\xt_{01}) + \mathcal{O}(D-4), \\
\end{split}
\end{equation}
where $\gamma_E$ is the Euler's constant and $ \widetilde{\mathcal{I}}_{\mathcal{V}}$ is the Fourier transformed version of the integral $\mathcal{I}_{\mathcal{V}}$ in \eq\nr{eq:IVsum} given by the following expression
\begin{equation}
\label{eq:IVsumFT}
\widetilde{\mathcal{I}}_{\mathcal{V}}(z,\xt_{01}) =  \widetilde{\mathcal{I}}_{\mathcal{V}_{\ref{diag:oneloopSEUPL} +\ref{diag:oneloopSEDOWNL}}}(z,\xt_{01})  +  \widetilde{\mathcal{I}}_{\mathcal{V}_{\ref{diag:vertexqbaremL} + \ref{diag:vertexqemL}}}(z,\xt_{01}) 
\end{equation}
with
\begin{equation}
\label{eq:J1int}
\begin{split}
\widetilde{\mathcal{I}}_{\mathcal{V}_{\ref{diag:oneloopSEUPL} +\ref{diag:oneloopSEDOWNL}}}(z,\xt_{01})  =  \int_{0}^{1} \frac{\ud \xi}{\xi}\biggl [-&\frac{2\log(\xi)}{(1-\xi)} + \frac{(1+\xi)}{2}\biggr ]\Biggl \{2K_0\left (\vert \xt_{01}\vert\sqrt{\overline{Q}^2+m^2}\right ) \\
& - K_0\left (\vert \xt_{01}\vert\sqrt{\overline{Q}^2+m^2 + \frac{(1-z)\xi}{1-\xi}m^2}\right ) - K_0\left (\vert \xt_{01}\vert\sqrt{\overline{Q}^2+m^2 + \frac{z\xi}{1-\xi}m^2}\right )\Biggr \},
\end{split}
\end{equation}
and
\begin{equation}
\label{eq:J2int}
\begin{split}
\widetilde{\mathcal{I}}_{\mathcal{V}_{\ref{diag:vertexqbaremL} + \ref{diag:vertexqemL}}}(z,\xt_{01})  = m^2\int_{0}^{1}\ud \xi\int_{0}^{1}\ud x \Biggl \{& \Biggl [K_0\left (\vert \xt_{01}\vert\sqrt{\overline{Q}^2+m^2}\right ) - K_0\left (\vert \xt_{01}\vert\sqrt{\frac{\overline{Q}^2+m^2}{1-x} + \kappa}\right )\Biggr ]\\
& \hspace{1.7cm} \times
\frac{C^{L}_{m}}{(1-\xi)(1-x)\biggl [x(1-\xi) + \frac{\xi}{(1-z)}\biggr ]\biggl [\frac{x\left (\overline{Q}^2 + m^2\right )}{(1-x)} + \kappa\biggr ]}\\
& \hspace{-0.3cm} + \Biggl [K_0\left (\vert \xt_{01}\vert\sqrt{\overline{Q}^2+m^2}\right ) - K_0\left (\vert \xt_{01}\vert\sqrt{\frac{\overline{Q}^2+m^2}{1-x} + \kappa'}\right )\Biggr ]\\
& \hspace{1.7cm} \times
\frac{\overline C^{L}_{m}}{(1-\xi)(1-x)\biggl [x(1-\xi) + \frac{\xi}{z}\biggr ]\biggl [\frac{x\left (\overline{Q}^2 + m^2\right )}{(1-x)} + \kappa'\biggr ]}
\Biggr \}.
\end{split}
\end{equation}
Here the coefficients   $C^{L}_{m}$ and  $\overline C^{L}_{m}$ are defined in \eqs\nr{eq:defClm} and~\nr{eq:defClmbar}, and  $\kappa$ and $\kappa'$ are defined as: 
\begin{equation}
\begin{split}
\kappa & = \frac{\xi m^2}{(1-\xi)(1-x)\biggl [x(1-\xi) + \frac{\xi}{(1-z)}\biggr ]}\biggl [\xi(1-x)+ x\left (1 - \frac{z(1-\xi)}{(1-z)}\right ) \biggr ],\\
\kappa' & = \frac{\xi m^2}{(1-\xi)(1-x)\biggl [x(1-\xi) + \frac{\xi}{z}\biggr ]}\biggl [\xi(1-x)+ x\left (1 - \frac{(1-z)(1-\xi)}{z}\right ) \biggr ].
\end{split}
\end{equation}

Interestingly, the expression for $\widetilde{\mathcal{I}}_{\mathcal{V}_{\ref{diag:vertexqbaremL}}}(z,\xt_{01})$ can be simplified by performing the following chain of changes of variables: $x\mapsto y\equiv \xi+(1-\xi)x$, $\xi\mapsto \eta \equiv \xi/y$, and finally $\eta\mapsto \chi\equiv z(1-\eta)$. One then arrives at
\begin{equation}
\label{eq:J2int_c_alt_1}
\begin{split}
\widetilde{\mathcal{I}}_{\mathcal{V}_{\ref{diag:vertexqbaremL}}}(z,\xt_{01})  = & m^2\int_{0}^{z}\frac{\ud \chi}{(1\!-\!\chi)}\;
 \frac{1}{\left[m^2+\chi(1\!-\!\chi)Q^2\right]}\int_{0}^{1}\ud y
\left[
\frac{2\chi}{y} +\frac{(2z\!-\!1)\chi(z\!-\!\chi)}{z(1\!-\!z)}
-\frac{y(z\!-\!\chi)^2}{z(1\!-\!z)}
\right]
\\
& 
\times
 \left\{K_0\left (\vert \xt_{01}\vert\sqrt{\overline{Q}^2+m^2}\right ) - K_0\left (\vert \xt_{01}\vert\sqrt{
\overline{Q}^2+m^2
+\frac{y(1\!-\!z)}{(1\!-\!y)(1\!-\!\chi)}\left[m^2+\chi(1\!-\!\chi)Q^2\right]
}\right )\right\}
 \, .
\end{split}
\end{equation}
The corresponding expression is obtained by the replacement $z\leftrightarrow (1\!-\!z)$, accompanied for convenience by the change of variable $\chi\leftrightarrow (1\!-\!\chi)$. This leads to
\begin{equation}
\label{eq:J2int_d_alt_1}
\begin{split}
\widetilde{\mathcal{I}}_{\mathcal{V}_{\ref{diag:vertexqemL}}}(z,\xt_{01})  = & m^2\int_{z}^{1}\frac{\ud \chi}{\chi}\;
 \frac{1}{\left[m^2+\chi(1\!-\!\chi)Q^2\right]}\int_{0}^{1}\ud y
\left[
\frac{2(1\!-\!\chi)}{y} +\frac{(2z\!-\!1)(1\!-\!\chi)(z\!-\!\chi)}{z(1\!-\!z)}
-\frac{y(z\!-\!\chi)^2}{z(1\!-\!z)}
\right]
\\
& 
\times
 \left\{K_0\left (\vert \xt_{01}\vert\sqrt{\overline{Q}^2+m^2}\right ) - K_0\left (\vert \xt_{01}\vert\sqrt{
\overline{Q}^2+m^2
+\frac{yz}{(1\!-\!y)\chi}\left[m^2+\chi(1\!-\!\chi)Q^2\right]
}\right )\right\}
 \, .
\end{split}
\end{equation}


This concludes our calculation of the one loop longitudinal photon to quark-antiquark LCWF for massive quarks, needed for the virtual correction to the DIS cross section. We will now proceed to the quark-antiquark-gluon final state needed for the radiative corrections. 


\section{Tree level gluon emission wave function}
\label{sec:qqbg}

\begin{figure}[tb!]
\centerline{
\includegraphics[width=6.4cm]{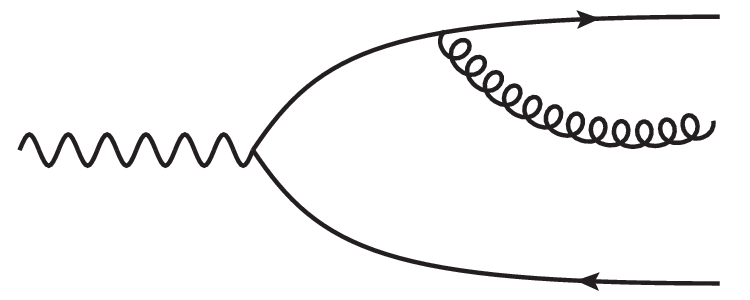}
\hspace*{1.7cm}
\includegraphics[width=6.4cm]{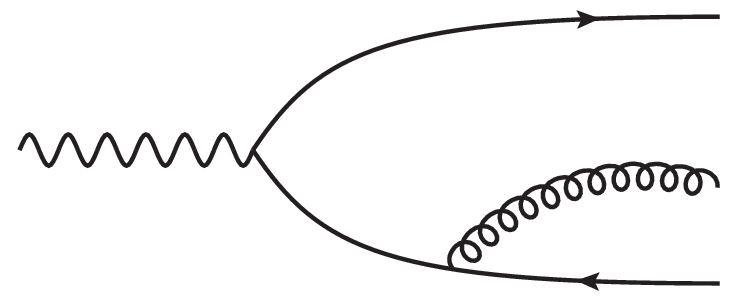}
\begin{tikzpicture}[overlay]
\draw [dashed] (-11.6,2.8) -- (-11.6,0);
\node[anchor=north] at (-11.6cm,-0.2cm) {$\ed_{q_{0'}\bar{q}_{1}}$};
\draw [dashed] (-9.2,2.8) -- (-9.2,0);
\node[anchor=north] at (-9.2,-0.2) {$\ed_{(q_{0}g_2)\bar{q}_{1}}$};
\node[anchor=west] at (-8.6,2.5) {$0,h_0,\alpha_0$};
\node[anchor=west] at (-8.6,1.4) {$2,\sigma,a$};
\node[anchor=west] at (-8.6,0.2) {$1,h_1,\alpha_1$};
\node[anchor=west] at (-13.6,0.9) {$q, \lambda $};
\node[anchor=south west] at (-12.3,2.1) {$0'$};
\node[anchor=east] at (-14.6,1.3) {$\gamma^{\ast}_{L}$};
\draw [dashed] (-3.2,2.8) -- (-3.2,0);
\node[anchor=north] at (-3.2cm,-0.2cm) {$\ed_{q_{0}\bar{q}_{1'}}$};
\draw [dashed] (-0.8,2.8) -- (-0.8,0);
\node[anchor=north] at (-0.8,-0.2) {$\ed_{q_{0} (g_2\bar{q}_{1})}$};
\node[anchor=west] at (-0.3,2.5) {$0,h_0,\alpha_0$};
\node[anchor=west] at (-0.3,1.1) {$2,\sigma,a$};
\node[anchor=west] at (-0.3,0.2) {$1,h_1,\alpha_1$};
\node[anchor=west] at (-5.3,0.9) {$q, \lambda $};
\node[anchor=south west] at (-4.4,0.3) {$1'$};
\node[anchor=east] at (-6.3,1.3) {$\gamma^{\ast}_{L}$};
\node[anchor=south west] at (-14cm,0cm) {\namediag{diag:qgqbarL}};
\node[anchor=south west] at (-7cm,0cm) {\namediag{diag:qqbargL}};
 \end{tikzpicture}
}
\rule{0pt}{1ex}
\caption{Tree level gluon emission diagrams \ref{diag:qgqbarL} and \ref{diag:qqbargL} contributing to the quark-antiquark-gluon component of the transverse virtual photon wave function at NLO. Imposing light cone three momentum conservation at each vertex gives for the diagram \ref{diag:qgqbarL}: $\kvec_{0'} = \kvec_0 + \kvec_2$ and $\qvec = \kvec_0 + \kvec_1 + \kvec_2$. Similarly for diagram\ref{diag:qqbargL}: $\kvec_{1'} = \kvec_1 + \kvec_2$ and $\qvec = \kvec_0 + \kvec_1 + \kvec_2$.}
\label{fig:qgqbarL}
\end{figure}
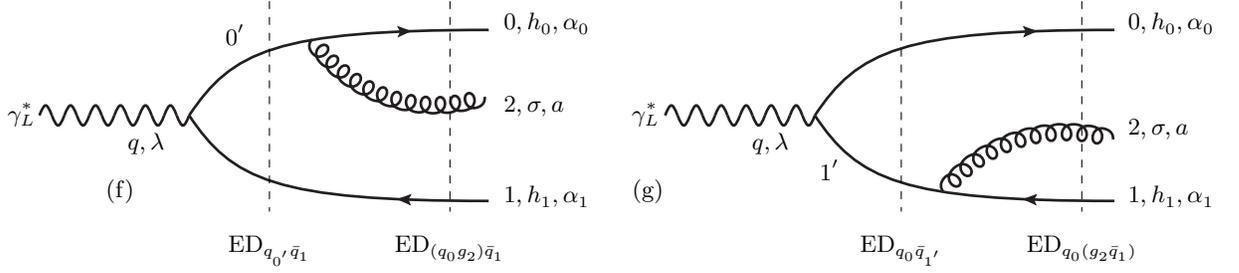

We then move to the tree level wave functions for gluon emission from a longitudinal photon state, which are needed for the full cross section at NLO. As in the massless case \cite{Beuf:2017bpd,Hanninen:2017ddy}, we need to calculate two gluon emission diagrams shown in \fig\ref{fig:qgqbarL}.

Applying the diagrammatic LCPT rules, we obtain for the diagram with gluon emission from the quark, diagram \ref{diag:qgqbarL}, the following expression in momentum space 
\begin{equation}
\label{eq:Lqgqbar}
\begin{split}
\Psi^{\gamma^{\ast}_{L}\rightarrow q\bar{q}g}_{\ref{diag:qgqbarL}} & = \int \dk_{0'} (2\pi)^{D-1}\delta^{(D-1)}(\bar{k}_{0'} - \kvec_2 - \kvec_0)\frac{gt^{a}_{\alpha_0\alpha_1}[\bar{u}(0)\epsl^{\ast}_{\sigma}(k_2)u(0')]ee_f(Q/q^+)[\bar{u}(0')\gamma^{+}v(1)]}{\ed_{q_{0'}\qvec_1}\ed_{(q_0g_2)\qvec_1}}   \\
& = \frac{+ee_fQgt^{a}_{\alpha_0\alpha_1}}{2q^+(k^+_0 + k^+_2)} \frac{\bar{u}(0)\epsl^{\ast}_{\sigma}(k_2)u(0')\bar{u}(0')\gamma^{+}v(1)}{\ed_{q_{0'}\qvec_1}\ed_{(q_0g_2)\qvec_1}},
\end{split}
\end{equation} 
where the energy denominators appearing in \eq\nr{eq:Lqgqbar} can be written as 
\begin{equation}
\begin{split}
\ed_{q_{0'}\qvec_1} & = \frac{\qt^2 - Q^2}{2q^+} - \left (\frac{\kt^2_{0'} + m^2}{2k^+_{0'}} + \frac{\kt^2_1 + m^2}{2k^+_1} \right ) \\
		 & = \frac{-q^+}{2k^+_1(k^+_0 + k^+_2)}\Biggl [\left (\kt_1 - \frac{k^+_1}{q^+}\qt\right )^2 + \overline{Q}^2_{\ref{diag:qgqbarL}} + m^2\Biggr ], \\
\end{split}
\end{equation}
\begin{equation}
\begin{split}
\ed_{(q_0g_2)\qvec_1} & = \frac{\qt^2 - Q^2}{2q^+} - \left (\frac{\kt^2_0 + m^2}{2k^+_0} + \frac{\kt^2_1 + m^2}{2k^+_1} + \frac{\kt^2_2}{2k^+_2} \right ) \\
& = \frac{-k^+_0}{2k^+_2(k^+_0 + k^+_2)}\Biggl [\left (\kt_2 - \frac{k^+_2}{k^+_0}\kt_0\right )^2 + \frac{q^+k^+_2}{k^+_0k^+_1}\Biggl \{\left (\kt_1 - \frac{k^+_1}{q^+}\qt\right )^2 + \overline{Q}^2_{\ref{diag:qgqbarL}} + m^2 + \lambda_{\ref{diag:qgqbarL}} m^2\Biggr \}\Biggr ],
\end{split}
\end{equation}
and the coefficients introduced in the denominators are defined as 
\begin{equation}
\label{eq:coeffqbarqg}
\begin{split}
\overline{Q}^2_{\ref{diag:qgqbarL}} & = \frac{k^+_1(q^+-k^+_1)}{(q^+)^2}Q^2,\quad\quad \lambda_{\ref{diag:qgqbarL}} = \frac{k^+_1k^+_2}{q^+k^+_0}.
\end{split}
\end{equation}
Similarly, for the diagram with emission from the antiquark \ref{diag:qqbargL}, we find 
\begin{equation}
\label{eq:Lqqbarg}
\begin{split}
\Psi^{\gamma^{\ast}_{L}\rightarrow q\bar{q}g}_{\ref{diag:qqbargL}} & = \int \dk_{1'} (2\pi)^{D-1}\delta^{(D-1)}(\bar{k}_{1'} - \kvec_2 - \kvec_1) \frac{gt^{a}_{\alpha_0\alpha_1}[\bar{u}(0)\gamma^{+}v(1')]ee_f(Q/q^+)[\bar{v}(1')\epsl^{\ast}_{\sigma}(k_2)v(1)]}{\ed_{q_0\qvec_{1'}}\ed_{q_0(g_2\qvec_1)}}\\ 
& = \frac{-ee_fQgt^{a}_{\alpha_0\alpha_1}}{2q^+(k^+_1 + k^+_2)} \frac{\bar{u}(0)\gamma^{+}v(1')\bar{v}(1')\epsl^{\ast}_{\sigma}(k_2)v(1)}{\ed_{q_0\qvec_{1'}}\ed_{q_0(g_2\qvec_1)}},
\end{split}
\end{equation} 
where the energy denominators in \eq\nr{eq:Lqqbarg} are given by
\begin{equation}
\begin{split}
\ed_{q_0\qvec_{1'}} & = \frac{\qt^2 - Q^2}{2q^+} - \left (\frac{\kt^2_{0} + m^2}{2k^+_{0}} + \frac{\kt^2_{1'} + m^2}{2k^+_{1'}} \right ) \\ 
	     & = \frac{-q^+}{2k^+_0(k^+_1 + k^+_2)}\Biggl [\left (\kt_0 - \frac{k^+_0}{q^+}\qt\right )^2 + \overline{Q}^2_{\ref{diag:qqbargL}} + m^2\Biggr ],\\
\end{split}
\end{equation}
\begin{equation}
\begin{split}
\ed_{q_0(g_2\qvec_1)} & = \frac{\qt^2 - Q^2}{2q^+} - \left (\frac{\kt^2_0 + m^2}{2k^+_0} + \frac{\kt^2_1 + m^2}{2k^+_1} + \frac{\kt^2_2}{2k^+_2} \right ) \\
& = \frac{-k^+_1}{2k^+_2(k^+_1 + k^+_2)}\Biggl [\left (\kt_2 - \frac{k^+_2}{k^+_1}\kt_1\right )^2 + \frac{q^+k^+_2}{k^+_0k^+_1}\Biggl \{\left (\kt_0 - \frac{k^+_0}{q^+}\qt\right )^2 + \overline{Q}^2_{\ref{diag:qqbargL}} + m^2 + \lambda_{\ref{diag:qqbargL}} m^2\Biggr \}\Biggr ],
\end{split}
\end{equation}
and the coefficients 
\begin{equation}
\label{eq:coeffqbarqgv2}
\begin{split}
\overline{Q}^2_{\ref{diag:qqbargL}} & = \frac{k^+_0(q^+-k^+_0)}{(q^+)^2}Q^2,\quad\quad \lambda_{\ref{diag:qqbargL}} = \frac{k^+_0k^+_2}{q^+k^+_1}.
\end{split}
\end{equation}

We can extract the transverse momentum dependence from the spinor and polarization vector structures in \eqs\nr{eq:Lqgqbar} and~\nr{eq:Lqqbarg} by using the  spinor matrix element decompositions given in \eq\nr{eq:decompv2}. This procedure gives
\begin{equation}
\begin{split}
\bar{u}(0)\epsl^{\ast}_{\sigma}(k_2)u(0') = \frac{k^+_0}{k^+_2(k^+_0 + k^+_2)}\Biggl \{\Biggl [\left (1 + \frac{k^+_2}{2k^+_0}\right )\delta^{ij}_{(D_s)}\bar{u}(0)\gamma^{+}u(0') & - \left (\frac{k^+_2}{4k^+_0}\right )\bar{u}(0)\gamma^{+}[\gamma^i,\gamma^j]u(0') \Biggr ][\kt^i_2 - \frac{k^+_2}{k^+_0}\kt^i_0]\\
& - \frac{m}{2}\left (\frac{k^+_2}{k^+_0}\right )^2\bar{u}(0)\gamma^{+}\gamma^{j}u(0') \Biggr \}\epst^{\ast j}_{\sigma}
\end{split}
\end{equation}
and
\begin{equation}
\begin{split}
\bar{v}(1')\epsl^{\ast}_{\sigma}(k_2)v(1) = \frac{k^+_1}{k^+_2(k^+_1 + k^+_2)}\Biggl \{\Biggl [\left (1 + \frac{k^+_2}{2k^+_1}\right )\delta^{ij}_{(D_s)}\bar{v}(1')\gamma^{+}v(1) & + \left (\frac{k^+_2}{4k^+_1}\right )\bar{v}(1')\gamma^{+}[\gamma^i,\gamma^j]v(1) \Biggr ][\kt^i_2 - \frac{k^+_2}{k^+_1}\kt^i_1]\\
& - \frac{m}{2}\left (\frac{k^+_2}{k^+_1}\right )^2\bar{v}(1')\gamma^{+}\gamma^{j}v(1) \Biggr \}\epst^{\ast j}_{\sigma}.
\end{split}
\end{equation}
Inserting the expressions above into \eqs\nr{eq:Lqgqbar} and \nr{eq:Lqqbarg}, we find for the sum of the diagrams \ref{diag:qgqbarL} and \ref{diag:qqbargL} the result
\begin{equation}
\label{eq:finalradLCWFlong}
\begin{split}
\Psi^{\gamma^{\ast}_{L}\rightarrow q\bar{q}g} = &\frac{4ee_fQgt^{a}_{\alpha_0\alpha_1}}{(q^+)^2} \Biggl \{ \frac{k^+_1(k^+_0+k^+_2)\bar{u}(0)\mathcal{M}^j_{\ref{diag:qgqbarL}}v(1)}{\biggl [\left (\kt_1 - \frac{k^+_1}{q^+}\qt\right )^2 + \overline{Q}^2_{\ref{diag:qgqbarL}} + m^2\biggr ]\biggl [\left (\kt_2 - \frac{k^+_2}{k^+_0}\kt_0\right )^2 + \frac{q^+k^+_2}{k^+_0k^+_1}\biggl \{\left (\kt_1 - \frac{k^+_1}{q^+}\qt\right )^2 + \overline{Q}^2_{\ref{diag:qgqbarL}} + m^2 + \lambda_{\ref{diag:qgqbarL}}m^2\biggr \}\biggr ]}\\
& - \frac{k^+_0(k^+_1+k^+_2)\bar{u}(0)\mathcal{M}^j_{\ref{diag:qqbargL}}v(1)}{\biggl [\left (\kt_0 - \frac{k^+_0}{q^+}\qt\right )^2 + \overline{Q}^2_{\ref{diag:qqbargL}} + m^2\biggr ]\biggl [\left (\kt_2 - \frac{k^+_2}{k^+_1}\kt_1\right )^2 + \frac{q^+k^+_2}{k^+_0k^+_1}\biggl \{\left (\kt_0 - \frac{k^+_0}{q^+}\qt\right )^2 + \overline{Q}^2_{\ref{diag:qqbargL}} + m^2 + \lambda_{\ref{diag:qqbargL}}m^2\biggr \}\biggr ]} \Biggr \}\epst_{\sigma}^{\ast j},
\end{split}
\end{equation}
where the Dirac structures $\mathcal{M}^j_{\ref{diag:qgqbarL}}$ and $\mathcal{M}^j_{\ref{diag:qqbargL}}$ are defined as:
\begin{equation}
\begin{split}
\label{eq:Mdiracqqbarg}
\mathcal{M}^j_{\ref{diag:qgqbarL}} & = \gamma^+\Biggl \{\biggl [\left (1 + \frac{k^+_2}{2k^+_0}\right )\delta^{ij}_{(D_s)} - \left (\frac{k^+_2}{4k^+_0}\right )[\gamma^i,\gamma^j] \biggr ][\kt^i_2 - \frac{k^+_2}{k^+_0}\kt^i_0] - \frac{m}{2}\left (\frac{k^+_2}{k^+_0}\right )^2\gamma^j\Biggr \},\\
\mathcal{M}^j_{\ref{diag:qqbargL}} & = \gamma^+\Biggl \{\biggl [\left (1 + \frac{k^+_2}{2k^+_1}\right )\delta^{ij}_{(D_s)} + \left (\frac{k^+_2}{4k^+_1}\right )[\gamma^i,\gamma^j] \biggr ][\kt^i_2 - \frac{k^+_2}{k^+_1}\kt^i_1] - \frac{m}{2}\left (\frac{k^+_2}{k^+_1}\right )^2\gamma^j\Biggr \}.
\end{split}
\end{equation}

\subsection{Fourier transform to coordinate space}
\label{subsec:FTqqbargL}

The Fourier transform to mixed space of the $\gamma^{\ast}_L \to q\bar qg$ LCWF was defined  by the expression~\nr{eq:FTqqbarg}. 
To simplify the Fourier transformation, we first make the following change of variables $(\kt_1,\kt_2) \mapsto (\Pt,\Kt)$ for the contribution coming from the diagram \ref{diag:qgqbarL} in \eq\nr{eq:finalradLCWFlong}
\begin{equation}
\begin{split}
\Pt & = -\kt_1 + \frac{k^+_1}{q^+}\qt,\\
\Kt & = \kt_2 - \frac{k^+_2}{k^+_0 + k^+_2}(\kt_0 + \kt_2) = \left (\frac{k^+_0}{k^+_0 + k^+_2}\right )\left[\kt_2 - \frac{k^+_2}{k^+_0}\kt_0 \right],
\end{split}
\end{equation}
and correspondingly the following change of variables $(\kt_0,\kt_2) \mapsto (\Pt,\Kt)$ for the contribution coming from the diagram \ref{diag:qqbargL} in \eq\nr{eq:finalradLCWFlong}:
\begin{equation}
\begin{split}
\Pt & = \kt_0 - \frac{k^+_0}{q^+}\qt, \\
\Kt & = \kt_2 - \frac{k^+_2}{k^+_1 + k^+_2}(\kt_1 + \kt_2) = \left (\frac{k^+_1}{k^+_1 + k^+_2}\right )\left[\kt_2 - \frac{k^+_2}{k^+_1}\kt_1 \right].
\end{split}
\end{equation}

Next, performing the integration in \eq\nr{eq:FTqqbarg} over the delta function of transverse momenta, we obtain the expression
\begin{equation}
\label{eq:1}
\widetilde{\Psi}^{\gamma^{\ast}_L\rightarrow q\bar{q}g}_{\ref{diag:qgqbarL}} =  e^{\frac{i\qt}{q^+}\cdot \left (k^+_0\xt_0 + k^+_1\xt_1 + k^+_2\xt_2 \right )}\mu^{2-\frac{D}{2}}\int \frac{\ud^{D-2}\Pt}{(2\pi)^{D-2}}\int \frac{\ud^{D-2}\Kt}{(2\pi)^{D-2}} e^{i\Pt \cdot \xt_{0+2;1}}e^{i\Kt \cdot \xt_{20}} \Psi^{\gamma^{\ast}_{L}\rightarrow q\bar{q}g}_{\ref{diag:qgqbarL}}
\end{equation}
and 
\begin{equation}
\label{eq:2}
\widetilde{\Psi}^{\gamma^{\ast}_L\rightarrow q\bar{q}g}_{\ref{diag:qqbargL}} =  e^{\frac{i\qt}{q^+}\cdot \left (k^+_0\xt_0 + k^+_1\xt_1 + k^+_2\xt_2 \right )}\mu^{2-\frac{D}{2}}\int \frac{\ud^{D-2}\Pt}{(2\pi)^{D-2}}\int \frac{\ud^{D-2}\Kt}{(2\pi)^{D-2}} e^{i\Pt \cdot \xt_{0;1+2}}e^{i\Kt \cdot \xt_{21}} \Psi^{\gamma^{\ast}_{L}\rightarrow q\bar{q}g}_{\ref{diag:qqbargL}},
\end{equation}
where the LCWF's in \eq\nr{eq:finalradLCWFlong} are written in terms of the new variables $\Pt$ and $\Kt$ and the following compact notation is introduced 
\begin{equation}
\label{eq:xtpnmdef}
\xt_{n+m;p} = -\xt_{p;n+m} = \left ( \frac{k^+_n\xt_n + k^+_m\xt_m}{k^+_n + k^+_m} \right ) - \xt_p
\end{equation}
with $\xt_{nm} = \xt_n - \xt_m$.

Making these simplifications, we can write the full Fourier transformed quark-antiquark-gluon LCWF as
\begin{equation}
\begin{split}
\widetilde{\Psi}^{\gamma^{\ast}_{L}\rightarrow q\bar{q}g} = t^{a}_{\alpha_0\alpha_1}e^{\frac{i\qt}{q^+}\cdot \left (k^+_0\xt_0 + k^+_1\xt_1 + k^+_2\xt_2 \right )}\widetilde{\psi}^{\gamma^{\ast}_{L}\rightarrow q\bar{q}g},
\end{split}
\end{equation}
where the reduced wavefunction reads
\begin{equation}
\label{eq:reducedLCWFqqbarg}
\begin{split}
\widetilde{\psi}^{\gamma^{\ast}_{L}\rightarrow q\bar{q}g}= 4ee_fQg\frac{k^+_0k^+_1}{(q^+)^2}&\Biggl \{\bar{u}(0)\gamma^+\biggl [\left (1 + \frac{k^+_2}{2k^+_0}\right )\delta^{ij}_{(D_s)} - \left (\frac{k^+_2}{4k^+_0}\right )[\gamma^i,\gamma^j] \biggr ] v(1)\mathcal{I}^{i}_{\ref{diag:qgqbarL}}\\
& - \bar{u}(0)\gamma^+\biggl [\left (1 + \frac{k^+_2}{2k^+_1}\right )\delta^{ij}_{(D_s)} + \left (\frac{k^+_2}{4k^+_1}\right )[\gamma^i,\gamma^j] \biggr ] v(1)\mathcal{I}^i_{\ref{diag:qqbargL}}\\
& - \frac{m}{2}\Biggl [\left (\frac{k^+_0}{k^+_0 + k^+_2}\right )\left (\frac{k^+_2}{k^+_0}\right )^2\mathcal{I}_{\ref{diag:qgqbarL}} - \left (\frac{k^+_1}{k^+_1 + k^+_2}\right )\left (\frac{k^+_2}{k^+_1}\right )^2\mathcal{I}_{\ref{diag:qqbargL}}  \Biggr ]\bar{u}(0)\gamma^+\gamma^jv(1)  \Biggr \}\epst^{\ast j}_{\sigma}.
\end{split}
\end{equation}
In the above expression, we have introduced  the $D$-dimensional Fourier integrals of the type  $\mathcal{I}^i$ and $\mathcal{I}$, which are defined in \eqs\nr{eq:Irankone} and \nr{eq:Iscalar}, respectively, in Appendix~\ref{app:FTSqqbargcase}.  For these integrals, the following compact notation has also been introduced
\begin{equation}
\begin{split}
\mathcal{I}^{i}_{\ref{diag:qgqbarL}} &= \mathcal{I}^{i}(\xt_{0+2;1},\xt_{20},\overline{Q}^2_{\ref{diag:qgqbarL}},\omega_{\ref{diag:qgqbarL}},\lambda_{\ref{diag:qgqbarL}}), \quad\quad \mathcal{I}_{\ref{diag:qgqbarL}} = \mathcal{I}(\xt_{0+2;1},\xt_{20},\overline{Q}^2_{\ref{diag:qgqbarL}},\omega_{\ref{diag:qgqbarL}},\lambda_{\ref{diag:qgqbarL}})\\
\mathcal{I}^{i}_{\ref{diag:qqbargL}} &= \mathcal{I}^{i}(\xt_{0;1+2},\xt_{21},\overline{Q}^2_{\ref{diag:qqbargL}},\omega_{\ref{diag:qqbargL}},\lambda_{\ref{diag:qqbargL}}),\quad\quad \mathcal{I}_{\ref{diag:qqbargL}} = \mathcal{I}(\xt_{0;1+2},\xt_{21},\overline{Q}^2_{\ref{diag:qqbargL}},\omega_{\ref{diag:qqbargL}},\lambda_{\ref{diag:qqbargL}})
\end{split}
\end{equation}
with the coefficients
\begin{equation}
\omega_{\ref{diag:qgqbarL}}  = \frac{q^+k^+_0k^+_2}{k^+_1(k^+_0 + k^+_2)^2}, \quad \omega_{\ref{diag:qqbargL}}  = \frac{q^+k^+_1k^+_2}{k^+_0(k^+_1 + k^+_2)^2}.
\end{equation}

We now have the full expressions for the gluon emission wavefunctions with the massive quarks. In addition, it is straightforward to check that in the massless quark limit the expression in \eq\nr{eq:reducedLCWFqqbarg} reduces to the massless quark result obtained in Refs.~\cite{Beuf:2017bpd,Hanninen:2017ddy}.

\section{The DIS cross section at NLO}
\label{sec:xs}

We now use the wave functions  to compute the DIS cross section at NLO 
in the dipole factorization framework. 

\subsection{Quark-antiquark contribution}

Let us first write down the $q\bar{q}$ contribution to the DIS cross section at NLO. Applying the formula for the cross section in \eq\nr{eq:crosssectionformula}, we find the following expression
\begin{equation}
\begin{split}
\sigma^{\gamma^{\ast}}_{L}\bigg\vert_{q\bar{q}} & = 2\nc\sum_{f}\int \frac{\ud k^+_0}{2k^+_0(2\pi)}\int \frac{\ud k^+_1}{2k^+_1(2\pi)}\frac{2\pi\delta(q^+-k^+_0 - k^+_1)}{2q^+}\int \ud^{D-2}\xt_0 \int \ud^{D-2}\xt_1\sum_{h_0,h_1}\vert \widetilde{\psi}^{\gamma^{\ast}_{L}\rightarrow q\bar{q}}\vert^2\mathrm{Re}[1-\mathcal{S}_{01}]\\
& =\frac{\nc}{(2\pi)4(q^+)^2}\sum_{f}\int_{0}^{1} \frac{\ud z}{z(1-z)}\int \ud^{D-2}\xt_0 \int \ud^{D-2}\xt_1\sum_{h_0,h_1}\vert \widetilde{\psi}^{\gamma^{\ast}_{L}\rightarrow q\bar{q}}\vert^2\mathrm{Re}[1-\mathcal{S}_{01}].\\
\end{split}
\end{equation}
At the accuracy we are working in here, i.e. up to terms $\mathcal{O}(\alpha_\text{em}\as^2)$, the wave function 
$\widetilde{\psi}^{\gamma^{\ast}_{L}\rightarrow q\bar{q}}$ should be taken as  $\widetilde{\psi}^{\gamma^{\ast}_{L}\rightarrow q\bar{q}}_{\nlo}$, neglecting the $\as^2$ contribution from the square of the loop corrections. 
Using the expressions for the LCWFs in \eqs\nr{eq:LOlcwfmixfact} and \nr{eq:finalNLOmixspacereduced}, it is straightforward to obtain 
\begin{equation}
\begin{split}
\sum_{h_0,h_1}\vert \widetilde{\psi}^{\gamma^{\ast}_{L}\rightarrow q\bar{q}}\vert^2 = &4\frac{4\alpha_{em}e^2_fQ^2}{2\pi}4(q^+)^2[z(1-z)]^3\left ( \frac{\sqrt{\overline{Q}^2 + m^2}}{2\pi\vert \xt_{01}\vert }\right )^{\frac{D}{2}-2}K_{\frac{D}{2} -2}\left (\vert \xt_{01}\vert\sqrt{\overline{Q}^2 + m^2}\right )\\
& \times \Biggl [\left ( \frac{\sqrt{\overline{Q}^2 + m^2}}{2\pi\vert \xt_{01}\vert }\right )^{\frac{D}{2}-2}K_{\frac{D}{2} -2}\left (\vert \xt_{01}\vert\sqrt{\overline{Q}^2 + m^2}\right )  + \left (\frac{\alpha_s\cf}{\pi}\right )\widetilde{\mathcal{V}}^{L} \Biggr ]  +  \mathcal{O}(\alpha_{em}\alpha_s^2),
\end{split}
\end{equation}
where  $\widetilde{\mathcal{V}}^{L}$ is given by \eq\nr{eq:VnloFTfinal}. Adding everything together, the $q\bar{q}$ contribution to the total cross section can be written as 
\begin{equation}
\label{eq:qqbarcsUVdiv}
\begin{split}
\sigma^{\gamma^{\ast}}_{L}\bigg\vert_{q\bar{q}} & =4\nc\alpha_{em}4Q^2\sum_{f}e_f^2\int_{0}^{1} \ud z  [z(1-z)]^2\int_{[\xt_0]}\int_{[\xt_1]} \left ( \frac{\sqrt{\overline{Q}^2 + m^2}}{2\pi\vert \xt_{01}\vert }\right )^{\frac{D}{2}-2}K_{\frac{D}{2} -2}\left (\vert \xt_{01}\vert\sqrt{\overline{Q}^2 + m^2}\right ) \\
& \times \Biggl [\left ( \frac{\sqrt{\overline{Q}^2 + m^2}}{2\pi\vert \xt_{01}\vert }\right )^{\frac{D}{2}-2}K_{\frac{D}{2} -2}\left (\vert \xt_{01}\vert\sqrt{\overline{Q}^2 + m^2}\right )  + \left (\frac{\alpha_s\cf}{\pi}\right )\widetilde{\mathcal{V}}^{L} \Biggr ]\mathrm{Re}[1-\mathcal{S}_{01}]  +  \mathcal{O}(\alpha_{em}\alpha_s^2). 
\end{split}
\end{equation}
Here, and in the following sections, we use the following the notation to denote $(D-2)$ and $2$-dimensional transverse coordinate integrations
\begin{equation}
\int_{[\xt_0]}\int_{[\xt_1]} = \int \frac{\ud^{D-2}\xt_0}{2\pi} \int \frac{\ud^{D-2}\xt_1}{2\pi},\quad\quad \int_{\xt_0}\int_{\xt_1} = \int \frac{\ud^{2}\xt_0}{2\pi} \int \frac{\ud^{2}\xt_1}{2\pi}.
\end{equation}

\subsection{Quark-antiquark-gluon contribution}

The $q\bar{q}g$ contribution to the DIS cross-section at NLO is given by the second term in \eq\nr{eq:crosssectionformula}. This simplifies to 
\begin{equation}
\label{eq:qqbargcs1}
\begin{split}
\sigma^{\gamma^{\ast}}_{L}\bigg\vert_{q\bar{q}g}  = 2\nc\cf &\sum_{f}\int_{0}^{\infty} \frac{\ud k^+_0}{2k^+_0(2\pi)}\int_{0}^{\infty} \frac{\ud k^+_1}{2k^+_1(2\pi)}\int_{0}^{\infty} \frac{\ud k^+_2}{2k^+_2(2\pi)}\frac{2\pi\delta(q^+-k^+_0 - k^+_1-k^+_2)}{2q^+}\\
&\times \int \ud^{D-2}\xt_0 \int \ud^{D-2}\xt_1\int \ud^{D-2}\xt_2\sum_{\sigma}\sum_{h_0,h_1}\vert \widetilde{\psi}^{\gamma^{\ast}_{L}\rightarrow q\bar{q}g}\vert^2\mathrm{Re}[1-\mathcal{S}_{012}].\\
\end{split}
\end{equation}
Using the expression in \eq\nr{eq:reducedLCWFqqbarg} for the gluon emission LCWF in mixed space, we obtain for the LCWF squared  the result 
\begin{equation}
\label{eq:emissionLCWFsquared}
\begin{split}
\sum_{\sigma}\sum_{h_0,h_1}&\vert \widetilde{\psi}^{\gamma^{\ast}_{L}\rightarrow q\bar{q}g}\vert^2 = \alpha_{em}e_f^2\alpha_{s}(4\pi)^2\frac{4Q^2}{(q^+)^4}2(2k^+_0)(2k^+_1) \mathcal{K}^L_{q\bar q g} +  \mathcal{O}(\alpha_{em}\alpha_s^2),
\end{split}
\end{equation}
where we have defined the function $\mathcal{K}_{q\bar q g}^L$ as 
\begin{equation}
\label{eq:Kfunction}
\begin{split}
\mathcal{K}_{q\bar q g}^L  =  &(k^+_1)^2\biggl [4k^+_0(k^+_0 + k^+_2) + (D_s-2)(k^+_2)^2 \biggr ]\vert \mathcal{I}^{i}_{\ref{diag:qgqbarL}}\vert^2 + (k^+_0)^2\biggl [4k^+_1(k^+_1 + k^+_2) + (D_s-2)(k^+_2)^2 \biggr ]\vert \mathcal{I}^{i}_{\ref{diag:qqbargL}}\vert^2\\
& -2k^+_0k^+_1\biggl [2k^+_1(k^+_0+k^+_2) + 2k^+_0(k^+_1+k^+_2) - (D_s-4)(k^+_2)^2 \biggr ]\mathrm{Re}\left (\mathcal{I}^{i}_{\ref{diag:qgqbarL}}\mathcal{I}^{i \ast}_{\ref{diag:qqbargL}} \right )\\
& + m^2(D_s-2)(k^+_2)^4\Biggl [\frac{(k^+_1)^2}{(k^+_0 + k^+_2)^2}\vert \mathcal{I}_{\ref{diag:qgqbarL}}\vert^2 + \frac{(k^+_0)^2}{(k^+_1 + k^+_2)^2}\vert \mathcal{I}_{\ref{diag:qqbargL}}\vert^2 - 2 \frac{k^+_0k^+_1}{(k^+_0 + k^+_2)(k^+_1 + k^+_2)}\mathrm{Re}\left (\mathcal{I}_{\ref{diag:qgqbarL}}\mathcal{I}^{\ast}_{\ref{diag:qqbargL}}  \right )  \Biggr ].
\end{split}
\end{equation}
The computation of individual terms in \eq\nr{eq:Kfunction} follows closely  the detailed derivation presented in the case of massless quarks. Therefore, for a detailed discussion, we refer the reader to our previous work \cite{Beuf:2017bpd, Hanninen:2017ddy}.

Finally, inserting the result in \eq\nr{eq:emissionLCWFsquared} into \eq\nr{eq:qqbargcs1}, we obtain 
\begin{equation}
\label{eq:qqbargunsub}
\begin{split}
\sigma^{\gamma^{\ast}}_{L}\bigg\vert_{q\bar{q}g} = 4\nc\alpha_{em}\left (\frac{\alpha_s\cf}{\pi}\right )\sum_{f}e_f^2 \frac{4Q^2}{(q^+)^4} \frac{(2\pi)^4}{2} & \int_{0}^{\infty} \ud k^+_0\int_{0}^{\infty} \ud k^+_1\int_{0}^{\infty} \frac{\ud k^+_2}{k^+_2}\frac{\delta(q^+-k^+_0-k^+_1-k^+_2)}{q^+}\\
&\times   \int_{[\xt_0]}\int_{[\xt_1]} \int_{[\xt_2]}   \mathcal{K}_{q\bar q g}^{L}\mathrm{Re}[1-\mathcal{S}_{012}] +  \mathcal{O}(\alpha_{em}\alpha_s^2).
\end{split}
\end{equation}

\subsection{UV subtraction}
\label{sec:UVsubtraction}

Since the UV renormalization of the coupling $g$ is not relevant at the accuracy of the present calculation, the remaining UV divergences have to cancel between the virtual $q\bar{q}$ and the real $q\bar{q}g$ contributions on the cross section level. Due to the complicated analytical structure of the gluon emission contribution in \eq\nr{eq:qqbargunsub}\footnote{This is also true in the massless case.}, the UV divergent phase-space integrals cannot be performed analytically for arbitrary dimension $D$.  Hence, it is desirable to understand the cancellation of UV divergences at the integrand level.

In the expression \eq\nr{eq:Kfunction}, the first and the second term are UV divergent when $\xt_2 \rightarrow \xt_0$ and $\xt_2 \rightarrow \xt_1$, respectively.  All the other terms  are UV finite and we can immediately take the limit $D = 4$ at the integrand level. In order to subtract the UV divergences, we will follow the same steps as presented in \cite{Hanninen:2017ddy, Beuf:2017bpd}. The general idea used in these works to subtract the UV divergences rely in the following property of Wilson lines at coincident transverse coordinate points
\begin{equation}
\begin{split}
\lim_{\xt\rightarrow \yt} \biggl [t^{b}U_F(\yt) \biggr ]U_A(\xt)_{ba} &= \biggl [U_F(\yt)t^{a} \biggr ],\\
\end{split}
\end{equation}
which implies that 
\begin{equation}
\lim_{\xt_2\rightarrow \xt_0} \mathcal{S}_{012} = \lim_{\xt_2\rightarrow \xt_1} \mathcal{S}_{012} =  \mathcal{S}_{01}.  
\end{equation}
Thus, the UV divergences in \eq\nr{eq:qqbargunsub} are subtracted by replacing the first and the second term with
\begin{equation}
\label{eq:subprocedure1}
\vert \mathcal{I}^{i}_{\ref{diag:qgqbarL}}\vert^2\mathrm{Re}[1-\mathcal{S}_{012}] \mapsto \Biggl \{\vert \mathcal{I}^{i}_{\ref{diag:qgqbarL}}\vert^2\mathrm{Re}[1-\mathcal{S}_{012}] - \vert \mathcal{I}^{i}_{\ref{diag:qgqbarL} \rm UV}\vert^2\mathrm{Re}[1-\mathcal{S}_{01}] \Biggr \} + \vert \mathcal{I}^{i}_{\ref{diag:qgqbarL} \rm UV}\vert^2\mathrm{Re}[1-\mathcal{S}_{01}] 
\end{equation}
\begin{equation}
\label{eq:subprocedure2}
\vert \mathcal{I}^{i}_{\ref{diag:qqbargL}}\vert^2\mathrm{Re}[1-\mathcal{S}_{012}] \mapsto \Biggl \{\vert \mathcal{I}^{i}_{\ref{diag:qqbargL}}\vert^2\mathrm{Re}[1-\mathcal{S}_{012}] - \vert \mathcal{I}^{i}_{\ref{diag:qqbargL} \rm UV}\vert^2\mathrm{Re}[1-\mathcal{S}_{01}] \Biggr \} + \vert \mathcal{I}^{i}_{\ref{diag:qqbargL} \rm UV}\vert^2\mathrm{Re}[1-\mathcal{S}_{01}],
\end{equation}
where the subtraction terms are given in terms of a single function $\mathcal{I}^{i}_{\rm UV}$:
\begin{equation}
\begin{split}
\mathcal{I}^{i}_{\ref{diag:qgqbarL} \rm UV} &=
\mathcal{I}^{i}_{\rm UV}  (\xt_{01},\xt_{20},\overline{Q}^2_{\ref{diag:qgqbarL}},\omega_{\ref{diag:qgqbarL}},\lambda_{\ref{diag:qgqbarL}}),
\\
\mathcal{I}^{i}_{\ref{diag:qqbargL} \rm UV} &=
\mathcal{I}^{i}_{\rm UV}  (\xt_{01},\xt_{21},\overline{Q}^2_{\ref{diag:qqbargL}},\omega_{\ref{diag:qqbargL}},\lambda_{\ref{diag:qqbargL}}).
\end{split}
\end{equation}
Now the function  $\mathcal{I}^{i}_{\rm UV}$ has to be a good UV approximation of the full integral, i.e. it must satisfy
\begin{equation}
\label{eq:subtractionintdef}
\lim_{\xt_2\rightarrow \xt_0}\mathcal{I}^{i} = 
 \lim_{\xt_2\rightarrow \xt_0}\mathcal{I}^{i}_{\rm UV} 
\end{equation}
from which it follows that:
\begin{equation}
\begin{split}
\lim_{\xt_2\rightarrow \xt_0}\mathcal{I}^{i}_{\ref{diag:qgqbarL}} &= 
 \lim_{\xt_2\rightarrow \xt_0}\mathcal{I}^{i}_{\ref{diag:qgqbarL} \rm UV},
\\
\lim_{\xt_2\rightarrow \xt_1}\mathcal{I}^{i}_{\ref{diag:qqbargL}} &= 
 \lim_{\xt_2\rightarrow \xt_1}\mathcal{I}^{i}_{\ref{diag:qqbargL} \rm UV} .
\end{split}
\end{equation}
It is important to note that there is no unique choice for the UV divergent subtraction in \eq\nr{eq:subtractionintdef}. The only requirement for the subtraction is that the UV divergence between virtual and real parts needs to cancel. Thus, it is sufficient for the subtraction to approximate the original integrals  by any function that has the same value in the UV coordinate limits (for any $D$). Because of this cancellation, the integrals of the expressions inside the curly brackets in \eqs\nr{eq:subprocedure1} and \nr{eq:subprocedure2} are finite, and one can safely take the limit $D_s = D = 4$ under the $\xt_2$ integral.

In an arbitrary dimension $D$, the integral $\mathcal{I}^{i}$ (see \eq\nr{eq:Irankonefinal}) is given by 
\begin{equation}
\label{eq:generalIi}
\mathcal{I}^i(\bt,\rt,\overline{Q}^2,\omega,\lambda) = \frac{i\mu^{2-D/2}}{2(4\pi)^{D-2}}\rt^i\int_{0}^{\infty} \ud u u^{1-D/2}e^{-u[\overline{Q}^2 + m^2]}e^{-\frac{\vert\bt\vert^2}{4u}}\int_{0}^{u/\omega}\ud t t^{-D/2}e^{-t\omega \lambda m^2}e^{-\frac{\vert\rt\vert^2}{4t}}.
\end{equation}
It is straightforward to see that to get the leading behavior in the limit $\vert\rt\vert^2 \rightarrow 0$ we can set $\lambda=0$. This leads us to 
\begin{equation}
\label{eq:uvfunctiongeneral}
\mathcal{I}^i(\bt,\rt,\overline{Q}^2,\omega)  \stackrel{\rt^2 \to 0}{=}
 \frac{i\mu^{2-D/2}}{2(4\pi)^{D-2}}\rt^i\left (\frac{\vert\rt\vert^2}{4}\right )^{1-D/2}\int_{0}^{\infty}\ud u u^{1-D/2}e^{-u[\overline{Q}^2 + m^2]}e^{-\frac{\vert\bt\vert^2}{4u}}\Gamma\left (\frac{D}{2}-1, \frac{\vert\rt\vert^2\omega}{4u}\right ),
\end{equation}
where we have suppressed the dependence on the variable $\lambda$ in the notation.

Now there are several possible ways of performing the UV subtraction. Using the exponential subtraction procedure
introduced in \cite{Hanninen:2017ddy}, we approximate the incomplete gamma function with
\begin{equation}
\label{eq:uvappr1}
\Gamma\left (\frac{D}{2}-1, \frac{\vert\rt\vert^2\omega}{4u}\right ) \mapsto \Gamma\left (\frac{D}{2}-1\right )e^{-\frac{\vert\rt\vert^2}{2\vert\bt\vert^2e^{\gamma_E}}},
\end{equation}
where the exponential is independent of $u$ allowing for an analytical calculation of the $u$-integral. This replacement has the correct behavior in the UV limit $\vert\rt\vert^2\to 0$, but also is regular in the  IR limit of large $\vert\rt\vert^2 \rightarrow \infty$. Another option would be to follow the polynomial subtraction scheme used in \cite{Beuf:2017bpd}  (see also discussion in Appendix~E of \cite{Hanninen:2017ddy}). Here the subtraction function is polynomial in $r$. This however, introduces a new IR divergence, which must be compensated with another subtraction. For the massive quark case, we present the derivation in the polynomial subtraction scheme in Appendix~\ref{app:subtraction}.

Proceeding with the exponential subtraction scheme we substitute \eq\nr{eq:uvappr1} into \eq\nr{eq:uvfunctiongeneral}. This gives an explicit expression for the UV approximation of the full integral
\begin{equation}
\label{eq:uvsubtexp}
\mathcal{I}^i_{\rm UV}(\bt,\rt,\overline{Q}^2,\omega)= \frac{i\mu^{2-D/2}}{4\pi^{D/2}}\rt^i (\vert\rt\vert^2 )^{1-D/2} \Gamma\left (\frac{D}{2}-1\right )e^{-\frac{\vert\rt\vert^2}{2\vert\bt\vert^2e^{\gamma_E}}}  \left ( \frac{\sqrt{\overline{Q}^2 + m^2}}{2\pi\vert \bt \vert }\right )^{\frac{D}{2}-2}K_{\frac{D}{2} -2}\left (\vert \bt\vert\sqrt{\overline{Q}^2 + m^2}\right ).
\end{equation}
In \eqs\nr{eq:subprocedure1} and~\nr{eq:subprocedure2} we will need the square of the UV approximation $\mathcal{I}^i_{\rm UV}$, which must be integrated over $\rt$ in $D-2$ dimensions. This integral can be performed using the following result
\begin{equation}
\label{eq:masterintegral}
\begin{split}
\frac{(\mu^2)^{2-\frac{D}{2}}\Gamma\left (\frac{D}{2}-1 \right )^2}{\pi^D}\int \ud^{D-2}\rt \,(\vert\rt\vert^2 )^{3-D}e^{-\frac{\vert\rt\vert^2}{\vert\bt\vert^2e^{\gamma_E}}}
 = \frac{1}{\pi^3}\Biggl \{&\frac{(4\pi)^{2-\frac{D}{2}}}{(2-\frac{D}{2})}\Gamma\left (3 - \frac{D}{2} \right ) + \log\left (\frac{\vert\xt_{01}\vert^2\mu^2}{4} \right ) \\
 & + 2\gamma_E + \mathcal{O}(D-4)\Biggr \}.
\end{split}
\end{equation}

\subsection{UV subtracted results}

Following the calculations in Sec.~\ref{sec:UVsubtraction}, we then obtain for the UV subtraction terms  
\begin{equation}
\label{eq:qqbaruvsubtexpf2v2}
\begin{split}
\sigma^{\gamma^{\ast}}_{L}&\bigg\vert^{\vert \ref{diag:qgqbarL}\vert^2_{\rm UV}}_{q\bar qg} = 4\nc\alpha_{em}\left (\frac{\alpha_s\cf}{\pi}\right )\sum_{f}e_f^2 \frac{4Q^2}{(q^+)^5} \int_{[\xt_0]}\int_{[\xt_1]} \left ( \frac{\sqrt{\overline{Q}^2 + m^2}}{2\pi\vert \xt_{01} \vert }\right )^{D-4}\biggl [K_{\frac{D}{2} -2}\left (\vert \xt_{01}\vert\sqrt{\overline{Q}^2 + m^2}\right )\biggr ]^2\\
&\times \int_{0}^{q^+} \ud k^+_1 (k^+_1)^2(q^+-k^+_1)^2\Biggl \{\biggl [-\frac{3}{4}-\log\left (\frac{k^+_{2,min}}{q^+-k^+_1}\right ) \biggr ]\biggl [ \frac{(4\pi)^{2-\frac{D}{2}} }{ (2-\frac{D}{2})} \Gamma \left (3-\frac{D}{2} \right ) + \log\left (\frac{\xt^2_{01}\mu^2}{4}\right )\\
& + 2\gamma_E \biggr  ] - \frac{1}{4}\frac{(D_s-4)}{(D-4)} \Biggr \}\mathrm{Re}[1-\mathcal{S}_{01}]+ \mathcal{O}(D-4)\\
\end{split}
\end{equation}
and 
\begin{equation}
\label{eq:qqbaruvsubtexpg2v2}
\begin{split}
\sigma^{\gamma^{\ast}}_{L}&\bigg\vert^{\vert \ref{diag:qqbargL}\vert^2_{\rm UV}}_{q\bar qg}  = 4\nc\alpha_{em}\left (\frac{\alpha_s\cf}{\pi}\right )\sum_{f}e_f^2 \frac{4Q^2}{(q^+)^5} \int_{[\xt_0]}\int_{[\xt_1]} \left ( \frac{\sqrt{\overline{Q}^2 + m^2}}{2\pi\vert \xt_{01} \vert }\right )^{D-4}\biggl [K_{\frac{D}{2} -2}\left (\vert \xt_{01}\vert\sqrt{\overline{Q}^2 + m^2}\right )\biggr ]^2\\
&\times \int_{0}^{q^+} \ud k^+_0 (k^+_0)^2(q^+-k^+_0)^2\Biggl \{\biggl [-\frac{3}{4}-\log\left (\frac{k^+_{2,min}}{q^+-k^+_0}\right ) \biggr ]\biggl [ \frac{(4\pi)^{2-\frac{D}{2}} }{ (2-\frac{D}{2})} \Gamma \left (3-\frac{D}{2} \right ) + \log\left (\frac{\xt^2_{01}\mu^2}{4}\right )\\
& + 2\gamma_E \biggr  ] - \frac{1}{4}\frac{(D_s-4)}{(D-4)} \Biggr \}\mathrm{Re}[1-\mathcal{S}_{01}] + \mathcal{O}(D-4).\\
\end{split}
\end{equation}
Next, introducing the same parametrization as in the $q\bar	q$ contribution $k^+_0 = zq^+$, $k^+_1 = (1-z)q^+$ and $k^+_{2,min} = \alpha q^+$, and changing the variables from $(k^+_1,k^+_0) \mapsto z$, the sum of \eqs\nr{eq:qqbaruvsubtexpf2v2} and \nr{eq:qqbaruvsubtexpg2v2} yields an expression for the UV divergent $q\bar{q}g$ subtraction contribution
\begin{equation}
\label{eq:uvsubtractionterm}
\begin{split}
\sigma^{\gamma^{\ast}}_{L}&\bigg\vert^{\vert \ref{diag:qgqbarL}\vert^2_{\rm UV} + \vert \ref{diag:qqbargL}\vert^2_{\rm UV}}_{q\bar qg}   = -4\nc\alpha_{em}4Q^2\left (\frac{\alpha_s\cf}{\pi}\right )\sum_{f}e_f^2  \int_{[\xt_0]}\int_{[\xt_1]} \left ( \frac{\sqrt{\overline{Q}^2 + m^2}}{2\pi\vert \xt_{01} \vert }\right )^{D-4}\biggl [K_{\frac{D}{2} -2}\left (\vert \xt_{01}\vert\sqrt{\overline{Q}^2 + m^2}\right )\biggr ]^2\\
&\times  \int_{0}^{1} \ud z [z(1-z)]^2\Biggl \{\biggl [\frac{3}{2} + \log\left (\frac{\alpha}{z}\right ) + \log\left (\frac{\alpha}{1-z}\right ) \biggr ]\biggl [  \frac{(4\pi)^{2-\frac{D}{2}} }{ (2-\frac{D}{2})} \Gamma \left (3-\frac{D}{2} \right ) + \log \left (\frac{\xt^2_{01} \mu^2}{4}\right )\\
& + 2\gamma_E  \biggr ]  + \frac{1}{2}\frac{(D_s-4)}{(D-4)}  \Biggr \}\mathrm{Re}[1-\mathcal{S}_{01}] + \mathcal{O}(D-4).\\
\end{split}
\end{equation}
This expression precisely cancels the scheme dependent UV-divergent part in the $q\bar{q}$ contribution in \eq\nr{eq:finalNLOmixspacereduced}. The remaining terms in \eq\nr{eq:qqbarcsUVdiv} are UV finite and regularization scheme independent.

In the case of the exponential subtraction scheme, the sum of two UV finite terms in \eqs\nr{eq:subprocedure1} and \nr{eq:subprocedure2} (inside the curly brackets) can be simplified to 
\begin{equation}
\label{eq:exptsubtqbarqg}
\begin{split}
&\sigma^{\gamma^{\ast}}_{L}\bigg\vert^{\vert \ref{diag:qgqbarL}\vert^2_{\rm fin} + \vert \ref{diag:qqbargL}\vert^2_{\rm fin}}_{q\bar qg}   =  4\nc\alpha_{em} 4Q^2\left (\frac{\alpha_s\cf}{\pi}\right )\sum_{f}e_f^2\int_{\xt_0}\int_{\xt_1}\int_{\xt_2} \int_{0}^{\infty} \ud k^+_0\int_{0}^{\infty} \ud k^+_1\int_{0}^{\infty} \frac{\ud k^+_2}{k^+_2}\frac{\delta(q^+-\sum_{i=0}^{2}k^+_i)}{(q^+)^5}\\
&\times \Biggl \{(k^+_1)^2\biggl [2k^+_0(k^+_0 + k^+_2) + (k^+_2)^2 \biggr ] \times \\
& \hspace{1.5cm} \times \biggl \{   \frac{\vert\xt_{20}\vert^2}{64}[\mathcal{G}_{\ref{diag:qgqbarL}}^{(1;2)}]^2  \mathrm{Re}[1-\mathcal{S}_{012}]  - \frac{e^{-\vert\xt_{20}\vert^2/(\vert\xt_{01}\vert^2 e^{\gamma_E})    }}{\vert\xt_{20}\vert^2} \biggl [K_0 \left (\vert \xt_{01}\vert \sqrt{\overline{Q}^2_{\ref{diag:qgqbarL}} + m^2} \right )\biggr ]^2   \mathrm{Re}[1-\mathcal{S}_{01}]   \biggr \} \\
& + (k^+_0)^2\biggl [2k^+_1(k^+_1 + k^+_2) + (k^+_2)^2 \biggr ] \times \\
& \hspace{1.5cm} \times \biggl \{   \frac{\vert\xt_{21}\vert^2}{64}[\mathcal{G}_{\ref{diag:qqbargL}}^{(1:2)}]^2  \mathrm{Re}[1-\mathcal{S}_{012}]  - \frac{e^{-\vert\xt_{21}\vert^2/(\vert\xt_{01}\vert^2 e^{\gamma_E})  }}{\vert\xt_{21}\vert^2} \biggl [K_0 \left (\vert \xt_{01}\vert \sqrt{\overline{Q}^2_{\ref{diag:qqbargL}} + m^2} \right )\biggr ]^2   \mathrm{Re}[1-\mathcal{S}_{01}]   \biggr \}\Biggr \}.
\end{split}
\end{equation}
Here we have introduced the notation: 
\begin{equation}
\label{eq:defGf}
\mathcal{G}_{\ref{diag:qgqbarL}}^{(n;m)} = \int_{0}^{\infty} \frac{\ud u}{u^n} e^{-u[\overline{Q}^2_{\ref{diag:qgqbarL}} + m^2]} e^{-\frac{\vert\xt_{0+2;1}\vert^2}{4u}}\int_{0}^{u/\omega_{\ref{diag:qgqbarL}}} \frac{\ud t}{t^m} e^{-t[\omega_{\ref{diag:qgqbarL}}\lambda_{\ref{diag:qgqbarL}}m^2]}e^{-\frac{\vert\xt_{20}\vert^2}{4t}}
\end{equation}
and
\begin{equation}
\label{eq:defGg}
\mathcal{G}_{\ref{diag:qqbargL}}^{(n:m)} = \int_{0}^{\infty} \frac{\ud u}{u^n} e^{-u[\overline{Q}^2_{\ref{diag:qqbargL}} + m^2]} e^{-\frac{\vert\xt_{0;1+2}\vert^2}{4u}}\int_{0}^{u/\omega_{\ref{diag:qqbargL}}} \frac{\ud t}{t^m} e^{-t[\omega_{\ref{diag:qqbargL}}\lambda_{\ref{diag:qqbargL}}m^2]}e^{-\frac{\vert\xt_{21}\vert^2}{4t}} 
\end{equation}
for the integrals that appear. These integrals could be seen as generalizations of the integral representation of Bessel functions that appear in the massless case. They could doubtlessly be transformed in many ways, but since these integrals are very rapidly converging at both small and large values of the integration variable, they should be well suited for numerical evaluation as is.
The corresponding result in the case of the polynomial subtraction scheme is written down in Appendix  \ref{app:subtraction}.

\section{Longitudinal photon cross section}
\label{sec:full}

We can now gather here the main result of our paper, which is the the longitudinal virtual photon total cross section at NLO with massive quarks. This cross section can be written as a sum of two UV finite terms
\begin{equation}
\label{eq:fullcslongitudinal}
\sigma^{\gamma^{\ast}}_{L} = \sigma^{\gamma^{\ast}}_{L}\bigg\vert^{\rm subt}_{q\bar{q}} +  \sigma^{\gamma^{\ast}}_{L}\bigg\vert^{\rm subt}_{q\bar{q}g} + \mathcal{O}(\alpha_{em}\alpha_s^2),
\end{equation}
where the first term in \eq\nr{eq:fullcslongitudinal} is the mass renormalized and the UV subtracted $q\bar{q}$ contribution, which is obtained by adding the UV subtraction term in \eq\nr{eq:uvsubtractionterm} into \eq\nr{eq:qqbarcsUVdiv}. This gives 
\begin{equation}
\begin{split}
 \sigma^{\gamma^{\ast}}_{L}\bigg\vert^{\rm subt}_{q\bar{q}}  = &  4\nc\alpha_{em}4Q^2\sum_{f}e_f^2\int_{0}^{1} \ud z \, [z(1-z)]^2\int_{\xt_0}\int_{\xt_1} \times\\
&\times  \Biggl \{\biggl [ 1 + \left (\frac{\alpha_s\cf}{\pi}\right ) \biggl \{ \frac{5}{2} - \frac{\pi^2}{3} + \log^2\left (\frac{z}{1-z}\right )   + \Omega_{\mathcal{V}}(\gamma;z) + L(\gamma;z) \biggr \} \biggr ]\biggl [K_0\left (\vert \xt_{01}\vert\sqrt{\overline{Q}^2 + m^2}\right )\biggr ]^2\\
& + \left (\frac{\alpha_s\cf}{\pi}\right )K_0\left (\vert \xt_{01}\vert\sqrt{\overline{Q}^2 + m^2}\right )\widetilde{\mathcal{I}}_{\mathcal{V}}(z,\xt_{01})  \Biggr \}\mathrm{Re}[1-\mathcal{S}_{01}],  
\end{split}
\end{equation}
where the functions $\Omega_{\mathcal{V}}(\gamma;z)$, $L(\gamma;z)$ and $\widetilde{\mathcal{I}}_{\mathcal{V}}(z,\xt_{01})$ are explicitly writen down in \eqs\nr{eq:GammaLm}, \nr{eq:Lfunction} and \nr{eq:IVsumFT}, respectively. The first term, not proportional to $\as$, is explicitly the known leading order cross section for massive quarks (see e.g.~\cite{Kowalski:2006hc}).

The second term in \eq\nr{eq:fullcslongitudinal} is the UV finite $q\bar{q}g$ contribution, which is obtained by replacing the first two terms in \eq\nr{eq:qqbargunsub} with the subtraction term derived in \eq\nr{eq:exptsubtqbarqg}. This can be simplified to 
\begin{equation}
\begin{split}
&\sigma^{\gamma^{\ast}}_{L}\bigg\vert^{\rm subt}_{q\bar{q}g}    =  4\nc\alpha_{em} 4Q^2\left (\frac{\alpha_s\cf}{\pi}\right )\sum_{f}e_f^2\int_{\xt_0}\int_{\xt_1}\int_{\xt_2} \int_{0}^{\infty} \ud k^+_0\int_{0}^{\infty} \ud k^+_1\int_{0}^{\infty} \frac{\ud k^+_2}{k^+_2}\frac{\delta(q^+-\sum_{i=0}^{2}k^+_i)}{(q^+)^5} \times\\
&\times \Biggl \{(k^+_1)^2\biggl [2k^+_0(k^+_0 + k^+_2) + (k^+_2)^2 \biggr ]\times \\
& \hspace{1.5cm} \times \biggl \{   \frac{\vert\xt_{20}\vert^2}{64}[\mathcal{G}_{\ref{diag:qgqbarL}}^{(1;2)}]^2  \mathrm{Re}[1-\mathcal{S}_{012}]  - \frac{e^{-\vert\xt_{20}\vert^2/(\vert\xt_{01}\vert^2 e^{\gamma_E})}}{\vert\xt_{20}\vert^2} \biggl [K_0 \left (\vert \xt_{01}\vert \sqrt{\overline{Q}^2_{\ref{diag:qgqbarL}} + m^2} \right )\biggr ]^2   \mathrm{Re}[1-\mathcal{S}_{01}]  \biggr \} \\
& + (k^+_0)^2\biggl [2k^+_1(k^+_1 + k^+_2) + (k^+_2)^2 \biggr ] \times \\
& \hspace{1.5cm} \times \biggl \{   \frac{\vert\xt_{21}\vert^2}{64}[\mathcal{G}_{\ref{diag:qqbargL}}^{(1;2)}]^2  \mathrm{Re}[1-\mathcal{S}_{012}]  - \frac{e^{-\vert\xt_{21}\vert^2/(\vert\xt_{01}\vert^2 e^{\gamma_E})}}{\vert\xt_{21}\vert^2} \biggl [K_0 \left (\vert \xt_{01}\vert \sqrt{\overline{Q}^2_{\ref{diag:qqbargL}} + m^2} \right )\biggr ]^2   \mathrm{Re}[1-\mathcal{S}_{01}]  \biggr \} \\
& - \frac{k^+_0k^+_1}{32} \biggl [k^+_1(k^+_0 + k^+_2) + k^+_0(k^+_1 + k^+_2) \biggr ] (\xt_{20}\cdot \xt_{21}) [\mathcal{G}_{\ref{diag:qgqbarL}}^{(1;2)}][\mathcal{G}_{\ref{diag:qqbargL}}^{(1;2)}]\mathrm{Re}[1-\mathcal{S}_{012}]\\
& + \frac{m^2}{16}(k^+_2)^4 \Biggl [\frac{k^+_1}{(k^+_0 + k^+_2)} [\mathcal{G}_{\ref{diag:qgqbarL}}^{(1;1)}]  - \frac{k^+_0}{(k^+_1 + k^+_2)}[\mathcal{G}_{\ref{diag:qqbargL}}^{(1;1)}] \Biggr ]^2
\mathrm{Re}[1-\mathcal{S}_{012}]\Biggr \},
\end{split}
\end{equation}
involving the generalized Bessel function integrals from \eqs\nr{eq:defGf} and~\nr{eq:defGg}. As discussed in more detail in Secs.~\ref{sec:qqbargmomspace} and \ref{subsec:FTqqbargL}, in the limit of zero quark mass these expressions reduce to the known results in Refs.~\cite{Beuf:2016wdz,Beuf:2017bpd,Hanninen:2017ddy}.

\section{Conclusions}
\label{sec:conc}

In conclusion, we have here calculated, we believe for the first time in the literature, the one-loop light cone wave function for a longitudinal photon splitting into a quark-antiquark pair including quark masses. Such a wavefunction is a central ingredient in any NLO calculation in the small-$x$ dipole factorization formulation for processes involving heavy quarks. In particular, while we concentrated here on the total cross section, this  also includes many possible diffractive or exclusive cross sections that will be important parts of the physics program at future DIS facilities.  

Our result includes a renormalization of the quark mass in a pole mass or on-shell scheme. The peculiarity of the longitudinal photon polarization state is that vertex correction type diagrams do not contribute to mass renormalization, since at tree level the longitudinal photon vertex does not have a term proportional to the quark mass. This issue will be different for the transverse polarization which we intend to return to in future work. There, one will have to address  the issue of the consistency in the mass renormalization between  the propagator and vertex correction diagrams.  We plan to revisit the issue of quark mass renormalization in LCPT 
much more thorougly in a separate paper. 

After obtaining the one loop LCWF we also  Fourier-transformed our result to mixed transverse coordinate--longitudinal momentum space. Combined with the tree level wavefunction for a quark-antiquark-gluon state this enabled us to obtain an explicit expression for the total longitudinal photon cross section in the dipole picture. The obtained cross section still has, just like the one for massless quarks, a high energy divergence when the longitudinal momentum of the gluon becomes small. This divergence (or large logarithm) needs to be further absorbed into a BK- or JIMWLK-evolution of the Wilson lines describing the target.  This factorization procedure, together with NLO renormalization group evolution, will need to be developed in a consistent way to confront our calculations with experimental data. We have not elaborated on this issue here, since this will proceed similarly (and have similar problematic issues) as for the case of massless quarks, discussed in previous works~\cite{Beuf:2014uia,Ducloue:2017ftk,Beuf:2017bpd,Beuf:2020dxl}.

The case of the transverse photon (virtual or real) is even more important for phenomenology, but much more complicated algebraically.
It will be addressed in a forthcoming separate  publication, where we will  calculate the  light cone wave function and the DIS cross section for transverse virtual photons. The combination of the two results will enable the calculation of a variety of DIS process cross sections in the dipole picture at NLO with massive quarks. In particular, this includes the  heavy quark structure function $F_{2}^c$, which could be expected to be crucial observable for the physics of gluon saturation at the future Electron-Ion Collider EIC~\cite{Accardi:2012qut,AbdulKhalek:2021gbh}.

\section*{Acknowledgments} 
This work was supported by the Academy of Finland, projects  321840 (T.L.) and 1322507 (R.P.), and  under the European Union's Horizon 2020 research and innovation programme by the European Research Council (ERC, grant agreements No. ERC-2015-CoG-681707 and ERC-2016-CoG-725369) and by the STRONG-2020 project (grant agreement No 824093. The content of this article does not reflect the official opinion of the European Union and responsibility for the information and views expressed therein lies entirely with the authors. 

\appendix

\section{Helicity decomposition for light cone vertices with massive quarks}
\label{app:vertexdecomp}

In this Appendix, we describe how to express QED/QCD emission vertices in the helicity basis by decomposing the spinor structure of given vertex to the light cone helicity non-flip and flip components.

Following the discussion presented in \cite{Hanninen:2017ddy}, the decomposition for the light cone vertex (without coupling and color structure) in the LC gauge can be expressed as \footnote{Here we again suppress the notation and write $\bar{\chi}(k_1,h_1) = \bar{\chi}(1)$ and $\omega(k_0,h_0) = \omega(0)$.}
\begin{equation}
\label{eq:decompv1}
\bar{\chi}(1)\epsl_{\lambda}(q)\omega(0) = \frac{\qt \cdot\epst_{\lambda}}{q^+}\bar{\chi}(1)\gamma^{+}\omega(0) - \epst^{i}_{\lambda}\bar{\chi}(1)\gamma^i\omega(0),
\end{equation}
where $\chi$ and $\omega$ can be either positive or negative massive spinors, i.e. $u$ or $v$. Note that we are now dealing with on-shell momenta and polarization vectors in the light cone gauge. We also use three-momentum conservation, with the appropriate signs depending on whether $\chi$ and $\omega$ are negative or positive energy spinors. The photon here is an incoming one with polarization vector $\varepsilon_{\lambda}(q)$, the corresponding expressions for an outgoing photon can be obtained by complex conjugation. To be explicit, momentum conservation means that $\qvec = \kvec_1+ \kvec_2$ for pair production $\chi(1)=v(1)$, $\omega(0)=u(0)$ or $\chi(1)=u(1)$, $\omega(0)=v(0)$. But for gauge boson absorption by a quark, $\chi(1)=u(1)$, $\omega(0)=u(0)$ and momentum conservation means $\qvec+\kvec_0= \kvec_1$. Vice versa, for gauge boson absorption by an antiquark, $\chi(1)=v(1)$ and $\omega(0)=v(0)$ momentum conservation means $\hat{q}+\hat{k}_1= \hat{k}_0$.

Using the Dirac equation and Clifford algebra, it is straightforward to show that \eq\nr{eq:decompv1} can be decomposed into three independent spinor structures
\begin{equation}
\label{eq:decompv2}
\begin{split}
\bar{\chi}(1)\epsl_{\lambda}(q)\omega(0) = &\biggl [\frac{\qt^j}{q^+} - \frac{\kt^j_0}{2k^+_0} - \frac{\kt^j_1}{2k^+_1} \biggr ]\epst^{j}_{\lambda}\bar{\chi}(1)\gamma^+\omega(0) - \biggl [\frac{\kt^i_0}{4k^+_0} - \frac{\kt^i_1}{4k^+_1} \biggr ]\epst^{j}_{\lambda}\bar{\chi}(1)\gamma^{+}[\gamma^i,\gamma^j]\omega(0)\\
& + \epst^{j}_{\lambda}\biggl [-\frac{1}{2k^{+}_0}\bar{\chi}(1)\gamma^+\gamma^j(\mp m)\omega(0)  + \frac{1}{2k^+_1}\bar{\chi}(1)(\mp m)\gamma^+\gamma^j\omega(0)\biggr ],
\end{split}
\end{equation}
where terms appearing in the first line are the light cone helicity non-flip components and terms appearing in the second line are the helicity flip components. The $(\mp)$ sign of the mass term  is determined from:
\begin{equation}
\begin{split}
(\mp m)\omega(0) =   \left\{
                \begin{array}{ll}
                   (-m) u(0)\\
                   (+m) v(0)\\
                  \end{array}
              \right.,
              \quad\quad
\bar{\chi}(1)(\mp m)=   \left\{
                \begin{array}{ll}
                   \bar{u}(1)(-m) \\
                   \bar{v}(1)(+m) \\
                  \end{array}
              \right.  .          
\end{split}
\end{equation}
Here, the obtained result in \eq\nr{eq:decompv2} is valid in arbitrary spacetime dimensions, but we emphasize that it relies on plus and transverse momentum conservation.


\section{Numerator for the quark self-energy diagram \ref{diag:oneloopSEUPL}}
\label{app:numa}

In this Appendix, we present some details for the calculation of the quark self-energy diagram \ref{diag:oneloopSEUPL}. The numerator for the vertex correction diagram \ref{diag:vertexqbaremL}  is discussed in the following section. 

The numerator for the diagram \ref{diag:oneloopSEUPL}, where sum over the internal helicities, gluon polarization and color is implicit, can be written as   
\begin{equation}
\label{eq:numaapp}
N^{L}_{\ref{diag:oneloopSEUPL}} = \frac{ee_fg^2Qt^{a}_{\alpha_0\bar{\alpha}_0}t^{a}_{\bar{\alpha}_0\alpha_1}}{q^+}\biggl [\bar{u}(0)\epsl_{\sigma}(k)u(0')\biggr ]\biggl [\bar{u}(0')\epsl^{\ast}_{\sigma}(k)u(0'')\biggr ]\biggl [\bar{u}(0'')\gamma^{+}v(1)\biggr ].
\end{equation}
Using the decompostion in \eq\nr{eq:decompv2} with kinematical variables as in \fig\ref{fig:selfenergyL}\ref{diag:oneloopSEUPL}, we can express the spinor structures inside the square brakets in the helicity basis as 
\begin{equation}
\label{eq:vertex1a}
\begin{split}
\bar{u}(0)\epsl_{\sigma}(k)u(0') = \frac{1}{k^+_0 \left (\frac{k^+}{k^+_0}\right )\left (1 -\frac{k^+}{k^+_0}\right ) }\Biggl \{\Biggl [\left (1-\frac{k^+}{2k^+_0}\right )\delta^{ij}_{(D_s)}\bar{u}(0)\gamma^{+}u(0') & + \left (\frac{k^+}{4k^+_0}\right )\bar{u}(0)\gamma^{+}[\gamma^i,\gamma^j]u(0')  \Biggr ]\Kt^i\\
& + \frac{m}{2}\left (\frac{k^+}{k^+_0}\right )^2\bar{u}(0)\gamma^+\gamma^ju(0')\Biggr \}\epst_{\sigma}^j
\end{split}
\end{equation}
and 
\begin{equation}
\label{eq:vertex2a}
\begin{split}
\bar{u}(0')\epsl^{\ast}_{\sigma}(k)u(0'') = \frac{1}{k^+_0 \left (\frac{k^+}{k^+_0}\right )\left (1 -\frac{k^+}{k^+_0}\right ) }\Biggl \{\Biggl [\left (1-\frac{k^+}{2k^+_0}\right )\delta^{kl}_{(D_s)}\bar{u}(0')\gamma^{+}u(0'') & - \left (\frac{k^+}{4k^+_0}\right )\bar{u}(0')\gamma^{+}[\gamma^k,\gamma^l]u(0'')  \Biggr ]\Kt^k\\
& - \frac{m}{2}\left (\frac{k^+}{k^+_0}\right )^2\bar{u}(0')\gamma^+\gamma^lu(0'')\Biggr \}\epst^{\ast l}_{\sigma},
\end{split}
\end{equation}
where the variable $\Kt$ is defined in \eq\nr{eq:KTdef}. Using the completeness relation for the spinors
\begin{equation}
\sum_{h'}u(0')\bar{u}(0') = \ksl'_0 + m, \quad\quad \sum_{h''}u(0'')\bar{u}(0'') = \ksl''_0 + m,
\end{equation}
and noting that $\gamma^+(\ksl'_0 + m)\gamma^+ = 2(k^+_0-k^+)\gamma^+$ and $\gamma^+(\ksl''_0 + m)\gamma^+ = 2k^+_0\gamma^+$, we obtain
\begin{equation}
\label{eq:numaappv2}
\begin{split}
N^{L}_{\ref{diag:oneloopSEUPL}} = \frac{4ee_fg^2Q\delta_{\alpha_0\alpha_1}\cf}{q^+\left (\frac{k^+}{k^+_0}\right )^2\left (1 -\frac{k^+}{k^+_0}\right )}\Biggl \{\Biggl [\left (1-\frac{k^+}{2k^+_0}\right )^2\delta^{ij}_{(D_s)}\delta^{kl}_{(D_s)}\bar{u}(0)\gamma^{+}v(1) & - \left (\frac{k^+}{4k^+_0}\right )^2\bar{u}(0)\gamma^{+}[\gamma^i,\gamma^j][\gamma^k,\gamma^l]v(1) \Biggr ]\Kt^i\Kt^k\\
& - \frac{m^2}{4}\left (\frac{k^+}{k^+_0} \right )^4 \bar{u}(0)\gamma^+\gamma^j\gamma^lv(1)\Biggr \}\epst_{\sigma}^j\epst^{\ast l}_{\sigma}.
\end{split}
\end{equation}
In \eq\nr{eq:numaappv2} the terms linear in the transverse integration variable $\Kt$ vanish due to rotational symmetry. The remaining tensor contractions are evaluated as follows: First, the tensor product of transverse momentum variable $\Kt$ is replaced with $\Kt^i\Kt^k\rightarrow \delta^{ik}_{(D)}\Kt^2/(D-2)$, which is true under the integration over the $(D-2)$-dimensional transverse momentum integral
\begin{equation}
\int\frac{\ud^{D-2}\Kt}{(2\pi)^{D-2}}\frac{\Kt^i\Kt^k}{\Kt^2 + \Delta_1} \mapsto  \frac{\delta^{ik}_{(D)}}{(D-2)}\int\frac{\ud^{D-2}\Kt}{(2\pi)^{D-2}}\frac{\Kt^2}{\Kt^2 + \Delta_1}.
\end{equation}
The gluon polarization vectors and the transverse gamma matrices appearing in \eq\nr{eq:numaappv2} are kept in $(D_s-2)$ dimension with $D_s > D$. Thus, summing over the helicity states of the gluon and performing the remaining tensor contractions yield
\begin{equation}
\label{eq:numaappv3}
\begin{split}
N^{L}_{\ref{diag:oneloopSEUPL}} = \frac{4ee_fg^2Q\delta_{\alpha_0\alpha_1}\cf}{q^+\left (\frac{k^+}{k^+_0}\right )^2\left (1 -\frac{k^+}{k^+_0}\right )}\bar{u}(0)\gamma^+v(1)\Biggl \{\Biggl [\left (1-\frac{k^+}{2k^+_0}\right )^2  + 4(D_s-3)\left (\frac{k^+}{4k^+_0}\right )^2\Biggr ]\Kt^2  + \frac{m^2}{4}(D_s-2)\left (\frac{k^+}{k^+_0} \right )^4 \Biggr \}.
\end{split}
\end{equation}
Finally, using the parametrization $k^+/k^+_0 = \xi$ and definition of the leading order QED photon splitting vertex in \eq\nr{eq:LOvertex}  gives
\begin{equation}
\label{eq:numaappvfinal}
N^{L}_{\ref{diag:oneloopSEUPL}} =  \frac{2g^2\cf}{\xi^2(1-\xi)}\delta_{\alpha_0\alpha_1}V^{\gamma^{\ast}\rightarrow q\bar{q}}_{h_0;h_1}\Biggl \{\biggl [1 + (1-\xi)^2\biggr ]\Kt^2  + m^2\xi^4 + \frac{(D_s-4)}{2}\xi^2\biggl [\Kt^2 + m^2\xi^2\biggr ] \Biggr \}.
\end{equation}

\section{Numerator for the vertex correction diagram \ref{diag:vertexqbaremL}}
\label{app:numc}

The numerator for the diagram \ref{diag:vertexqbaremL}, where again sum over the internal helicities, gluon polarization and color is implicit, can be written as  
\begin{equation}
\label{eq:numcapp}
N^{L}_{\ref{diag:vertexqbaremL}} = \frac{-ee_fg^2Qt^{a}_{\alpha_0\bar{\alpha}_0}t^{a}_{\bar{\alpha}_0\alpha_1}}{q^+}\biggl [\bar{u}(0)\epsl_{\sigma}(k)u(0')\biggr ]\biggl [\bar{u}(0')\gamma^{+}v(1')\biggr ]\biggl [\bar{v}(1')\epsl^{\ast}_{\sigma}(k)v(1)\biggr ].
\end{equation}
Using the decomposition in \eq\nr{eq:decompv2} with the kinematical variables as in \fig\ref{fig:vertexL}\ref{diag:vertexqbaremL}, we find 
\begin{equation}
\label{eq:vertex1c}
\begin{split}
\bar{u}(0)\epsl_{\sigma}(k)u(0') = \frac{1}{k^+_0 \left (\frac{k^+}{k^+_0}\right )\left (1 -\frac{k^+}{k^+_0}\right ) }\Biggl \{\Biggl [\left (1-\frac{k^+}{2k^+_0}\right )\delta^{ij}_{(D_s)}\bar{u}(0)\gamma^{+}u(0') & + \left (\frac{k^+}{4k^+_0}\right )\bar{u}(0)\gamma^{+}[\gamma^i,\gamma^j]u(0')  \Biggr ]\Kt^i\\
& + \frac{m}{2}\left (\frac{k^+}{k^+_0}\right )^2\bar{u}(0)\gamma^+\gamma^ju(0')\Biggr \}\epst_{\sigma}^j
\end{split}
\end{equation}
and 
\begin{equation}
\label{eq:vertex2c}
\begin{split}
\bar{v}(1')\epsl^{\ast}_{\sigma}(k)v(1) = \frac{1}{k^+_1 \left (\frac{k^+}{k^+_1}\right )\left (1 + \frac{k^+}{k^+_1}\right ) }\Biggl \{\Biggl [\left (1+\frac{k^+}{2k^+_1}\right )\delta^{kl}_{(D_s)}\bar{v}(1')\gamma^{+}v(1) & + \left (\frac{k^+}{4k^+_1}\right )\bar{v}(1')\gamma^{+}[\gamma^k,\gamma^l]v(1)  \Biggr ]\Rt^k\\
& - \frac{m}{2}\left (\frac{k^+}{k^+_1}\right )^2\bar{v}(1')\gamma^+\gamma^lv(1)\Biggr \}\epst^{\ast l}_{\sigma},
\end{split}
\end{equation}
where we have introduced the variable 
\begin{equation}
\label{eq:Rtdef}
\Rt = \Kt + \frac{k^+q^+}{k^+_0k^+_1}\Pt.
\end{equation}
Substituting \eqs\nr{eq:vertex1c} and \nr{eq:vertex2c} into \nr{eq:numcapp} and following the same steps as in \ref{app:numa}, we obtain the following result
\begin{equation}
\begin{split}
N^{L}_{\ref{diag:vertexqbaremL}}  = &-2\delta_{\alpha_0\alpha_1}V^{\gamma^{\ast}_{L}\rightarrow q\bar{q}}_{h_0;h_1}(g^2\cf)\Biggl \{a_{\ref{diag:vertexqbaremL}}\delta^{ik}_{(D_s)}\Kt^i\left (\Kt^{k} + \frac{k^+q^+}{k^+_0k^+_1}\Pt^{k}\right )+ \frac{(k^+)^2}{k^+_0k^+_1}m^2\biggl [ 1 + \frac{(D_s-4)}{2}\biggr ] \Biggr \}\\
& + 2\delta_{\alpha_0\alpha_1}\frac{ee_fQ}{q^+}(g^2\cf) m\bar{u}(0)\gamma^+\gamma^iv(1)\frac{q^+}{k^+_0}\Biggl \{b_{\ref{diag:vertexqbaremL}}\Kt^i - \frac{k^+}{k^+_0}c_{\ref{diag:vertexqbaremL}}\Pt^i\Biggr \},
\end{split}
\end{equation}
where we have defined the coefficients $a_{\ref{diag:vertexqbaremL}}, b_{\ref{diag:vertexqbaremL}}$ and $c_{\ref{diag:vertexqbaremL}}$ as 
\begin{equation}
\begin{split}
a_{\ref{diag:vertexqbaremL}} & = \frac{k^+_0k^+_1}{(k^+)^2} + \frac{(k^+_0-k^+)(k^+_1+k^+)}{(k^+)^2} - \frac{(D_s-4)}{2},\\
b_{\ref{diag:vertexqbaremL}} & = \frac{k^+_0}{k^+_1} - 1 - \frac{k^+}{k^+_1}\biggl [1 + \frac{(D_s-4)}{2}\biggr ],\\
c_{\ref{diag:vertexqbaremL}} & = 1 + \frac{k^+}{k^+_1}\biggl [1 + \frac{(D_s-4)}{2}\biggr ].
\end{split}
\end{equation}

Finally, by using the parametrization $k^+_0 = zq^+$, $k^+_1 = (1-z)q^+$ and $k^+ = \xi k^+_0$ gives
\begin{equation}
\label{numcappfinal}
\begin{split}
N^{L}_{\ref{diag:vertexqbaremL}} = &-2\delta_{\alpha_0\alpha_1}V^{\gamma^{\ast}_{L}\rightarrow q\bar{q}}_{h_0;h_1}(g^2\cf)\Biggl \{\frac{1}{\xi^2}\Biggl [\frac{(1-z)}{z} + \frac{(1-\xi)(1-z(1-\xi))}{z} - \frac{(D_s-4)}{2}\xi^2\Biggr ]\\
& \times \delta^{ik}_{(D_s)}\Kt^i\left (\Kt^{k} + \frac{\xi}{1-z}\Pt^{k}\right ) + \frac{z\xi^2}{(1-z)}m^2 f_{(D_s)} \Biggr \}\\
& +2\delta_{\alpha_0\alpha_1}\frac{ee_fQ}{q^+}(g^2\cf) m\bar{u}(0)\gamma^+\gamma^iv(1)\frac{1}{z}\Biggl \{\Biggl [\frac{z}{1-z} - 1 - \frac{z\xi}{1-z}f_{(D_s)}\Biggr ]\Kt^i - \xi\biggl [1 + \frac{z\xi}{1-z}f_{(D_s)}\biggr ]\Pt^i\Biggr \},
\end{split}
\end{equation}
where $f_{(D_s)} = 1 + (D_s-4)/2$.

\section{Fourier transform integrals for the quark-antiquark Fock state}
\label{app:FTSqqbarcase}

In this Appendix, we present the relevant integrals that are needed to calculate the Fourier transformed $\gamma^{\ast}_L \rightarrow q\bar q$ LCWF’s in mixed space up to NLO. To begin with, let us first consider the following general integral
\begin{equation}
\label{eq:generalFTint}
 \int\frac{\ud^{D-2}\Pt}{(2\pi)^{D-2}}\frac{e^{i\Pt\cdot \xt_{01}}}{[\Pt^2 + \Delta^2]},
\end{equation}
where $\xt_{01} = \xt_0 - \xt_1$ and $\Pt^2 + \Delta^2 >0$. The standard technique to compute these type of integrals is to first introduce a Schwinger parametrisation for each denominator
\begin{equation}
\label{eq:SParam}
\frac{1}{A^{\beta}} = \frac{1}{\Gamma(\beta)}\int_{0}^{\infty}\ud t t^{\beta - 1} e^{-t A}, \quad A,\beta >0.  
\end{equation}
Applying this to the general integral in \eq\nr{eq:generalFTint}, we obtain 
\begin{equation}
\label{eq:generalFTintSP}
 \int\frac{\ud^{D-2}\Pt}{(2\pi)^{D-2}}\frac{e^{i\Pt\cdot \xt_{01}}}{[\Pt^2 + \Delta^2]} = \int\frac{\ud^{D-2}\Pt}{(2\pi)^{D-2}} e^{i\Pt\cdot \xt_{01}}\int_{0}^{\infty} \ud t e^{-t[\Pt^2 + \Delta^2]}.
\end{equation}
The expression above appears in a simple Gaussian form in $\Pt$, and hence, we can perform integral over the  $(D-2)$-dimensional transverse momentum space by using the Gaussian integral
\begin{equation}
\label{eq:GausInt}
\int \frac{\ud^n \yt}{(2\pi)^n} \exp \biggl [-\sum_{i,j=1}^{n} a_{ij}y_i y_j \biggr ]\biggl [\sum_{i=1}^{n}b_i y_i\biggr ] = \sqrt{\frac{1}{(4\pi)^{n}\det a}}\exp \biggl [ {\frac{1}{4}(a^{-1})_{ij}b_ib_j} \biggr ],
\end{equation}
where $\yt = (y_1,y_2,\ldots,y_n)$, $\mathbf{b} = (b_1,b_2,\ldots, b_n)$ and $a_{ij}$ is a symmeric, non-singular and positive defined $n\times n$ matrix. This leads to the result
\begin{equation}
\label{eq:generalFTintSPG}
 \int\frac{\ud^{D-2}\Pt}{(2\pi)^{D-2}}\frac{e^{i\Pt\cdot \xt_{01}}}{[\Pt^2 + \Delta^2]} = (4\pi)^{1-\frac{D}{2}} \int_{0}^{\infty} \ud t\, t^{1 - \frac{D}{2}}e^{-t \Delta^2}e^{-\frac{\vert \xt_{01}\vert^2}{4t}}.
\end{equation}
Finally, the remaining one-dimensional integral can be performed by using the general formula
\begin{equation}
\int_{0}^{\infty} \ud t\, t^{\beta -1}e^{-tA} e^{-\frac{B}{t}} = 2\left (\frac{B}{A} \right )^{\beta/2} K_{-\beta}(2\sqrt{AB}), \quad A,B > 0,    
\end{equation}
where $K_{-\beta}(x)$ is the modified Bessel function of the second kind. Note that for positive $x$ values $K_{\beta}(x)$ is an analytic function of $\beta$. In addition, $K_{\beta}(x)$ is even in $\beta$, i.e. $K_{\beta}(x) = K_{-\beta}(x)$. This gives the result
\begin{equation}
\label{eq:generalFTintSPfinal}
 \int\frac{\ud^{D-2}\Pt}{(2\pi)^{D-2}}\frac{e^{i\Pt\cdot \xt_{01}}}{[\Pt^2 + \Delta^2]} = \frac{1}{2\pi}\left (\frac{\Delta}{2\pi\vert \xt_{01}\vert} \right )^{\frac{D}{2}-2}K_{\frac{D}{2}-2}\left (\vert \xt_{01}\vert \Delta\right ).
\end{equation}
Note also that if the general integral in \eq\nr{eq:generalFTint} contains  a logarithmic function, one can first use the relation $\log(A) = \lim_{\alpha \rightarrow 0} \partial_{\alpha} A^{\alpha}$ and then apply the Schwinger parametrization formula in \eq\nr{eq:SParam}.

Let us then use these results by considering first the general integral in \eq\nr{eq:generalFTintSPfinal} with $\Delta = \overline{Q}^2 + m^2$. In this case, the result reads
\begin{equation}
\label{eq:F1}
 \int\frac{\ud^{D-2}\Pt}{(2\pi)^{D-2}}\frac{e^{i\Pt\cdot \xt_{01}}}{[\Pt^2 + \overline{Q}^2 + m^2]} = \frac{1}{(2\pi)}\left (\frac{\sqrt{\overline{Q}^2 + m^2}}{2\pi\vert \xt_{01}\vert} \right )^{\frac{D}{2}-2}K_{\frac{D}{2}-2}\left (\vert \xt_{01}\vert \sqrt{\overline{Q}^2 + m^2}\right ).
\end{equation}
For the NLO contribution, we also need (in addition to the result in \eqs\nr{eq:F1}) the following set of  transverse Fourier integrals in $D = 4$:
\begin{equation}
\label{eq:F2}
\begin{split}
\int\frac{\ud^{D-2}\Pt}{(2\pi)^{D-2}}\frac{e^{i\Pt\cdot \xt_{01}}}{[\Pt^2 + \overline{Q}^2 + m^2]}\log  \left ( \frac{\Pt^2 + \overline{Q}^2 + m^2}{\overline{Q}^2 + m^2}\right ) & =  \frac{1}{(2\pi)}\left (\frac{\sqrt{\overline{Q}^2 + m^2}}{2\pi\vert \xt_{01}\vert} \right )^{\frac{D}{2}-2}K_{\frac{D}{2}-2}\left (\vert \xt_{01}\vert \sqrt{\overline{Q}^2 + m^2}\right )\\
 & \times \Biggl \{-\frac{1}{2}\log\left (\frac{\vert\xt_{01}\vert^2\left (\overline{Q}^2 + m^2\right )}{4}\right ) + \Psi_0(1) + \mathcal{O}(D-4)\Biggr \},
\end{split}
\end{equation}
\begin{equation}
\label{eq:F3}
\begin{split}
 \int\frac{\ud^2\Pt}{(2\pi)^2}e^{i\Pt\cdot \xt_{01}}\biggl [\frac{1}{\Pt^2 + \overline{Q}^2 + m^2} & - \frac{1}{\Pt^2 + \overline{Q}^2 + m^2 + \frac{\xi(1-z)}{(1-\xi)}m^2}\biggr ]\\
 & = \frac{1}{(2\pi)}\Biggl \{K_0\left (\vert \xt_{01}\vert \sqrt{\overline{Q}^2 + m^2}\right ) - K_0\left (\vert \xt_{01}\vert \sqrt{\overline{Q}^2 + m^2 + \frac{\xi(1-z)}{(1-\xi)}m^2}\right ) \Biggr \},
\end{split}
\end{equation}
\begin{equation}
\label{eq:F4}
\begin{split}
 \int\frac{\ud^2\Pt}{(2\pi)^2}& \frac{e^{i\Pt\cdot \xt_{01}}}{[\Pt^2 + \overline{Q}^2 + m^2] [x(1-x)\Lt^2 + (1-x)\Delta_1 + x\Delta_2]}\\
 & =  \frac{1}{(2\pi)}\Biggl \{K_0\left (\vert \xt_{01}\vert \sqrt{\overline{Q}^2 + m^2}\right ) - K_0\left (\vert \xt_{01}\vert \sqrt{\frac{\overline{Q}^2 + m^2}{1-x} + \kappa}\right ) \Biggr \}\frac{(1-\xi)^{-1}(1-x)^{-1}}{\biggl [x(1-\xi) + \frac{\xi}{(1-z)}\biggr ]\biggl [\frac{x(\overline{Q}^2 + m^2)}{(1-x)} + \kappa\biggr ]}.
\end{split}
\end{equation}
Here, the coefficient $\Psi_0(1) = -\gamma_E$ (Euler's constant) and the variables $(\xi,z,x) \in [0,1]$, $\Delta_1,\Delta_2 >0$ and  $\Lt^2 = (1-\xi)^2\Pt^2 > 0$. The coefficient $\kappa$ is defined as 
\begin{equation}
\kappa = \frac{\xi m^2}{(1-\xi)(1-x)\biggl [x(1-\xi) + \frac{\xi}{(1-z)}\biggr ]}\biggl [\xi(1-x) + x\left (1 - \frac{z(1-\xi)}{(1-z)}\right )\biggr ].
\end{equation}

\section{Fourier transform integrals for the quark-antiquark-gluon Fock state}
\label{app:FTSqqbargcase}

For the gluon emission diagrams from a longitudinal photon state, we need to calculate the following two Fourier integrals
\begin{equation}
\label{eq:Irankone}
\mathcal{I}^i(\bt,\rt,\overline{Q}^2,\omega,\lambda) = \mu^{2-D/2}\int\frac{\ud^{D-2}\Pt}{(2\pi)^{D-2}}\int\frac{\ud^{D-2}\Kt}{(2\pi)^{D-2}} \frac{\Kt^ie^{i\Pt\cdot \bt}e^{i\Kt\cdot \rt}}{\biggl [\Pt^2 + \overline{Q}^2 + m^2 \biggr ]\biggl [\Kt^2 + \omega\left (\Pt^2 + \overline{Q}^2 + m^2 + \lambda m^2\right )\biggr ]}
\end{equation}
and
\begin{equation}
\label{eq:Iscalar}
\mathcal{I}(\bt,\rt,\overline{Q}^2,\omega,\lambda) = \mu^{2-D/2}\int\frac{\ud^{D-2}\Pt}{(2\pi)^{D-2}}\int\frac{\ud^{D-2}\Kt}{(2\pi)^{D-2}} \frac{e^{i\Pt\cdot \bt}e^{i\Kt\cdot \rt}}{\biggl [\Pt^2 + \overline{Q}^2 + m^2 \biggr ]\biggl [\Kt^2 + \omega\left (\Pt^2 + \overline{Q}^2 + m^2 + \lambda m^2\right )\biggr ]}.
\end{equation}
First, using the Schwinger parametrisation, \eq\nr{eq:SParam}, for the denominators appearing in \eqs\nr{eq:Irankone} and \nr{eq:Iscalar}, and then performing the remaining transverse momentum integrals by using the $(D-2)$-dimensional Gaussian integral \eq\nr{eq:GausInt} yields
\begin{equation}
\label{eq:Irankonev2}
\mathcal{I}^i(\bt,\rt,\overline{Q}^2,\omega,\lambda) = \frac{i\mu^{2-D/2}}{2(4\pi)^{D-2}}\rt^i\int_{0}^{\infty}\ud t t^{-D/2}e^{-t\omega \lambda m^2}e^{-\frac{\vert\rt\vert^2}{4t}}\int_{0}^{\infty}\frac{\ud s}{(s + t\omega)^{D/2-1}}e^{-(s+t\omega)[\overline{Q}^2 + m^2]}e^{-\frac{\vert\bt\vert^2}{4(s+t\omega)}}
\end{equation}
and 
\begin{equation}
\label{eq:Iscalarv2}
\mathcal{I}(\bt,\rt,\overline{Q}^2,\omega,\lambda) = \frac{\mu^{2-D/2}}{(4\pi)^{D-2}}\int_{0}^{\infty}\ud t t^{1-D/2}e^{-t\omega \lambda m^2}e^{-\frac{\vert\rt\vert^2}{4t}}\int_{0}^{\infty}\frac{\ud s}{(s + t\omega)^{D/2-1}}e^{-(s+t\omega)[\overline{Q}^2 + m^2]}e^{-\frac{\vert\bt\vert^2}{4(s+t\omega)}}.
\end{equation}
By making the change of variables $u = s + t\omega$ and then changing the order of integration leads to the result
\begin{equation}
\label{eq:Irankonefinal}
\mathcal{I}^i(\bt,\rt,\overline{Q}^2,\omega,\lambda) = \frac{i\mu^{2-D/2}}{2(4\pi)^{D-2}}\rt^i\int_{0}^{\infty} \ud u u^{1-D/2}e^{-u[\overline{Q}^2 + m^2]}e^{-\frac{\vert\bt\vert^2}{4u}}\int_{0}^{u/\omega}\ud t t^{-D/2}e^{-t\omega \lambda m^2}e^{-\frac{\vert\rt\vert^2}{4t}}
\end{equation}
and 
\begin{equation}
\label{eq:Iscalarfinal}
\mathcal{I}(\bt,\rt,\overline{Q}^2,\omega,\lambda) = \frac{\mu^{2-D/2}}{(4\pi)^{D-2}}\int_{0}^{\infty} \ud uu^{1-D/2}e^{-u[\overline{Q}^2 + m^2]}e^{-\frac{\vert\bt\vert^2}{4u}} \int_{0}^{u/\omega}\ud t t^{1-D/2}e^{-t\omega \lambda m^2}e^{-\frac{\vert\rt\vert^2}{4t}}.
\end{equation}

In an arbitrary dimension $D$, the integral over $t$ would then give a dependent of incomplete Gamma function, preventing us from  expressing the final result in terms of familiar special functions, e.g. the modified Bessel functions. These forms, however, serve as a sufficient starting point for deducing the appropriate UV subtractions in Sec.~\ref{sec:UVsubtraction}.

\section{Useful integrals}
\label{app:logintegrals}

In this Appendix, we present the results the integrals $\mathcal{I}_{\xi;1}, \mathcal{I}_{\xi;2}$ and $\mathcal{I}_{\xi;3}$, appearing in the calculation of \eq\nr{eq:sumvceAv3}. These integrals are given by:
\begin{equation}
\label{eq:xiint1}
\mathcal{I}_{\xi;1} = \int^{1}_{\alpha/z}  \frac{\ud \xi}{\xi}\biggl [-\log\left (\frac{\Delta_2}{\mu^2}\right )\biggr ] = \log\left (\frac{\alpha}{z}\right )\log\left (\frac{\overline{Q}^2 + m^2}{\mu^2}\right ) + \mathrm{Li}_2\left (\frac{1}{1-\frac{1}{2z}(1-\gamma)} \right ) + \mathrm{Li}_2\left (\frac{1}{1-\frac{1}{2z}(1+\gamma)} \right ),
\end{equation}
\begin{equation}
\label{eq:xiint2}
\begin{split}
\mathcal{I}_{\xi;2} = \int^{1}_{\alpha/z} \ud \xi\biggl [-\log\left (\frac{\Delta_2}{\mu^2}\right )\biggr ] = -\log\left (\frac{\overline{Q}^2 + m^2}{\mu^2}\right ) + 2 &- \frac{1}{z}\gamma\log\left (\frac{1+\gamma}{1+\gamma - 2z}\right ) - \frac{1}{2z}(\gamma-1 )\log\left (\frac{\overline{Q}^2+m^2}{m^2}\right ),
\end{split}
\end{equation}
and
\begin{equation}
\label{eq:xiint3}
\begin{split}
\mathcal{I}_{\xi;3} = \int^{1}_{\alpha/z} \ud \xi \xi\biggl [-\log\left (\frac{\Delta_2}{\mu^2}\right )\biggr ] = & -\frac{1}{2}\log\left (\frac{\overline{Q}^2 + m^2}{\mu^2}\right ) + \frac{3}{2}-\frac{1}{2z} - \frac{1}{z}\left (1-\frac{1}{2z}\right )\gamma\log\left (\frac{1+\gamma}{1+\gamma - 2z}\right )\\
& - \frac{1}{2z}\left (1-\frac{1}{2z}\right ) (\gamma-1 )\log\left (\frac{\overline{Q}^2+m^2}{m^2}\right ) - \frac{m^2}{2z^2Q^2}\log\left (\frac{\overline{Q}^2+m^2}{m^2}\right ).
\end{split}
\end{equation}
In the above expressions, we have used the notation $\gamma = \sqrt{1 + 4m^2/Q^2}$, introduced in \eq\nr{eq:xizeros}.



\section{Detailed derivation of the subtraction terms in the polynomial subtraction scheme}
\label{app:subtraction}

Following the subtraction procedure introduced in \cite{Beuf:2017bpd},  the correct UV behaviour of \eq\nr{eq:uvfunctiongeneral} could also obtained by simply replacing the incomplete gamma function with 
\begin{equation}
\label{eq:uvappr2a}
\Gamma\left (\frac{D}{2}-1, \frac{\vert\rt\vert^2\omega}{4u}\right ) \mapsto \Gamma\left (\frac{D}{2}-1\right ).
\end{equation}
This leads to the UV approximation 
\begin{equation}
\label{eq:uvsubtpoly}
\mathcal{I}^i_{\rm UV}(\bt,\rt,\overline{Q}^2) =  \frac{i(\mu^2)^{2-D/2}}{4\pi^{D/2}}\rt^i (\vert\rt\vert^2 )^{1-D/2} \Gamma\left (\frac{D}{2}-1\right )\left ( \frac{\sqrt{\overline{Q}^2 + m^2}}{2\pi\vert \bt \vert }\right )^{\frac{D}{2}-2}K_{\frac{D}{2} -2}\left (\vert \bt\vert\sqrt{\overline{Q}^2 + m^2}\right )
\end{equation}
of \eq\nr{eq:uvfunctiongeneral}.
This approximation has a Coulomb tail at large distances in $\rt$, leading to an IR divergence which is absent when using the original expression from \eq\nr{eq:uvfunctiongeneral}. Hence, in \cite{Beuf:2017bpd}, an extra IR subtraction has been included, in order to turn the Coulomb tail into a dipole tail, thus avoiding the appearance of the unphysical IR divergence.

All in all, in this scheme, the UV subtraction procedure can be written as
\begin{equation}
\begin{split}
\vert \mathcal{I}^{i}_{\ref{diag:qgqbarL}}\vert^2\mathrm{Re}[1-\mathcal{S}_{012}] 
\mapsto 
\Biggl \{& \vert \mathcal{I}^{i}_{\ref{diag:qgqbarL}}\vert^2\mathrm{Re}[1-\mathcal{S}_{012}] 
\\
& 
- \Biggl [\vert \mathcal{I}^{i}_{\rm UV}(\xt_{01},\xt_{20},\overline{Q}^2_{\ref{diag:qgqbarL}})\vert^2 
- \mathrm{Re}\left (\mathcal{I}^{\ast i}_{\rm UV}(\xt_{01},\xt_{20},\overline{Q}^2_{\ref{diag:qgqbarL}})\, 
\mathcal{I}^{i}_{\rm UV}(\xt_{01},\xt_{21},\overline{Q}^2_{\ref{diag:qgqbarL}})\right )
\Biggr ]\mathrm{Re}[1-\mathcal{S}_{01}]\Biggr \}
\\
& 
+  \Biggl [\vert \mathcal{I}^{i}_{\rm UV}(\xt_{01},\xt_{20},\overline{Q}^2_{\ref{diag:qgqbarL}})\vert^2 
- \mathrm{Re}\left (\mathcal{I}^{\ast i}_{\rm UV}(\xt_{01},\xt_{20},\overline{Q}^2_{\ref{diag:qgqbarL}})\, 
\mathcal{I}^{i}_{\rm UV}(\xt_{01},\xt_{21},\overline{Q}^2_{\ref{diag:qgqbarL}})\right )
\Biggr ]\mathrm{Re}[1-\mathcal{S}_{01}], 
\end{split}
\end{equation}
\begin{equation}
\begin{split}
\vert \mathcal{I}^{i}_{\ref{diag:qqbargL}}\vert^2\mathrm{Re}[1-\mathcal{S}_{012}] 
\mapsto 
\Biggl \{& \vert \mathcal{I}^{i}_{\ref{diag:qqbargL}}\vert^2\mathrm{Re}[1-\mathcal{S}_{012}] 
\\
& 
- \Biggl [\vert \mathcal{I}^{i}_{\rm UV}(\xt_{01},\xt_{21},\overline{Q}^2_{\ref{diag:qqbargL}})\vert^2 
- \mathrm{Re}\left (\mathcal{I}^{\ast i}_{\rm UV}(\xt_{01},\xt_{20},\overline{Q}^2_{\ref{diag:qqbargL}})\, 
\mathcal{I}^{i}_{\rm UV}(\xt_{01},\xt_{21},\overline{Q}^2_{\ref{diag:qqbargL}})\right )
\Biggr ]\mathrm{Re}[1-\mathcal{S}_{01}]\Biggr \}
\\
& 
+  \Biggl [\vert \mathcal{I}^{i}_{\rm UV}(\xt_{01},\xt_{21},\overline{Q}^2_{\ref{diag:qqbargL}})\vert^2 
- \mathrm{Re}\left (\mathcal{I}^{\ast i}_{\rm UV}(\xt_{01},\xt_{20},\overline{Q}^2_{\ref{diag:qqbargL}})\, 
\mathcal{I}^{i}_{\rm UV}(\xt_{01},\xt_{21},\overline{Q}^2_{\ref{diag:qqbargL}})\right )
\Biggr ]\mathrm{Re}[1-\mathcal{S}_{01}] .
\end{split}
\end{equation}

In the case of the polynomial subtraction scheme, we hence obtain
\begin{equation}
\label{eq:poltsubtqbarqg}
\begin{split}
\sigma^{\gamma^{\ast}}_{L}\bigg\vert^{\vert \ref{diag:qgqbarL}\vert^2_{\rm fin} + \vert \ref{diag:qqbargL}\vert^2_{\rm fin}}_{q\bar qg}   & =  4\nc\alpha_{em} 4Q^2\left (\frac{\alpha_s\cf}{\pi}\right )\sum_{f}e_f^2\int_{\xt_0}\int_{\xt_1}\int_{\xt_2} \int_{0}^{\infty} \ud k^+_0\int_{0}^{\infty} \ud k^+_1\int_{0}^{\infty} \frac{\ud k^+_2}{k^+_2}\frac{\delta(q^+-\sum_{i=0}^{2}k^+_i)}{(q^+)^5}\\
&\times \Biggl \{(k^+_1)^2\biggl [2k^+_0(k^+_0 + k^+_2) + (k^+_2)^2 \biggr ] \biggl \{   \frac{\vert\xt_{20}\vert^2}{64}[\mathcal{G}_{\ref{diag:qgqbarL}}^{(1;2)}]^2   \mathrm{Re}[1-\mathcal{S}_{012}] \\
& - \frac{\xt_{20}}{\vert\xt_{20}\vert^2}\cdot \left ( \frac{\xt_{20}}{\vert\xt_{20}\vert^2} -  \frac{\xt_{21}}{\vert\xt_{21}\vert^2} \right ) \biggl [K_0 \left (\vert \xt_{01}\vert \sqrt{\overline{Q}^2_{\ref{diag:qgqbarL}} + m^2} \right )\biggr ]^2   \mathrm{Re}[1-\mathcal{S}_{01}]           \biggr \} \\
&\quad + (k^+_0)^2\biggl [2k^+_1(k^+_1 + k^+_2) + (k^+_2)^2 \biggr ] \biggl \{   \frac{\vert\xt_{21}\vert^2}{64}[\mathcal{G}_{\ref{diag:qqbargL}}^{(1;2)}]^2   \mathrm{Re}[1-\mathcal{S}_{012}] \\
& -  \frac{\xt_{21}}{\vert\xt_{21}\vert^2}\cdot \left ( \frac{\xt_{21}}{\vert\xt_{21}\vert^2} -  \frac{\xt_{20}}{\vert\xt_{20}\vert^2} \right ) \biggl [K_0 \left (\vert \xt_{01}\vert \sqrt{\overline{Q}^2_{\ref{diag:qqbargL}} + m^2} \right )\biggr ]^2   \mathrm{Re}[1-\mathcal{S}_{01}]           \biggr \}\Biggr \}.
\end{split}
\end{equation}
while all of the other contributions to the cross section are the same as in the exponential subtraction scheme.

\section{$\mathcal{S}^L$ and the Pauli form factor}
\label{app:formfactorSL}

In this appendix, a cross-check of our results is provided, by comparison with the literature.
The usual parametrization of the $\gamma e^{-} e^{+}$ or $\gamma q \bar{q}$ vertex function in QED and/or QCD (based on Lorentz and gauge invariance, and discrete symmetries such as parity) can be written as
\begin{align}
\Gamma^{\mu}(q)
= & F_D(q^2/m^2)\gamma^{\mu} + F_P(q^2/m^2) \frac{q_{\nu}}{2m} i \sigma^{\mu\nu}
\label{1PI_vertex_generic}
\end{align}
with the Dirac and Pauli form factors, and $\sigma^{\mu\nu} \equiv (i/2)\, [\gamma^{\mu},\gamma^{\nu}]$. The relation \eqref{1PI_vertex_generic} relies on energy and momentum conservation at the vertex, and requires the two external fermion lines to be on mass shell. The photon virtuality $q^2$ is thus the only scale, apart from the fermion mass $m$ and QCD non-perturbative scales.

In the case of a $\gamma_L^*\rightarrow q\bar{q}$ splitting, only the $\mu=+$ component contributes in Light-Cone gauge, due to the longitudinal polarization vector. Including the spinors for the outgoing quark of momentum $k_0$ and antiquark of momentum $k_1$, one obtains
\begin{align}
\overline{u}(0) \Gamma^{+}(q)v(1)
= & F_D(q^2/m^2)\;  \overline{u}(0)\gamma^{+}v(1) + F_P(q^2/m^2)\,  \frac{(-1)}{4m}\; \overline{u}(0)[\gamma^{+},\slashed{q}]v(1)
\, .
\label{1PI_vertex_gammaL_to_qqbar}
\end{align}
Using momentum conservation $k_0^{\mu}+k_1^{\mu}=q^{\mu}$, one can rewrite Eq.~\eqref{1PI_vertex_gammaL_to_qqbar} after some algebra as
\begin{align}
\overline{u}(0) \Gamma^{+}(q)v(1)
= & \left[ F_D(q^2/m^2)+\frac{(q^+)^2}{4k_0^+ k_1^+}\, F_P(q^2/m^2)\right]\;  \overline{u}(0)\gamma^{+}v(1) 
+ F_P(q^2/m^2)\,  \frac{(k_0^+\!-\!k_1^+)}{4m}\;
\left[\frac{\kt_0^j}{k_0^+} \!-\!\frac{\kt_1^j}{k_1^+}\right]\; \overline{u}(0)\gamma^{+}\gamma^{j} v(1)
\nonumber\\
= & \left[ F_D(q^2/m^2)+\frac{F_P(q^2/m^2)}{4z(1\!-\!z)} \right]\;  \overline{u}(0)\gamma^{+}v(1) 
+ F_P(q^2/m^2)\,  \frac{(2z\!-\!1)}{4z(1\!-\!z)m}\: \Pt^j\; \overline{u}(0)\gamma^{+}\gamma^{j} v(1)
\, ,
\label{1PI_vertex_gammaL_to_qqbar_2}
\end{align}
using the same notations $\Pt$ and $z$ as in the rest of the present article (see Eq.~\eqref{eq:PtdefandQ}). Moreover, using momentum conservation $k_0^{\mu}+k_1^{\mu}=q^{\mu}$ and the on-shell conditions $k_0^2=k_1^2=m^2$, the photon virtuality $q^2$ can be expressed in terms of $\Pt$ and $z$ as 
\begin{align}
q^2 = 2m^2 +2k_0^{\mu} {k_1}_{\mu}  = \frac{\Pt^2 + m^2}{z(1 - z)}.
\label{timelike_photon_virt_FS}  
\end{align}

On the other hand, in section \ref{sec:loop}, the initial-state LCWF for $\gamma_L^*\rightarrow q\bar{q}$ as been calculated at NLO accuracy in QCD. It has the general form
\begin{equation}
\begin{split}
 \Psi^{\gamma_L^{*}\rightarrow q\bar{q}}_{\nlo} = 
\delta_{\alpha_{0}\alpha_{1}}
  \frac{e e_f}{(\ed_{\lo})} \frac{Q}{q^+}
	\Bigg\{ \overline{u}(0) \gamma^+ v(1) 
	\left[1+ \frac{\alpha_s \cf}{2\pi} {\cal V}^{L} \right] 
	+ \frac{(q^+)^2}{2 k_0^+ k_1^+} m \Pt^j 
	\overline{u}(0) \gamma^+ \gamma^j v(1)
	\left[\frac{\alpha_s \cf}{2\pi}\right]{\cal S}^{L}
	\Bigg\}
\label{WF_L_generic_1}
\end{split}
\end{equation}
with ${\cal V}^{L}$ and ${\cal S}^{L}$ collecting the helicity non-flip and helicity flip contributions respectively. One recognizes the same two Dirac structures  in Eqs.~\eqref{1PI_vertex_gammaL_to_qqbar_2} and \eqref{WF_L_generic_1}. Apart from the normalization, there is however a major difference between the vertex function \eqref{1PI_vertex_gammaL_to_qqbar_2} and the LFWF \eqref{WF_L_generic_1} : only the $+$ and transverse components of the momentum is conserved in the splitting in the LFWF. Due to the absence of the conservation of the $-$ component of the momentum, the relation \eqref{timelike_photon_virt_FS} is not valid for the LFWF, so that ${\cal V}^{L}$ and ${\cal S}^{L}$ a priori depend on $q^2$, $\Pt^2$ and $z$ independently. 
Indeed, the relation \eqref{timelike_photon_virt_FS} is equivalent to $\Pt^2 + m^2+\overline{Q}^2=0$, meaning $(\ed_{\lo})=0$. 

For that reason, the two coefficients ${\cal V}^{L}$ and ${\cal S}^{L}$ contain more information than the Dirac and Pauli form factors, and can be related to them only when imposing the relation \eqref{timelike_photon_virt_FS}. In such a way, one obtains the constraints
\begin{align}
\left[\frac{\alpha_s \cf}{2\pi}\right] \left.\frac{2 m^2}{(2z\!-\!1)}\; {\cal S}^{L}\right|_{\Pt^2= -\overline{Q}^2 -m^2}& = F_P(q^2/m^2)
\label{FP_SL_correspondence}
\\
\left[\frac{\alpha_s \cf}{2\pi}\right] \left.\left[{\cal V}^{L}- \frac{m^2}{2 z (1\!-\!z) (2z\!-\!1)}\, {\cal S}^{L}\right]\right|_{\Pt^2= -\overline{Q}^2 -m^2} & = F_D(q^2/m^2)-1 
\label{FD_VL_SL_correspondence} 
,
\end{align}
which can be used to cross-check our results for the $\gamma_L^*\rightarrow q\bar{q}$ LCWF at NLO with massive quarks. Note that ${\cal V}^{L}$ and ${\cal S}^{L}$ depend on $q^2$, $\Pt^2$ and $z$ independently (and $m^2$), and we are then imposing one single relation between them, whereas the form factors depend only $q^2/m^2$, so that the $z$ dependence has to drop. Due to this observation, the relations \eqs\nr{FP_SL_correspondence} and \nr{FD_VL_SL_correspondence} impose very strong constraints on ${\cal V}^{L}$ and ${\cal S}^{L}$.

The form factor ${\cal S}^{L}$ receives contributions only from the two non-instantaneous vertex correction diagrams \fig\ref{fig:vertexL}\ref{diag:vertexqbaremL} and \ref{diag:vertexqemL}, and by symmetry one has
\begin{align}
&  {\cal S}^{L} = {\cal S}^{L}_{\ref{diag:vertexqbaremL}} + {\cal S}^{L}_{\ref{diag:vertexqemL}} = {\cal S}^{L}_{\ref{diag:vertexqbaremL}} -\Big(z \leftrightarrow (1\!-\!z) \Big) .
\label{SL_antisym}
\end{align}
From the explicit calculation of diagram \ref{diag:vertexqbaremL} (see section \ref{sec:vertexcor}), one gets
\begin{align}
{\cal S}^{L}_{\ref{diag:vertexqbaremL}} & = 2z \int_{0}^{1} d\xi\, (1\!-\!\xi) \left\{[2z\!-\!1\!-\!z\xi]\; \frac{\Pt^j\, \mathcal{B}^j}{\Pt^2} 
+[z\!-\!1\!-\!z\xi]\; \xi\, \mathcal{B}_0 \right\} + O(D\!-\!4)
\label{SL1_1}.
\end{align}
Note that the expression \eq\nr{SL1_1} for ${\cal S}^{L}_{\ref{diag:vertexqbaremL}}$ is fully finite, both in the UV and at $\xi=0$, which is expected since ${\cal S}^{L}$ is absent at tree level.

Using the Feynman parametrization, one can write $\mathcal{B}_0$ and $\mathcal{B}^j$ as
\begin{align}
\left[ \begin{array}{c}
\mathcal{B}_0  \\
\mathcal{B}^j
\end{array} \right]
& =  \int_{0}^{1} d x \;
\left[ \begin{array}{c}
1  \\
-x\, \Lt^j
\end{array} \right]
\frac{1}{\left[x(1\!-\!x)\Lt^2 +(1\!-\!x) \Delta_1 + x \Delta_2\right]}
\label{Feyn_param}
\, .
\end{align}
Furthermore, using the change of variable $x\mapsto y=\xi + (1\!-\!\xi)x$, these expressions can be simplified into
\begin{align}
\left[ \begin{array}{c}
\mathcal{B}_0  \\
 \frac{\Pt^j\, \mathcal{B}^j}{\Pt^2} 
\end{array} \right]
& = \frac{1}{(1\!-\!\xi)} \int_{\xi}^{1} d y \;
\left[ \begin{array}{c}
1  \\
(y\!-\!\xi)
\end{array} \right]
\frac{1}{\left\{y^2 m^2 + \left(y+\frac{z \xi}{(1\!-\!z)}\right) \left[
(1\!-\!y)\left(\Pt^2 \!+\!\overline{Q}^2 \!+\!m^2\right) +(y\!-\!\xi) \overline{Q}^2 
\right] 
\right\}}
\label{Feyn_param_better}
\, .
\end{align}
Inserting \eq\nr{Feyn_param_better} into \eq\nr{SL1_1} and changing the order of integrations one finds
\begin{equation}
\begin{split}
{\cal S}^{L}_{\ref{diag:vertexqbaremL}} & = 2z  \int_{0}^{1} d y \int_{0}^{y} d\xi\, \frac{
\{[2z\!-\!1\!-\!z\xi]\; (y\!-\!\xi)
+[z\!-\!1\!-\!z\xi]\; \xi\}
}{\left\{y^2 m^2 + \left(y+\frac{z \xi}{(1\!-\!z)}\right) \left[
(1\!-\!y)\left(\Pt^2 \!+\!\overline{Q}^2 \!+\!m^2\right) +(y\!-\!\xi) \overline{Q}^2 
\right] 
\right\}} + O(D\!-\!4)
\\
& = 2z  \int_{0}^{1} d y\, y \int_{0}^{1} d\eta\, \frac{
\{[z\!-\!1\!-\!z\eta y]
+z(1\!-\!\eta)\}
}{\left\{y m^2 + \left(1+\frac{z \eta}{(1\!-\!z)}\right) \left[
(1\!-\!y)\left(\Pt^2 \!+\!\overline{Q}^2 \!+\!m^2\right) +y(1\!-\!\eta) \overline{Q}^2 
\right] 
\right\}} + O(D\!-\!4),
\label{SL1_2}
\end{split}
\end{equation}
where in the second line we have again performed the change of variable $\xi\mapsto \eta = \frac{\xi}{y}$.

Up to this point, we have considered ${\cal S}^{L}_{\ref{diag:vertexqbaremL}}$ in the general kinematics relevant for the LCWF. Let us now impose $\Pt^2= -\overline{Q}^2 -m^2$ in order to study the correspondence with the Pauli form factor. In that case, the integral over $y$ becomes polynomial, and one obtains
\begin{equation}
\begin{split}
\left.{\cal S}^{L}_{\ref{diag:vertexqbaremL}}\right|_{\Pt^2= -\overline{Q}^2 -m^2} 
& =  
z  \int_{0}^{1} d\eta\, \frac{
\{2(2z\!-\!1)\!-\!3z\eta\}
}{\left\{m^2 -z(1\!-\!\eta) \left[1- z(1\!-\!\eta)\right] q^2
\right\}} + O(D\!-\!4)
\\
& = \int_{0}^{z} d\chi\, \frac{
\{z\!-\!2\!+\!3\chi\}
}{\left\{m^2 -\chi (1\!-\!\chi) q^2\right\}} + O(D\!-\!4)
\\
& = \frac{1}{2}\int_{0}^{1} d\chi\, \frac{1}{\left\{m^2 -\chi (1\!-\!\chi) q^2\right\}}
\bigg\{
(z\!-\!2\!+\!3\chi)\; \theta(z\!-\!\chi) 
+ (z\!+\!1\!-\!3\chi)\; \theta(\chi\!-\!1\!+\!z) 
\bigg\}
+ O(D\!-\!4)
\label{SL1_FF_1}
\end{split}
\end{equation}
using the change of variable $\eta \mapsto \chi= z (1- \eta)$, and then symmetrizing the result with respect to 
$\chi \leftrightarrow 1-\chi$. 

Including the contribution of the diagram \ref{diag:vertexqemL} as prescribed by \eq\nr{SL_antisym}, one finally gets%
\begin{align}
\left.{\cal S}^{L}\right|_{\Pt^2= -\overline{Q}^2 -m^2} 
& =  \frac{(2z\!-\!1)}{2}\int_{0}^{1} d\chi\, \frac{1}{\left\{m^2 -\chi (1\!-\!\chi) q^2\right\}}
+ O(D\!-\!4)
 ,
\label{SL_FF}
\end{align}
and hence from \eq\nr{FP_SL_correspondence}
\begin{align}
F_P(q^2/m^2)
& =   \left[\frac{\alpha_s\, C_F}{2\pi}\right]\;   
\int_{0}^{1} d\chi\, \frac{m^2}{\left\{m^2 -\chi (1\!-\!\chi) q^2\right\}}
+ O(D\!-\!4)
 . 
\label{FP_expr_space}
\end{align}
In order to have an expression valid also in the time-like case, one can restore the $i0$ by looking at the relative sign of $q^2$ and $i0$ in the energy denominators at the beginning of the calculation.
Then,
\begin{align}
F_P(q^2/m^2)
& =   \left[\frac{\alpha_s\, \cf}{2\pi}\right]\;   
\int_{0}^{1} d\chi\, \frac{m^2}{\left\{m^2 -\chi (1\!-\!\chi) q^2 - i0\right\}}
+ O(D\!-\!4)
 . 
\label{FP_expr}
\end{align}
This is indeed the known result for the Pauli form factor at one loop in QCD, which is identical to the QED result~\cite{Schwinger:1951nm,Peskin:1995ev} up to the replacement
$\alpha_s\, C_F \leftrightarrow \alpha_{em}\, e_f^2$.

\bibliography{spires}
\bibliographystyle{JHEP-2modlong}

\end{document}